\documentclass[%
 aip,
 amsmath,amssymb,
 reprint,%
]{revtex4-1}

\usepackage{graphicx}
\usepackage{dcolumn}
\usepackage{bm}

\usepackage[utf8]{inputenc}
\usepackage[T1]{fontenc}
\usepackage{mathptmx}
\usepackage{etoolbox}

\usepackage{amsmath, amssymb} 
\usepackage{lineno}           
\usepackage{booktabs}         
\usepackage{multirow}         
\usepackage{float}            
\usepackage{tabularx}         
\usepackage{tikz}             

\newcolumntype{Y}{>{\centering\arraybackslash}X}
\AtBeginDocument{\DeclareMathAlphabet{\mathbf}{OT1}{cmr}{bx}{n}}

\usepackage{xcolor}
\definecolor{newcolor}{rgb}{.8,.349,.1}

\definecolor{darkgreen}{rgb}{0.0, 0.5, 0.0}
\definecolor{darkred}{rgb}{0.55, 0.0, 0.0}

\usepackage{subfig}
\usepackage{makecell}
\usepackage[linesnumbered, ruled]{algorithm2e}
\usepackage{algpseudocode}
\SetKwRepeat{Do}{do}{while}%

\usepackage[utf8]{inputenc}
\usepackage[T1]{fontenc}
\usepackage{CJKutf8}

\usepackage[bottom,norule,hang]{footmisc}
\usepackage{hyperref}         
\hypersetup{
	colorlinks   = true,
	urlcolor     = blue,
	linkcolor    = blue,
	citecolor    = red
}
\newcommand{\blueText}[1]{{\color{black}#1}}
\makeatletter
\def\@email#1#2{%
 \endgroup
 \patchcmd{\titleblock@produce}
  {\frontmatter@RRAPformat}
  {\frontmatter@RRAPformat{\produce@RRAP{*Corresponding Author: {#2}}}\frontmatter@RRAPformat}
  {}{}
}%
\makeatother

\begin{document}

\captionsetup{justification=raggedright,singlelinecheck=false} 

\preprint{AIP/123-QED}

\title[ ]{Predicting unsteady incompressible fluid dynamics with finite volume informed neural network}

\author{Tianyu Li\begin{CJK*}{UTF8}{gbsn}(李天宇)\end{CJK*}}
\affiliation{
School of Computer Science, Sichuan University, Chengdu, 610065, China
}
\altaffiliation[Also at ]{Tianfu Engineering-oriented Numerical Simulation \& Software Innovation Center, Sichuan University, Chengdu, 610207, China}

\author{Shufan Zou\begin{CJK*}{UTF8}{gbsn}(邹舒帆)\end{CJK*}}
\affiliation{College of Aerospace Science and Engineering, National University of Defense Technology, Changsha, 410000, China}

\author{Xinghua Chang\begin{CJK*}{UTF8}{gbsn}(常兴华)\end{CJK*}\textsuperscript{*}}
\affiliation{
School of Computer Science, Sichuan University, Chengdu, 610065, China
}
\email{cxh\_cardc@126.com}

\author{Laiping Zhang\begin{CJK*}{UTF8}{gbsn}(张来平)\end{CJK*}}
\affiliation{
Unmanned Systems Research Center, National Innovation Institute of Defense Technology, Beijing, 100071, China
}

\author{Xiaogang Deng\begin{CJK*}{UTF8}{gbsn}(邓小刚)\end{CJK*}}
\affiliation{
Academy of Military Sciences, Beijing, 100190, China
}

\date{\today}

\begin{abstract}
The rapid development of deep learning has significant implications for the advancement of Computational Fluid Dynamics (CFD). Currently, most pixel-grid-based deep learning methods for flow field prediction exhibit significantly reduced accuracy in predicting boundary layer flows and poor adaptability to geometric shapes. Although Graph Neural Network (GNN) models for unstructured grids based unsteady flow prediction have better geometric adaptability, these models suffer from error accumulation in long-term predictions of unsteady flows. More importantly, fully data-driven models often require extensive training time, greatly limiting the rapid update and iteration speed of deep learning models when facing more complex unsteady flows. Therefore, this paper aims to balance the demands for training overhead and prediction accuracy by integrating physical constraints based on the finite volume method into the loss function of the graph neural network. Additionally, it incorporates a twice-massage aggregation mechanism inspired by the extended stencil method to enhance the unsteady flow prediction accuracy and geometric shape generalization ability of the graph neural network model on unstructured grids. We focus particularly on the model's predictive accuracy within the boundary layer. Compared to fully data-driven methods, our model achieves better predictive accuracy and geometric shape generalization ability in a shorter training time.
\end{abstract}

\maketitle


\section{Introduction}
\label{sec:Introduction}

Computational Fluid Dynamics (CFD) has become an important tool for the analysis of flow phenomena and the prediction of mechanical properties. Traditional CFD methods predict flow fields by iteratively solving a set of discretized partial differential equations. For practical engineering problems, this requires a significant amount of computational resources. Particularly for unsteady viscous flow problems, to more accurately capture the flow characteristics of the boundary layer, the demand for computational resources is much greater compared to solving steady-state problems. With the rapid development of deep learning, the CFD community has also benefited greatly, such as the significant efficiency improvements in flow field prediction using neural network models on pixel-based uniform grids \cite{kochkov2021machine}. However, it still faces the enormous challenges of failing to meet the practical engineering applications and the lack of physical interpretability. In response to the issue of lacking physical interpretability, among many deep neural network-based methods, Physics-Informed Neural Networks (PINNs) that do not require grid discretization \cite{raissi2019physics, rao_physics-informed_2020, jin_nsfnets_2021, li2021physics, wandel_spline-pinn_2022} have shown considerable potential in solving Partial Differential Equations (PDEs), including the Navier-Stokes equations. PINNs integrate physical laws into the neural network within the deep learning framework using automatic differentiation techniques, and they perform well even in cases of scarce or non-existent observable data \cite{arzani2021uncovering}. However, PINNs based on automatic differentiation generate many large tensors during backward propagation through the chain rule \cite{baydin2018automatic}, leading to excessive training overhead when dealing with large-scale problems.

\begin{figure*}
\centering
\begin{minipage}{\linewidth}
    \centering
    \includegraphics[width=1\textwidth]{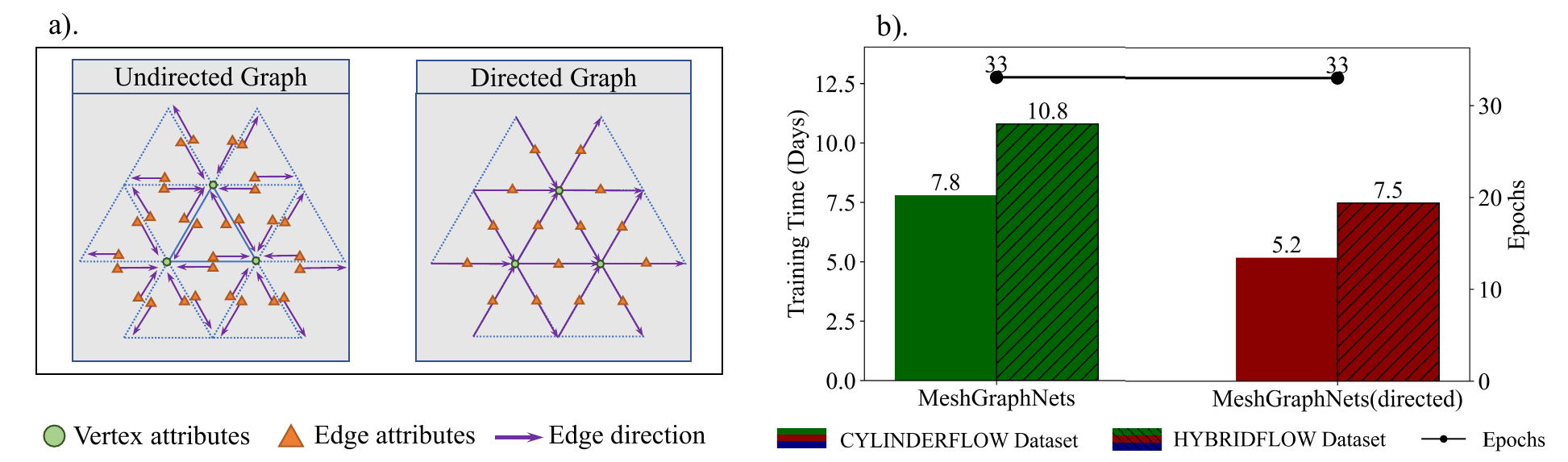}
    \end{minipage}
    \caption{ a). Two different transformation methods for unstructured grids, undirected graph(left) and directed graph(right); b). Training time consumption and epochs used for training. We assess the time-consumption metric of MeshGraphNets on a workstation equipped with a single V100S GPU. All the configurations of two MGN models are followed by \cite{pfaff2020learning}. (Batch size=2, Message Passing Layer=15, Hidden-size=128, total training steps=$10^{8}$), we also define every $6 \times 10^{5}$ training samples the model encountered as a single epoch. Because there are 1000 cases and each case has 600 time steps in both CYLINDERFLOW and HYBRIDFLOW datasets. FVGN and all subsequent models related to it are based on a unidirectional edge setup, or in other words, a directed graph.
}
    \label{intro-rmse-comp}
\end{figure*}

To further reduce the excessive training overhead, PINNs can draw inspiration from the numerical discretization methods of CFD, making them not solely reliant on automatic differentiation tools to obtain the derivatives of physical fields. Adopting numerical differentiation can significantly reduce the computational overhead of solving physical field derivatives compared to automatic differentiation \cite{ranade2021discretizationnet}. Depending on the type of spatial discretization grid used, numerical methods can be divided into Finite Difference Methods (FDM) based on structured grids \cite{wandel2020learning, wandel_teaching_2021, gao_phygeonet_2021}, as well as Finite Volume Methods (FVM) and Finite Element Methods (FEM) suitable for unstructured grids \cite{gao_physics-informed_2022, gao2022finite}. For complex practical engineering problems, unstructured grids offer better geometric flexibility. The Finite Volume Method (FVM) \cite{moukalled2016finite} is one of the most commonly used numerical discretization methods for unstructured grids, and integrating it into neural networks is increasingly gaining attention \cite{praditia_finite_2021, karlbauer_composing_2022, ranade2021discretizationnet}. However, most of the aforementioned research works are limited to verifying effectiveness on uniform pixel-like grids, which often cannot adapt to complex geometries and reduce prediction accuracy near the boundary layer for viscous flow problems. This deviates from the original intention of FVM to adapt to arbitrarily shaped grid cells. Therefore, to enhance the engineering application capabilities of neural network models, it is urgent to expand their handling abilities for non-uniform, unstructured grids.

It is well-known that compared to methods based on Convolutional Neural Networks (CNNs) \cite{chen_towards_2021, wandel2020learning, wandel_teaching_2021, gao_phygeonet_2021, brahmachary2023unsteady}, Graph Neural Networks (GNNs) have a natural adaptability to unstructured grids. GNNs take graph structures composed of nodes and edges as input and learn the feature representation of nodes and edges through a message-passing mechanism \cite{gilmer_neural_nodate, battaglia2018relational}. GNNs, with their adaptability and scalability, are widely applied in various fields, including social network analysis, recommendation systems, and knowledge graphs, where they have shown tremendous potential \cite{zhou_graph_2020}. Therefore, PDE-solving methods based on graph neural networks have become one of the current popular research directions \cite{horie_physics-embedded_2023, li_graph_2022, gao_physics-informed_2022, peng2023physics}. Also, numerous scholars have developed purely data-driven methods based on graph neural networks \cite{sanchez2018graph, sanchez2020learning, pfaff2020learning, seo2019physics, han2022predicting, vaswani_attention_nodate, horie_physics-embedded_2023, chen2021graph, he2022flow}. However, these purely data-driven studies often need to trade off between accuracy and training overhead. Especially when predicting unsteady flow fields \cite{sanchez2020learning, pfaff2020learning, han2022predicting}, they usually require long training times to achieve reliable prediction accuracy. Moreover, as the number of grids or training cases increases, the total training overhead can dramatically rise. At the same time, since purely data-driven methods hardly introduce any physical inductive bias, they lead to weak geometric generalization in most models. These issues significantly hinder the application of graph neural network models in complex and varied practical engineering problems.

Reducing the training overhead of Graph Neural Network (GNN) models can be quite straightforward by reducing the size of the model's hidden space or using fewer training rounds (Fig.\ref{intro-rmse-comp}a). Taking MeshGraphNets (MGN) \cite{pfaff2020learning} as an example, which is a purely data-driven graph neural network model capable of predicting unsteady flow fields. It constructs by tiling multiple GN (short for "Graph Network") blocks\cite{battaglia2018relational}, and encodes unstructured grids into a MultiGraph (undirected graph), as illustrated in Fig.\ref{intro-rmse-comp}a), where physical field-related variables are placed on the vertices of the graph, and the adjacency relationships of grid edges are transformed into the adjacency relationships of the graph, with the undirected graph satisfying permutation invariance \cite{battaglia2018relational}. To reduce training costs, one can convert the bidirectional edge attributes in MeshGraphNets into unidirectional edge attributes, or in other words, transform the undirected graph into a directed graph. At the same time, randomly swapping the direction of each edge in the graph during the training process can maintain permutation invariance. This can significantly reduce the size of edge features, thereby improving training efficiency, as shown in Fig.\ref{intro-rmse-comp}b) and Sec.\ref{Message Passing Process Comparsion}. In particular, for a standard MeshGraphNets model with a hidden space size set to 128, the actual hidden space size for edge attributes is 256. The method of unidirectional edges can still maintain the hidden space size of the edge attributes at 128. However, it's evident that directly reducing the model's hidden space size will inevitably lead to a decrease in the model's final prediction accuracy (see Tab.\ref{tab: cylinder flow result comparison}).

In summary, regarding the issue of physical interpretability, the method combining PINNs with numerical discretization loss functions has achieved a good balance between accuracy and computational cost. However, for the engineering application capability, the uniform grid used at the current stage has considerable limitations. The data-driven GNN adapted for unstructured grids has higher computational costs and poorer physical interpretability. Reducing the computational cost simply by decreasing the size of the hidden space leads to significant accuracy loss, which is unacceptable. Therefore, to reduce computational costs and training difficulty while improving physical interpretability and meeting complex engineering application requirements, this study attempts to introduce a physics-constrained method based on the finite volume method into MeshGraphNets with directed edges, complemented by a twice message aggregation mechanism inspired by extended stencil method\cite{moukalled2016finite} to enhance the performance of directed MeshGraphNets. Particularly, in the message-passing layer of MeshGraphNets, the Node-MLP (Multi-layer Perceptron, i.e., fully connected network) structure is very similar to the convolutional kernels in convolutional neural networks, both featuring shared parameters \cite{battaglia2018relational}. Compared with MeshGraphNets, our Finite Volume Graph Network (FVGN) aggregates edge features to neighboring vertices and then to the cell centers in the message-passing layer, allowing the receptive field of the Cell-MLP in FVGN to increase quadratically (see Sec.\ref{Message Passing Process Comparsion}), thereby significantly improving the final performance of the model.

The aforementioned design not only compensates for the accuracy decline caused by reducing the hidden space size and training epochs but also maintains high training efficiency. This study is roughly divided into 4 parts: Sec.\ref{Problem Setup Methodology} to Sec.\ref{Training Strategy} introduce how to construct the loss function based on FVM and the overall forward process of FVGN, training details, etc. Sec.\ref{Comparison of methods} details the twice message aggregation method and compares it with the message aggregation process of MeshGraphNets. Sec.\ref{Dataset and Data Preprocessing} presents the two datasets used to validate the methods in this study. The final flow field prediction results can be found in Sec.\ref{Results of FVGN and MGN} and Sec.\ref{Simulations of unsteady flow}. Furthermore, the generalization capability of FVGN is verified in Sec.\ref{Generalisation capability}.

\section{Problem Setup \& Methodology}\label{Problem Setup Methodology}
 
In this section, we will present a brief overview of the unsteady incompressible Navier-Stokes (N-S) equations. 
We first describe how the finite volume method is used to transform these differential equations into their integral forms. This transformation is integral to the formulation of a constrained loss function, a critical component in our approach. The integration of the momentum equation, a pivotal step in the Finite Volume Graph Network (FVGN) methodology, is discussed in detail. Furthermore, we explain how we use a graph to represent the unsteady flow field based on unstructured grids. Concluding this section, we present the physics-constrained loss function, a novel aspect of FVGN, which is a significant contribution of our study.

\subsection{Governing Equations \& Graph Representation}\label{Graph Represtation}
In this work, we focus on the 2-dimensional (2D) Incompressible Navier-Strokes (NS) equations, which read, 
\begin{equation}\label{momtem}
\frac{\partial \mathbf{u} }{\partial t} + (u\cdot \bigtriangledown )\mathbf{u}=
-\frac{1}{\rho} \bigtriangledown p+
\nu \bigtriangledown ^{2}\mathbf{u}+\mathbf{f} \qquad \text{in} \quad \Omega,
\end{equation}
\begin{equation}\label{continus_difference}
\bigtriangledown \cdot \mathbf{u} =0 \qquad \text{in} \quad \Omega,
\end{equation}
where, $t$ is the time, $\mathbf{u}(\mathbf{x},t)=(u,v)$  are $x,y$-direction velocity, $p$ is pressure, $\rho$ is the density, and $\mu$ is dynamics viscosity.
$\Omega$ and $\Gamma$ are the fluid domain and the boundary, respectively.
$Re=\dfrac{U_{ref} D_{def}}{\nu}$ is the Reynolds number defined by a characteristic length $D_{def}$, reference velocity $U_{ref}$ and kinematic viscosity ${\nu} = \mu /\rho $.

FVM works for equations in conservation form.
It can be shown via the divergence theorem\cite{moukalled2016finite} that transforms the volume integral of $\phi$ over a single cell into the surface integral over the cell's faces. Therefore we can use conservation laws to form the governing equations in an integral form.
And Eq.(\ref{momtem}) and Eq.(\ref{continus_difference}) can be integrated and applied the divergence theorem lead to formulas as follows,
\begin{equation}\label{NS-integration-continus}
\int_{V}^{} \bigtriangledown \cdot \mathbf{u} dV=\int_{S} \mathbf{u} \cdot \mathbf{n} \mathrm{d} S=0
\end{equation}
\begin{equation}\label{NS-integration-x-mom}
\frac{\partial}{\partial t} \int_{V} u \mathrm{~d} V=-\oint_{S} u \mathbf{u} \cdot \mathbf{n} \mathrm{d} S-\frac{1}{\rho} \oint_{S} p  n_{x} \mathrm{~d} S+\nu \oint_{S} \nabla u \cdot \mathbf{n} \mathrm{d} S 
\end{equation}
\begin{equation}\label{NS-integration-y-mom}
\frac{\partial}{\partial t} \int_{V} v \mathrm{~d} V=-\oint_{S} v \mathbf{u} \cdot \mathbf{n} \mathrm{d} S-\frac{1}{\rho} \oint_{S} p  n_{y} \mathrm{~d} S+\nu \oint_{S} \nabla v \cdot \mathbf{n} \mathrm{d} S
\end{equation}
\begin{equation}\label{eq5}
\mathbf{J} \equiv \mathbf{A}-\mathbf{D}=\mathbf{u}\mathbf{u}-\nu \nabla \mathbf{u}
\end{equation}
Eq.(\ref{NS-integration-continus}) represents the continuity equation, Eq.(\ref{NS-integration-x-mom}) and Eq.(\ref{NS-integration-y-mom}) stands for momentum equation in $x$ and $y$ direction. More specifically, we define flux $\mathbf{J}\equiv\mathbf{A}-\mathbf{D}$, $\mathbf{A}=\mathbf{u}_f\mathbf{u}_f$ and $\mathbf{D}=\bigtriangledown \mathbf{u}$. Meanwhile, in the above integral, the form of the pressure term is obtained by employing the divergence theorem in a broader sense\cite{wu_vortical_2015}, thereby transforming the pressure gradient into the integral form of pressure on the face of the control volume.

\begin{figure*}
\centering
\begin{minipage}{\linewidth}
    
    \centering
    \includegraphics[width=0.95\textwidth]{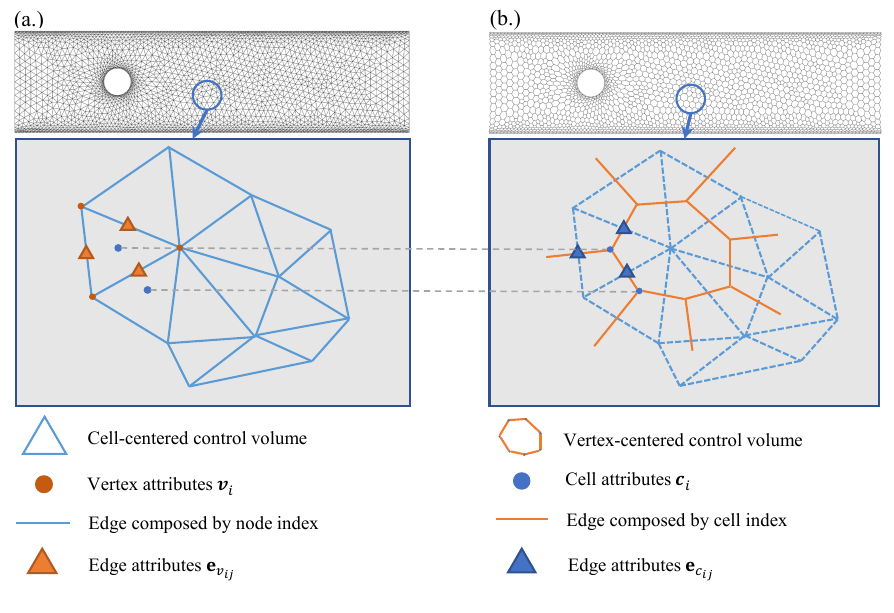}
    \end{minipage}
    \caption{a.) Cell-centered grids; b.) Vertex-centered grids, where the subscript $i, j$ denotes the index number of the cell or vertex }
    \label{graph_representation}
\end{figure*}

The original grid generated by traditional numerical solvers can be viewed as a graph $G_{v}=(V, E_{v})$, where $\mathbf{v}_i \in V$ corresponds to grid nodes and $e_{v_{ij}} \in E_{v}$ corresponds to grid edges (Fig.~\ref{graph_representation} a.). This is known as cell-centered transformation.
And, we can also use the vertex-centered graph $G_{c}=(C, E_{c})$ to represent the grid.  $\mathbf{c}_i \in C$ are grid elements and also where physical variables are placed, while $e_{c_{ij}} \in E_{c}$ are edge attributes subscript by two neighboring cells indices (Fig.~\ref{graph_representation} b.). 
These two types of variable placement methods are commonly used in traditional finite volume methods\cite{moukalled2016finite}. The computational accuracy may vary with the choice of cell-centered or vertex-centered grids. 
However, in this work, we can take advantage of both mesh types simultaneously to enhance the efficiency of graph neural networks during the message-passing process. 
In our model, an assumption, that is $E_{v}$ and $E_{c}$ share the same edge attributes, is made.
Therefore, in the latter section, we only use $E_{c}$ as the edge attribute of both $G_{v}$ and $G_{c}$.
For further details, please refer to Sec.~\ref{Message Passing Process Comparsion}.

\subsection{Physics Constrained Loss}\label{sec Physics Constrained Loss}
As described above, in Eq.(\ref{NS-integration-continus}), we define the continuity equation's flux as $\mathbf{H}=\mathbf{u}_{f}=[u,v]$.
Also in Eq.(\ref{NS-integration-x-mom}) and Eq.(\ref{NS-integration-y-mom}) the advection flux $\mathbf{A}=\mathbf{u}_{f}\mathbf{u}_{f}$ was defined.
For the transient term, we used the first-order time-difference method $\partial \mathbf{u}/\partial t =\left(\mathbf{u}^{t+dt}-\mathbf{u}^{t}\right)/\left(\bigtriangleup t\right)$ to discretize. 
The diffusion flux is defined as $D=[\partial u/\partial x, \partial v/\partial x, \partial u/\partial y, \partial v/\partial y]^{T}$ ,which means $\nabla{u}=[\partial u/\partial x,\partial u/\partial y]^{T},\nabla{v}=[\partial v/\partial x, \partial v/\partial y]^{T}$ in Eq.(\ref{NS-integration-x-mom}) and Eq.(\ref{NS-integration-y-mom}). 
The last pressure flux $p_{f}$ was also involved. 

If there exist numerical solutions, which have ranges of time steps $\mathbf{U}=\{\mathbf{u}_{0}, \mathbf{u}_{1}, \mathbf{u}_{2}...\mathbf{u}_{k}\}$, we could use $\bigtriangleup \mathbf{u}_{k} = \mathbf{u}_{k}-\mathbf{u}_{k-1}$ as the supervision of Eq.\eqref{NS-integration-x-mom} and Eq.\eqref{NS-integration-y-mom}'s ride hand side at each training time steps, eventually, after a few simple steps of moving items, Eq.(\ref{eq physics Constrained Loss continus}) and Eq.(\ref{eq physics Constrained Loss momtem}) were composed. 

\begin{align}
\label{eq physics Constrained Loss continus}
L_{\text{continuity}} = \left\| \sum_{f \in \text{cell}_i} \mathbf{H} \cdot \mathbf{n} \Delta l \right\|_2
\end{align}

\begin{align}
\label{eq physics Constrained Loss momtem}
\begin{split}
L_{\text{momentum}} &= \left\| (\mathbf{u}^{t+dt} - \mathbf{u}^{t}) - (\mathbf{u}_{\text{target}}^{t+dt} - \mathbf{u}_{\text{target}}^{t}) \right\|_2 \\
&= \bigg\| \frac{\Delta t}{\Delta V} \bigg[ -\sum_{f \in \text{cell}_i} \mathbf{A} \cdot \mathbf{n} \Delta l - \frac{1}{\rho} \sum_{f \in \text{cell}_i} p \mathbf{n} \Delta l \\
& \quad + \nu \sum_{f \in \text{cell}_i} \mathbf{D} \cdot \mathbf{n} \Delta l \bigg] - (\mathbf{u}_{\text{target}}^{t+dt} - \mathbf{u}_{\text{target}}^{t}) \bigg\|_2
\end{split}
\end{align}

Eq.\eqref{eq physics Constrained Loss momtem} represents a complete momentum residual in the form of a loss function. $\Delta V$ is the cell area, $\mathbf{n}$ is the normal vector pointing outside of the cell at face. $\Delta l$ is the face area or edge length in 2D.
In practice, we possess the target values for the transient term $\partial \mathbf{u}/\partial t  \approx   \bigtriangleup \mathbf{u}/\bigtriangleup t $, as well as the target values for \textbf{H}, \textbf{A}, and $p$ on the edge of $G_{cell}$.
For example, $\textbf{H} = \mathbf{u}_f$ are directly obtained from the solver's solution at the cell center through interpolation, and the same applies for $\mathbf{A}=\mathbf{u}_f\mathbf{u}_f$ and $p$.
It is worth noting that we transform the diffusion term into a more concise form, denoted as $\mathbf{Q} =\nu \mathbf{D} \cdot \mathbf{n}\bigtriangleup l =\left [  q_x,q_y \right ]^{T}$, which alters the meaning of Eq.\eqref{eq physics Constrained Loss momtem} to let the model learn a compensation term that conforms to the momentum equation for the diffusion term. Thus we no longer necessarily need to employ the Green-Gaussian method or least square method\cite{moukalled2016finite} to obtain the cell's gradient. 
To ensure the accurate calculation of the convective term flux and the pressure term flux at the element faces, we also introduced additional supervision terms, $L_{face_\mathbf{u}}$ and $L_{face_P}$ for the $\mathbf{H}$, $\mathbf{A}$, and $p$ variables on the control volume faces respectively. The interpolation process mentioned above can be implemented by a simple geometric averaging method to interpolate variables at grid vertices or cell centers to the face center. Although this inevitably introduces additional errors, $\bigtriangleup \mathbf{u}_k$ can be a high-precision variable. Therefore, Eq.\eqref{eq physics Constrained Loss momtem} can ensure that the compensation term correctly satisfies the momentum equation.

\begin{figure*}
\begin{minipage}{\linewidth}
    \includegraphics[width=1\textwidth]{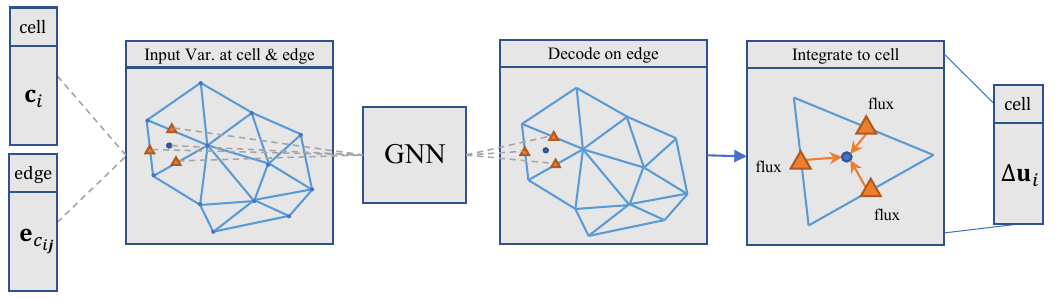}
    \end{minipage}
    \caption{Sketch of Finite Volume Graph Network (FVGN) Forward Process in this study. Cell attribute input: $\mathbf{c}_i = \left [\mathbf{u}_i,\mathbf{\eta}_i \right ]$, where $\mathbf{u}_i=[u,v]$ is the velocity vector, $\mathbf{\eta}$ is a one-hot vector that specified node type; Edge attribute input: $\mathbf{e}_{c_{ij}}=[\mathbf{u}_{i}-\mathbf{u}_{j},\mathbf{x}_{i}-\mathbf{x}_{j},\left \|  \mathbf{x}_{i}-\mathbf{x}_{j} \right \|]$, where $\mathbf{x}$ is grid coordinates; And the decoder output at edge is $\left[ \mathbf{u}_f,p_f, q_x,q_y \right ]$}
    \label{FVGN-foward-process}
\end{figure*}

In our practice, compared to directly predicting $\mathbf{u}^{t+dt}$ or $\bigtriangleup \mathbf{u}$, the loss function in this form can constrain the model's search space using the residual of the physical equation, reduces the prediction error in a single prediction step, thus alleviating the issue of error accumulation during long-term predictions. Therefore, in order to accurately predict the flux at the face of the control volume, we introduce the following two additional supervision items:

\begin{equation}\label{face_U}
L_{face_\mathbf{u}}=\left\|\mathbf{u}_{target\_face}^{t+dt}-\mathbf{u}_{f}\right\|_{2}   
\end{equation}
\begin{equation}\label{face_P}
L_{face_p}=\left\|p^{t+dt}_{target\_face}-p_{f}\right\|_{2}
\end{equation}

Combining the described loss terms, we obtain the following loss function:
\begin{equation}\label{total_loss_function}
    L=\alpha L_{c}+\beta L_{m}+\gamma L_{fu}+\lambda L_{fp}
\end{equation}
where $\alpha, \beta, \gamma,\lambda$ are hyperparameters that weigh the contributions of the different loss terms. 
Note that the weighting coefficients in the loss functions Eq.(\ref{total_loss_function}) play a crucial role in the training process. However, choosing appropriate weights for Eq(\ref{total_loss_function}) is generally very tedious. On the one hand, the optimal values of $\alpha, \beta, \gamma,\lambda$ are problem-dependent and we cannot fix them for different flows. Since We only trained in two datasets, we froze the hyperparameters for all training cases, more details can be found in Sec.\ref{Training Strategy}.

\subsection{Finite Volume Graph Network}\label{Finite Volum Graph Network}
We describe the task as an auto-regressive process, which means the learned model maps the current cell-centered state $ G^t_c(C^{t}, E_{c}) $ to the next state $G^{t+dt}_c(C^{t+dt}, E_c)$ with graph nodes features $ \mathbf{c}_i \in C^{t}$, graph edges features $\mathbf{e}_{ij} \in E_c$.
Our goal is to train a learned model that can extract the changing tendency at the current mesh state, 
which is the advection term flux $\mathbf{A}$, compensation term flux $\mathbf{Q}$, pressure term flux $p$ depicted in Fig.(\ref{FVGN-foward-process}) at every edge at next time step,  and integrate it to acquire the next time mesh state. Therefore, we propose 
\large F\normalsize INITE\large V\normalsize OLUME\large G\normalsize RAPH\large N\normalsize ETWORK (FVGN), which is a graph neural network model with an Encode-Process-Decode architecture \cite{pfaff2020learning,battaglia2018relational,sanchez2020learning},  followed by a space-integrate-layer(SIL in short).
Fig.(\ref{FVGN-foward-process}) shows a visual scheme of the FVGN architecture. In addition, we present the pseudocode for the entire forward process in Sec.\ref{TRAINING PROCESS TRAINING NOISE}.

\noindent$\textbf{Encoder}$ \quad The encoder has 2 parts, cell attribute encoder $\phi^{c}$ and edge attribute encoder $\phi^{e}$.The encoder only encodes the current vertex-centered mesh state $G_c(C^{t}, E_c)$ into a directed graph $\mathcal{G}_{c}(\mathbf{c}^{\prime}, \mathbf{e}_{c_{ij}}^{\prime})$, which mentioned in Sec.(\ref{graph_representation}).
Then each cell center has hidden features $\mathbf{c}^{\prime}$ with the size of 128. Also, the edge features are the same as the cell center dimension. Additionally, the input of $\phi^{c}$ is the velocity field $\mathbf{u}^{t}=[u,v]$ in x-dir and y-dir at time steps $t$ and a one-hot vector that specified node type (boundary node or interior node). The input of $\phi^{e}$ is the difference of each edge neighbor cell \cite{brandstetter2022message,seo2019physics}, edge length, and the difference between the relative coordinates of the verteices forming each edge, which can be expressed as $\mathbf{e}_{c_{ij}}^{t} = \phi^{e}(\mathbf{u}_{i}-\mathbf{u}_{j},\mathbf{x}_{i}-\mathbf{x}_{j},\left \|  \mathbf{x}_{i}-\mathbf{x}_{j} \right \|),$. Both $\phi^{c}$ and $\phi^{e}$ are implemented with a two linear layer with SiLU\cite{elfwing2018sigmoid} activation function MLPs and output with a LayerNorm. Given that we employ a directed graph where each edge has a single direction, either $i \to j$ or $j \to i$, it follows that the evaluation of $\mathbf{u}_{i}-\mathbf{u}_{j},\mathbf{x}_{i}-\mathbf{x}_{j}$ will yield differing outcomes during training and testing. This discrepancy may adversely impact the generalizability of the model, or in other words, this method doesn't satisfy permutation invaiance\cite{battaglia2018relational}. To address the issue of determining the correct subtraction order, we employ a straightforward approach: randomly shuffle the connection direction during training, effectively meaning that the indices $i$ and $j$ are randomly exchanged while maintaining an arbitrary connection direction during testing. Subsequent experiments demonstrate the effectiveness of this method, and at the same time, it significantly reduces the computational cost of the model compared to MeshGraphNets.

\begin{figure*}
\centering
\begin{minipage}{\linewidth}
    \includegraphics[width=1\textwidth]{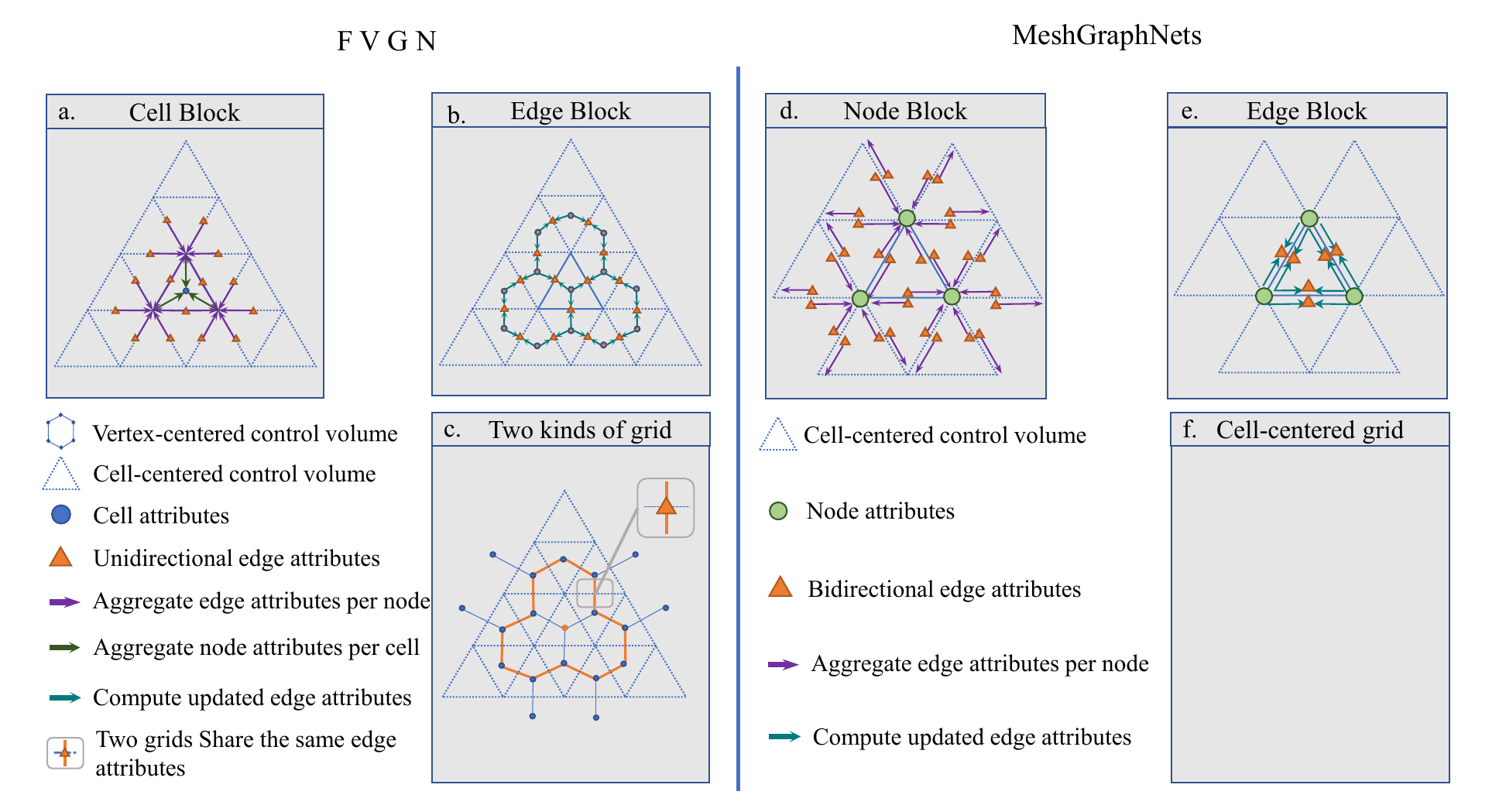}
    \end{minipage}
    \caption{Message Passing Process Comparsion}
    \label{Message Passing Process Comparsion}
\end{figure*}

\noindent$\textbf{Processor}$ \quad The processor consists of M identical message passing layers, which generalize GraphNetwork blocks \cite{sanchez2018graph} hereinafter referred to short for GN blocks. 
Each GN block consists of a cell block and an edge block, depicted in Fig.(\ref{Message Passing Process Comparsion}). Each block contains a separate set of network parameters,
and is applied in sequence to the output of the previous block, updating the encoded edge features $\mathbf{e}_{c_{ij}}^{\prime}$ and cell features $\mathbf{c}_{i}^{\prime}$ to $\overline{\mathbf{e}}_{c_{ij}}^{\prime}$, $\overline{\mathbf{c}}^{\prime}$ respectively by
\begin{equation}\label{processor}
\begin{array}{l}  

  \quad \overline{\mathbf{v}}_{i}^{\prime} \leftarrow \frac{1}{N} \left(
\sum {}  \mathbf{e}_{c_{ij}}^{\prime} \right) 

 \quad \overline{\mathbf{c}}_{i}^{\prime} \leftarrow \phi^{cp}\left(\mathbf{c}^{\prime}_{i}, 
\sum_{v \in cell_i}^{} \overline{\mathbf{v}}_{i}^{\prime},\right)\\ 

   \quad  \quad  \quad  \quad  \quad  \quad \overline{\mathbf{e}}_{c_{ij}}^{\prime} \leftarrow \phi^{ep}\left(\mathbf{e}_{c_{i j}}^{\prime}, 
\overline{\mathbf{c}}^{\prime}_{i}, \overline{\mathbf{c}}^{\prime}_{j}\right) \\ 
 
\end{array} 
\end{equation}
In Eq.\eqref{processor}, $N$ is the number of edges that share the same vertex.
\par
Our model has significant differences in the message aggregating process inside cell blocks compared to MeshGraphNets, which is our model takes message aggregating edge feature per node (as the $\rho^{e\rightarrow v}$ depicted in Fig.\ref{Message Passing Process Comparsion} or Eq.\ref{processor}) first and then passes the information to the cell center(as the $\rho^{v\leftarrow c}$ depicted in Fig.\ref{Message Passing Process Comparsion} or Eq.\ref{processor}), updates current cell attributes through $Cell-MLP$(w.r.t multi-layer perception, akin fully connected network) $\phi^{cp}$. Then, concatenate the two neighboring cell attributes with edge attributes and pass it to the $Edge-MLP$ $\phi^{ep}$ to update edge features.
The reason why we design this process is that if we consider the $Cell-MLP$ analogy of cell center as the convolution kernel of CNN, then for the MeshGraphNets on the cell-centered graph $G_c$, the perceptual field of its $Node-MLP$ can be considered to be linearly increasing (detailed in Fig.\ref{Message Passing Process Comparsion} right).
After adopting our method, each cell-centered $Cell-MLP$ inside FVGN can always get the features of the farther edges than the $Node-MLP$ in MeshGraphNets, and this farther distance can be considered as a squared increase of the perceptual field, which also undoubtedly improves the efficiency of message transmission. Our proposed method is also validated in the later experiments in Sec.\ref{Comparison of methods}.
\par
To more specifically clarify how the process works, we first named our method "twice-message-aggregating" and "one-step-message-aggregating" for MeshGraphNets.
As shown in Fig.(\ref{Message Passing Process Comparsion}), the encoded graph $\mathcal{G}_c(\mathbf{c}^{\prime}_{i}, \mathbf{e}_{c_{ij}}^{\prime})$ enter first message passing layer, which process is summing together all edge features that share the same vertex. The first process will use the adjacency relationship (composed of node indices) stored in $G_{v}(V, E_v)$ Fig.(\ref{graph_representation}. a). The output, denoted as $\overline{v'}_{i}$, will enter the second message aggregating process, which passes the vertex information that makes up a single cell to the cell center, and the second process uses the adjacency relationship stored in $G_{c}(C, E_c)$ Fig.(\ref{graph_representation},b). At the end of the Cell block process, the aggregated features will concatenate $c_{i}^{\prime}$ pass through an $Cell-MLP$, we denoted it as $\phi^{cp}$ to update cell attributes.
Finally, inside the remaining $Edge Block$ will concatenate the two neighboring cells' features with the current edge features, and pass them through another $MLP$, we denoted it as $Edge-MLP$ or $\phi^{ep}$, output $\overline{\mathbf{e}}^{\prime}_{c_{ij}}$.
\\
\\
\noindent$\textbf{Decoder And Spatial Integration Layer}$ \quad The decoder in FVGN was mainly composed of a 2 hidden layer, 128 hidden sizes, SiLU activation function $MLP$ $\phi^{d}$. And $\phi^{d}$ are not applied with LayerNormalization(LN) layers compared with aforementioned $MLP$. The decoder takes on the task of decoding the latent edge features after the message passing layer into true fluxes on edges. It is pretty similar to a traditional numerical solver that finds the higher-order derivative over each control volume face, these derivatives are summed to approximate a specific flux value. So, the decoder decodes the latent $derivatives$ into physical domain $\mathbf{A}, \mathbf{Q}, p$, which stands for $Advection$ $flux$ term, $compensation$ $flux$ term, and $Pressure$ $flux$ term at every control volume's face.
\par
The next step is to integrate and sum these fluxes to the cell center according to the Eq.\eqref{eq physics Constrained Loss momtem}. This process is inspired by the divergence theorem, we found that the conventional Finite Volume method is highly cost in flux calculation. Therefore, the approximation of flux on edge by the neural network will surely achieve the acceleration effect compared to traditional FVM-based solvers. We use such a graph neural network to approximate the physical variables, which means the output of the decoder is an approximation of physics variables on the control volume's face. Then, as stated at the beginning of Sec.\ref{Finite Volum Graph Network}, the right-hand side term of the momentum equation can be obtained by multiplying the physical quantities or gradients at each control volume face by the corresponding surface vector and subsequently summing them up. We refer to this operation as the Spatial Integration Layer (SIL) because such an Integration operation happens during the forward process of our proposed neural network model.  After approximating the right-hand side, we use a simple first-order difference to approximate the time derivative, $\frac{\partial \mathbf{u} }{\partial t}  = \frac{u^{t+dt}-u^{t}}{\bigtriangleup t}+O(u)$. By feeding an initial time step of the velocity field, during training/inferencing, FVGN extracts the flux on every edge/face, integrates them to the cell center, and receives $\bigtriangleup \mathbf{u}$, which can be expressed as

\begin{equation}
\begin{split}
\mathbf{u} ^{t+dt} = \mathbf{u} ^{t} + 
\frac{\bigtriangleup t}{\bigtriangleup  S}
\left.  \left [  -\sum_{f \in cell_{i}}^{}\mathbf{A}\cdot \mathbf{n}\bigtriangleup l \right. \right. \\
\left. \left. -\frac{1}{\rho }(\sum_{f \in cell_{i}}^{}p \mathbf{n}\bigtriangleup l)+ (\sum_{f \in cell_{i}}^{}\mathbf{Q}) \right ]\right.
\end{split}
\end{equation}



According to the traiditional CFD method, our approach is similar to the explicit scheme, which is strictly constrained by the CFL number when taking the stability into consideration. However, as can be seen from equations Eq.\eqref{face_U} and Eq.\eqref{face_P}, both our $A$ and $p$ use the values at time $t+dt$ for constraints. When the dataset was obtained using implicit schemes combined with the substep iteration method, our model could maintain good stability and accuracy through repeated training without taking stability conditions into consideration.
The stability analysis for traditional CFD seems not suitable for our model.
In addition, the stability and accuracy performance of our model for long-term predictions is also verified by subsequent numerical results.

\subsection{Comparison of Methods}\label{Comparison of methods}
In summary, as shown in Fig.\ref{Message Passing Process Comparsion}, we can observe that within the Cell Block, FVGN performs two message aggregation operations. The first operation involves aggregating the edge features to the two adjacent endpoints, while the second operation aggregates the vertex features to the cell centers. This process occurs on a cell-centered graph, utilizing the edge set $E_{v}$ and vertex set $V$ derived from the cell-centered graph $G_v$. Subsequently, in the Edge Block, we concatenate the features of each edge with the features of the two adjacent cells and pass them through the MLP of the Edge Block to compute the updated features of the current edge. This process occurs within a vertex-centered graph $G_c$, where the edge set $E_c$ is composed of the indices of the two neighboring cells of the current edge. We also assume that the cell-centered and the vertex-centered graph share the same edge features. 

In contrast, for MeshGraphNets, message passing only occurs on the cell-centered graph $G_v$, and each Node Block and Edge Block can always only aggregate features of neighboring entities. The "receptive field" of the Node Block can be considered to be increasing linearly. But for FVGN, when viewed from the vertex-centered perspective, the "receptive field" of the Cell Block can be considered to be increasing quadratically. A demonstration can be observed in Fig.\ref{Message Passing Process Comparsion}c., where it is evident that if only a single type of graph, such as the vertex-centered graph, is employed, the cell attributes (\begin{tikzpicture}[baseline=-0.6ex]\node[fill=orange, circle, inner sep=1.5pt] {};\end{tikzpicture} ) in the graph $G_c$ would be unable to acquire information from the (\begin{tikzpicture}[baseline=-0.6ex]\draw[orange, line width=1pt] (0,0) -- (0.5,0);\end{tikzpicture}). However, when utilizing a method that simultaneously employs two types of graph, the (\begin{tikzpicture}[baseline=-0.6ex]\node[fill=orange, circle, inner sep=1.5pt] {};\end{tikzpicture} ) can obtain features from edges beyond a two-hop distance. After multiple iterations of message-passing, each cell attribute like the (\begin{tikzpicture}[baseline=-0.6ex]\node[fill=orange, circle, inner sep=1.5pt] {};\end{tikzpicture} ) is able to access features from neighbors two hops away during each message-passing step. Consequently, our approach can be regarded as a method for aggregating higher-order neighbor messages \cite{chen_measuring_2020}, thereby enhancing the efficiency of message-passing within the model.

\subsection{Training Strategy And Stabilizing Training Method}\label{Training Strategy}
\begin{figure*}
\begin{minipage}{\linewidth}
    \includegraphics[width=1\textwidth]{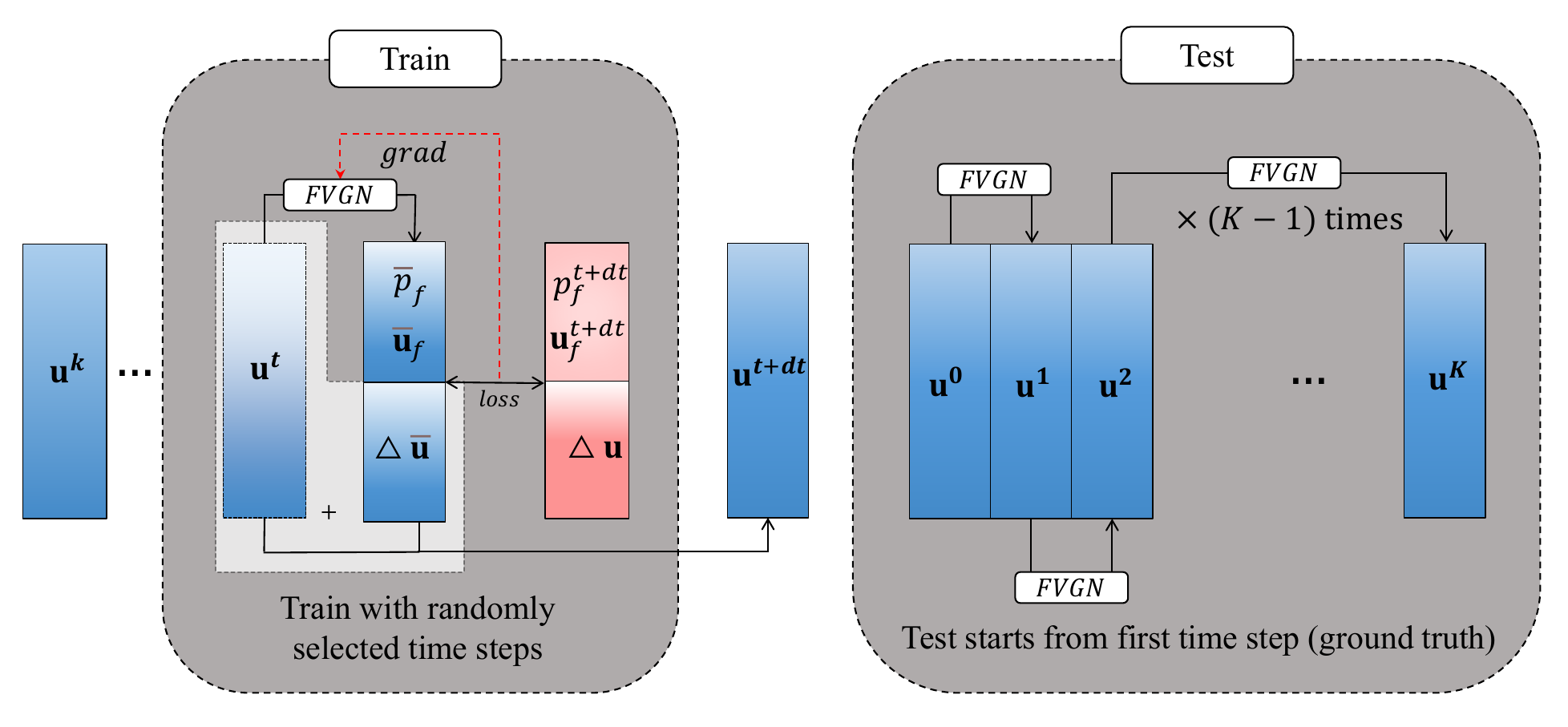}
    \end{minipage}
    \caption{One-step training Strategy, The term $\mathbf{u}_{t}$ denotes the velocity field, while the blue region, denoted as $\bigtriangleup \overline{\mathbf{u}}$, represents the deviation in the velocity field integrated by neural network output. The variables $\overline{p}_{f}$ and $\overline{\mathbf{u}}_{f}$ correspond to the predicted values on the edge. In contrast, the red-colored values stand for the ground truth or actual values. At the same time, it is only necessary to provide the velocity field results after the flow field initialization or the velocity field results before the onset of instability at the initial moment.}
    \label{training strategy}
\end{figure*}

We trained our model in Next-Step mode\cite{pfaff2020learning, sanchez2020learning} and also used the CYLINDERFLOW dataset provided by MeshGraphNets. 
The Next-Step process of FVGN can be described as $\overline{\mathbf{u}}^{t+1} = \mathcal{A} _{NN}(\mathbf{u}^{t})$(depicted in Fig.\ref{training strategy} left). It is pretty straightforward that our model can directly output the flow field at the next time step.
But we also follow the method in \cite{rubanova2021constraint}, that is, at a certain time step, the physical field consists of a tendency showing how the physical field will evolve. 
It is much easier for a Neural Network to extract such tendencies and feed these tendencies to an explicit physical equation (Eq.\eqref{momtem}) to receive the next time state of the input field, compared to those directly predicting the next time state method. Therefore, our method can be considered highly similar to an implicit format, because the right-hand side of Eq.\eqref{momtem} are all composed of variables at $t+dt$ (as shown in Eq.\eqref{face_U} and Eq.\ref{face_P}). This is reflected in our use of values at time $t+dt$ when reconstructing the flux value of the control volume faces. Finally, we also used the noise injection method\cite{pfaff2020learning, sanchez2020learning} to stabilize the rollout process. The noise scale used in the latter sections can be found in Sec.\ref{TRAINING PROCESS TRAINING NOISE}.

\noindent$\textbf{Learning rate and Hyperparameters}$ \quad The weight of each loss in the Eq.\eqref{total_loss_function} is particularly important. We choose hyperparameters of $\alpha=1,\beta=10,\gamma=1,\lambda=1$ for all the data sets used in this paper, 
with an initial learning rate of $1\times10^{-3}$ and a total number of training rounds of 14k (We define the following: For every accumulation of 1,000 training samples, one training rounds is completed within the model's training process. It should be noted that its definition differs from that of an epoch, where an epoch is defined as one epoch for every 600,000 time-step samples encountered by the model.).
Additionally, When the training round reaches 21\% (i.e., the 3000th round), the learning rate decreases to $1\times10^{-4}$. Then, when the training epoch reaches 50\%, we initiate an exponential decay of the learning rate until it reaches the minimum learning rate of $1\times10^{-6}$. We determine the parameters in this way because for a fully data-driven approach, the results of the numerical solver must be incompressible, so we can directly set $\alpha$ to 0 for the continuity equation, and because the momentum equation plays the most important role in the autoregressive prediction process of the whole network, we adjust the weight of the momentum equation to be very large, and the loss on the control volume face is the direct prediction value attached after each forward therefore, we can interpret the above process as the velocity-pressure field on the control volume face is determined by the momentum equation. In Sec.\ref{tab: hybrid dataset result comparison}, we will discuss the detail of the rationale for setting the parameter $\alpha=0$ . Throughout the subsequent sections, unless otherwise specified, we employed the aforementioned hyperparameters to configure and train the FVGN model.
\begin{table*}
    \caption{Dataset used for training comparison}
    \label{tab: dataset comparison}
    \begin{ruledtabular}
    \begin{tabular}{ccccccccc}
    \textbf{Dataset} &
      \textbf{\begin{tabular}[c]{@{}c@{}}Case\\ amount\end{tabular}} &
      \textbf{\begin{tabular}[c]{@{}c@{}}Time\\ steps\end{tabular}} &
      \textbf{\begin{tabular}[c]{@{}c@{}}Elements\\ (avg.)\end{tabular}} &
      \textbf{\begin{tabular}[c]{@{}c@{}}Geo.\\ shape\end{tabular}} &
      \textbf{\begin{tabular}[c]{@{}c@{}}Sample\\ ratio\end{tabular}} &
      \textbf{\begin{tabular}[c]{@{}c@{}}Attack \\ angle\\ (start,\\ step,\\ end)\end{tabular}} &
      \textbf{\begin{tabular}[c]{@{}c@{}}Re\\ number \\range\end{tabular}} &
      \begin{tabular}[c]{@{}c@{}}$\mathbf{\bigtriangleup  t} $\\  s\end{tabular} \\ \hline
    \begin{tabular}[c]{@{}c@{}}CYLINDERFLOW\\ \end{tabular} &
      1000 &
      600 &
      \begin{tabular}[c]{@{}c@{}}about \\ 3500\end{tabular} &
      \begin{tabular}[c]{@{}c@{}}cylinder \\ (only)\end{tabular} &
      1 &
      - &
      \textless{}10,250\textgreater{} &
      0.01 \\
    \begin{tabular}[c]{@{}c@{}}HYBRIDFLOW\\ \end{tabular} &
      1000 &
      600 &
      \begin{tabular}[c]{@{}c@{}}about \\ 7500\end{tabular} &
      \begin{tabular}[c]{@{}c@{}}cylinder, \\ airfoil, \\ rect.\end{tabular} &
      8:1:1 &
      \begin{tabular}[c]{@{}c@{}}(-30°, \\ 15°, \\ 30°)\end{tabular} &
      \textless{}200,1000\textgreater{} &
      0.01
    \end{tabular}
    \end{ruledtabular}
\end{table*}

\section{Dataset and Data Preprocessing}\label{Dataset and Data Preprocessing}

In this study, we used two datasets to train and evaluate the performance of the FVGN model. The first dataset is the CYLINDERFLOW dataset from \cite{pfaff2020learning}. This dataset consists of 1,000 cases, each containing 600 time steps of solver results. The dataset features only a single geometry type - a cylinder but includes various Reynolds numbers, cylinder sizes, and cylinder positions. The Reynolds number range for the training set is \{10, 230\}. Consequently, this dataset includes both steady-state cases and unsteady cases, with a ratio of approximately 3:2. And the dataset's initial moment at time $t=0$ is the result of the solver performing only the initial calculation of the flow field. Further details can be found in the first row of Tab.\ref{tab: dataset comparison}.

Additionally, we introduced another HYRBRIDFLOW dataset. Like the first dataset, the HYRBRIDFLOW dataset also consists of 1,000 cases, each containing 600 time steps of solver results. While creating this dataset, we exported the results from time steps [15,614] using the solver to form the dataset. Essentially, cases at the $15th$ time step in the HYBRIDFLOW dataset are at a critical state just before the flow becomes unstable. The HYRBRIDFLOW dataset includes not only cylindrical geometries but also airfoils and rectangular geometries, with a ratio of approximately 8:1:1. The airfoil cases also feature different angles of attack, ranging from <$-30^\circ, 30^\circ$> and with only shape of NACA0012. 

The Reynolds number range of the HYRBRIDFLOW dataset is between 200 and 1,000, with an average cell count between <6,000, 8,000> for the cases. This dataset, compared to the previous cylindrical flow dataset, has a significantly broader Reynolds number range and increased mesh density, thus posing greater learning challenges for the model. We expect the FVGN to achieve better performance on the HYRBRIDFLOW dataset while maintaining a consistent total number of learnable parameters with MeshGraphNets. Specific details about the HYRBRIDFLOW dataset can be found in the second row of Tab.\ref{tab: dataset comparison}.

The Tab.\ref{tab: dataset comparison} presents several important metrics for the two distinct datasets. The CYLINDERFLOW dataset (indicated in the first row of the table), exclusively comprises cases involving a cylindrical obstacle without the inclusion of any other geometrical shapes. However, the position, size, and inflow velocity (i.e., Reynolds number) of the cylindrical computational domain vary randomly within a certain range. On the other hand, the HYBRIDFLOW dataset(indicated in the second row of the Tab.\ref{tab: dataset comparison}) encompasses cases with three different geometrics as obstacles, namely cylinder, NACA0012 airfoil, and rectangular column. Simultaneously, this dataset also contains various combinations of position, size, and Reynolds number parameters, akin to the CYLINDERFLOW dataset. Moreover, for airfoil cases, the angle of attack is subject to variation within a certain range, adding another layer of complexity.

\noindent$\textbf{Data preprocessing}$ \quad In the entire training process of FVGN, numerous parameters related to mesh properties are employed, such as cell area, the length of cell edges or face areas, and the unit normal vectors of each control volume face. All these relevant numerical values needed in the finite volume method calculations are precomputed during the dataset generation process. Therefore, the data preprocessing of FVGN can be roughly divided into two steps. The first step is to compute the area, edge length, and unit normal vector of each cell from the original cell-centered graph $G_{v}(V, E_{v})$ (this study only involves triangular meshes). Then, the cell-centered graph $G_{v}(V, E_{v})$ is used to compute and generate the adjacency relationship of neighboring cells within vertex-centered graph $G_{c}(C, E_{v})$, along with edge lengths (face areas), cell areas, and $G_{c}(C,E_c)$ unit normal vectors, are written into the dataset before training. The second step involves real-time statistical analysis of the mean and variance of the features on the vertex and edge in $G_c$ input to the neural network during the training process. The obtained mean and variance values are then applied to normalize the input features. Similarly, for each training step, the mean and variance of the input features are accumulated through serval training steps. The introduction of this physics-informed feature normalization\cite{pfaff2020learning, thuerey_deep_2020} can significantly improve the prediction accuracy of the model. And finally, the normalization process on the flow field can be expressed as $\mathbf{u}^{ \prime }=\frac{\mathbf{u}-\overline{\mathbf{u}}}{\hat{\mathbf{u}}}$,$\hat{\mathbf{u}}$ stands for the variance of velocity field $\mathbf{u}$,$\overline{\mathbf{u}}$ stands for the mean of the velocity field, and $p^{ \prime }=\frac{p-\overline{p}}{\hat{p}}$, $\hat{p}$ stands for the variance of the pressure field, $\overline{p}$ stands for the mean value of the pressure field.

\begin{figure*}
\begin{minipage}{\linewidth}
    \includegraphics[width=1\textwidth]{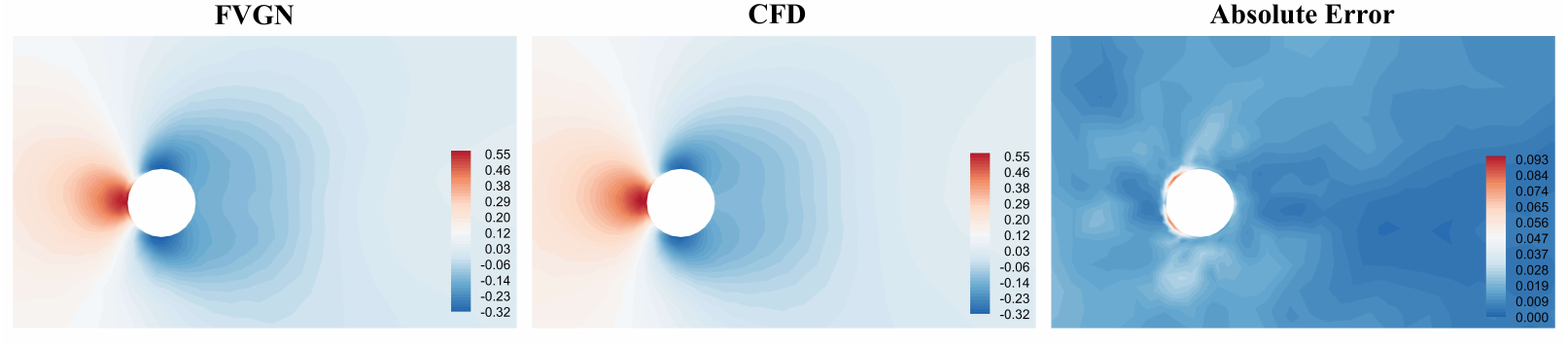}
    \end{minipage}
    \caption{Steady state cylinder flow Re=41, pressure filed prediction at the $t=600$ time step}
    \label{steady_cylinder_field}
\end{figure*}

\begin{figure*}
\begin{minipage}{\linewidth}
    \includegraphics[width=1\textwidth]{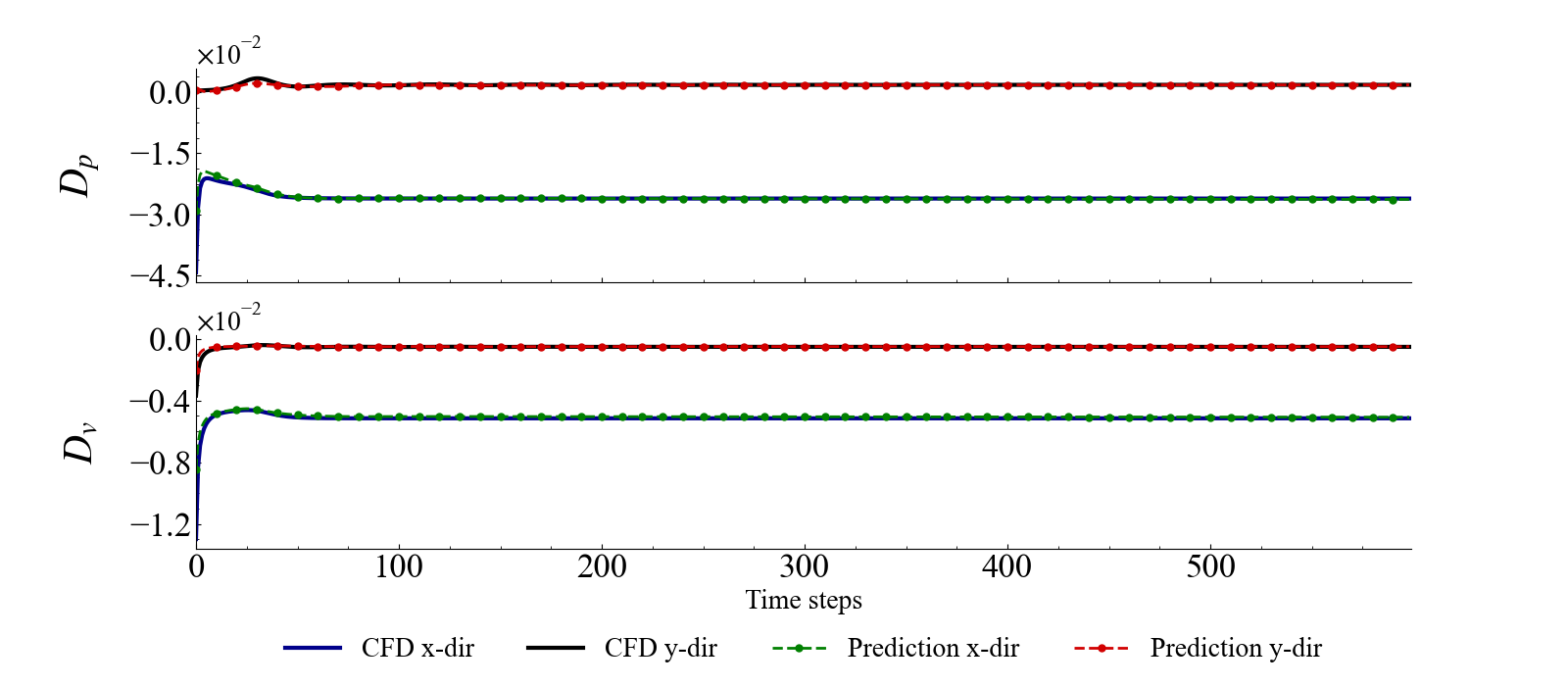}
    \end{minipage}
    \caption{Steady state cylinder flow Re=41, Viscousity Force $D_v = \int_{\partial B } \tau_{w} \, dS
$ and Pressure Force $D_p = \int_{\partial B } p \mathbf{n} \, dS
$ prediction Comparison  over 600 rollout time}
    \label{steady_cylinder_pressure_force}
\end{figure*}

\section{Results }\label{Results of FVGN and MGN}
we compare the results with Relative Mean Square Error, referred to as $\operatorname{RMSE}\left(\hat{\mathbf{u}}^{\text{prediction}}, \mathbf{u}^{\text{truth}}\right) = \frac{\sum\left(\hat{u}_{i}^{\text{prediction}} - u_{i}^{\text{truth}}\right)^{2}}{\sum\left(u_{i}^{\text{truth}}\right)^{2}}$. The Tab.\ref{tab: cylinder flow result comparison} and Tab.\ref{tab: hybrid dataset result comparison} present the test results of FVGN on limited training data and encoding only unidirectional edge features. In the column for training time consumption, we also display the training time overhead of MeshGraphNets on the two training sets. The implementation code is derived from the \href{GitHub homepage}{https://github.com/deepmind/deepmind-research/tree/master/meshgraphnets}, and we maintain the same hyperparameters as MeshGraphNets, such as 15 layers of message passing, a 128-size latent vector, and $10^{6}$ training steps. We modified the output of MeshGraphNets based on the original code so that it can predict the values of the pressure field(the original code can be only used to predict the velocity field). The modified MeshGraphNets code can be found at \href{GitHub homepage}{https://github.com/Litianyu141/My-CODE}.

\subsection{Simulations of Steady flow}\label{Simulations of steady flow}
In this section, we primarily showcase the application of FVGN in predicting steady flow fields under unsteady processes. Moreover, the following two examples are derived from model training and testing conducted on the CYLINDERFLOW dataset. It's worth noting that on this dataset, FVGN only utilized data from the first 300 time steps, whereas both MGN and MGN (directed) employed the full 600 time steps for training.
\begin{table}[H]
\caption{Result comparison in all rollout time steps of steady cylinder flow, The values of $C_l$ and $C_d$ can be derived from the following formulas $C_l = L / (0.5 \times \rho \times V^{2} \times A)$, $C_d = D / (0.5 \times \rho \times V^{2} \times A)$}
\label{tab: steady cylinder flow result comparison}
\centering
\begin{ruledtabular}
\begin{tabular}{ccc|cc}
\multirow{2}{*}{Model} &
  \multicolumn{2}{c|}{\begin{tabular}[c]{@{}c@{}}RMSE$\times10^{-3}$\\ \textless{}0,600\textgreater{}\end{tabular}} &
  \multirow{2}{*}{$C_d$} &
  \multirow{2}{*}{$C_l$} \\
                                                         & UV             & P              &        &        \\ \hline
FVGN                                                     & \textbf{0.063} & 6.098          & 1.537 & -0.165 \\
\begin{tabular}[c]{@{}c@{}}MGN\\ (directed)\end{tabular} & 0.511          & 10.632         & 2.282  & -0.139 \\
MGN                                                      & 0.295          & \textbf{6.928} & 1.581  & -0.172 \\ \hline
CFD                                                      & -              & -              & 1.522 & -0.167
\end{tabular}
 \end{ruledtabular}
\end{table}

In this case study, the Reynolds number is 41, and the number of mesh cells is 3732. As can be seen from Fig.\ref{steady_cylinder_field}, for a steady flow field, it has a simpler flow field structure, thus the pressure prediction error of FVGN is mainly concentrated near the boundary layer. Moreover, as seen from Fig.\ref{steady_cylinder_pressure_force}, in this steady flow situation, FVGN is able to maintain very stable prediction results over a long period. At the same time, as seen from Tab.\ref{tab: steady cylinder flow result comparison}, it has a lower RMSE and more accurate lift and drag coefficient predictions compared to MGN and MGN (directed). However, it is important to note that although the pressure field prediction error of FVGN is mainly concentrated on the surface of the cylinder, which to some extent affects the accuracy of the lift and drag coefficient predictions, this does not affect the overall long-term prediction error of the model. In other words, the errors at the boundary layer do not affect the global prediction accuracy through message passing. Overall, for steady cases, the aforementioned models are able to achieve good results. However, considering that even on the CYLINDERFLOW dataset, the average number of grid cells is only around 3500, and there is no significant difference in the number of parameters between FVGN and MGN, FVGN still achieves higher prediction accuracy. This provides effective proof of the validity of the methods proposed in this paper.

\begin{figure*}
\centering
\begin{minipage}{\linewidth}
    \centering
    \includegraphics[width=1\textwidth]{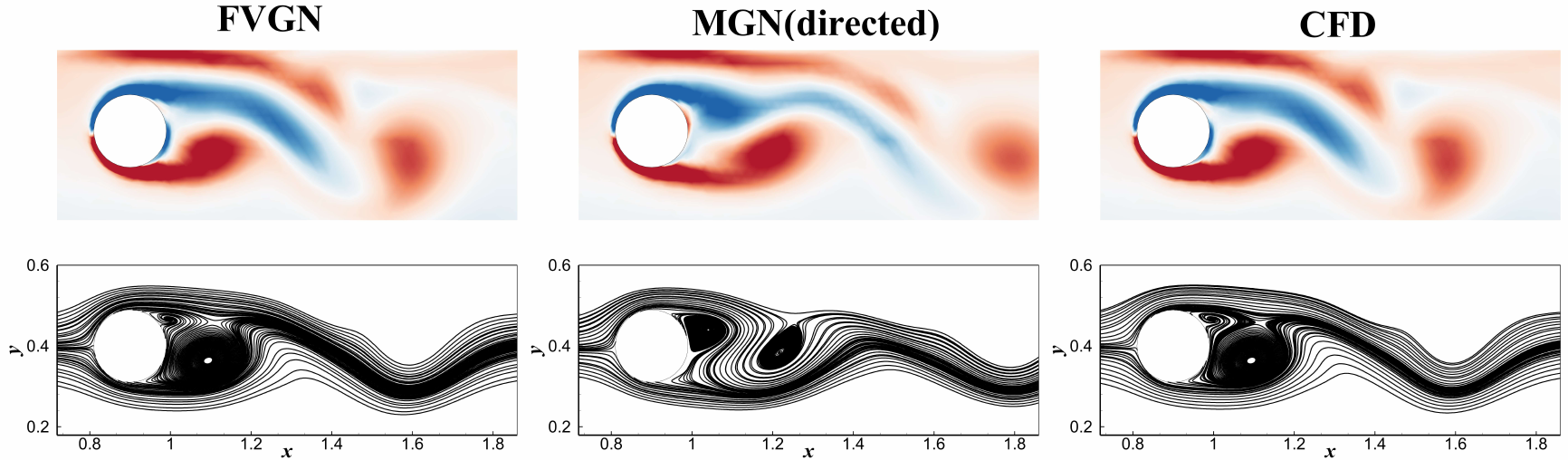}
    \end{minipage}
    \caption{Cylinder, Re=320, at t=400, Comparison between FVGN and MGN(directed) in Z-vorticity and streamline. The value range of Z-vorticity is [50,-50]. As can be seen above the hidden size reduced MGN(directed) got the wrong vortex shedding phase}
    \label{unsteady_cylinder_Z-Vorticity_re=320}
\end{figure*}

\begin{figure*}
\begin{minipage}{\linewidth}
    \includegraphics[width=0.9\textwidth]{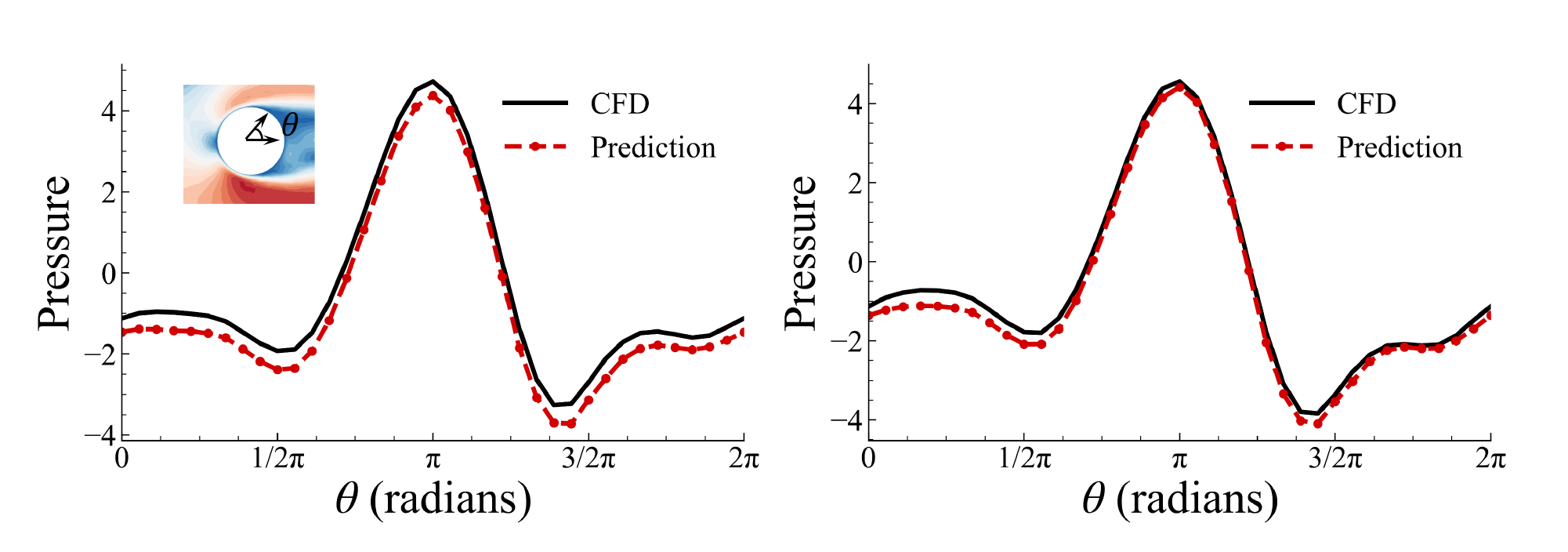}
    \end{minipage}
    \caption{Cylinder, Re=320, left: Pressure distribution of cylinder surface at t=400; right: Pressure distribution of cylinder surface at t=600}
    \label{unsteady_cylinder_pressure_distribution}
\end{figure*}

\begin{figure*}
\begin{minipage}{\linewidth}
    \includegraphics[width=0.9\textwidth]{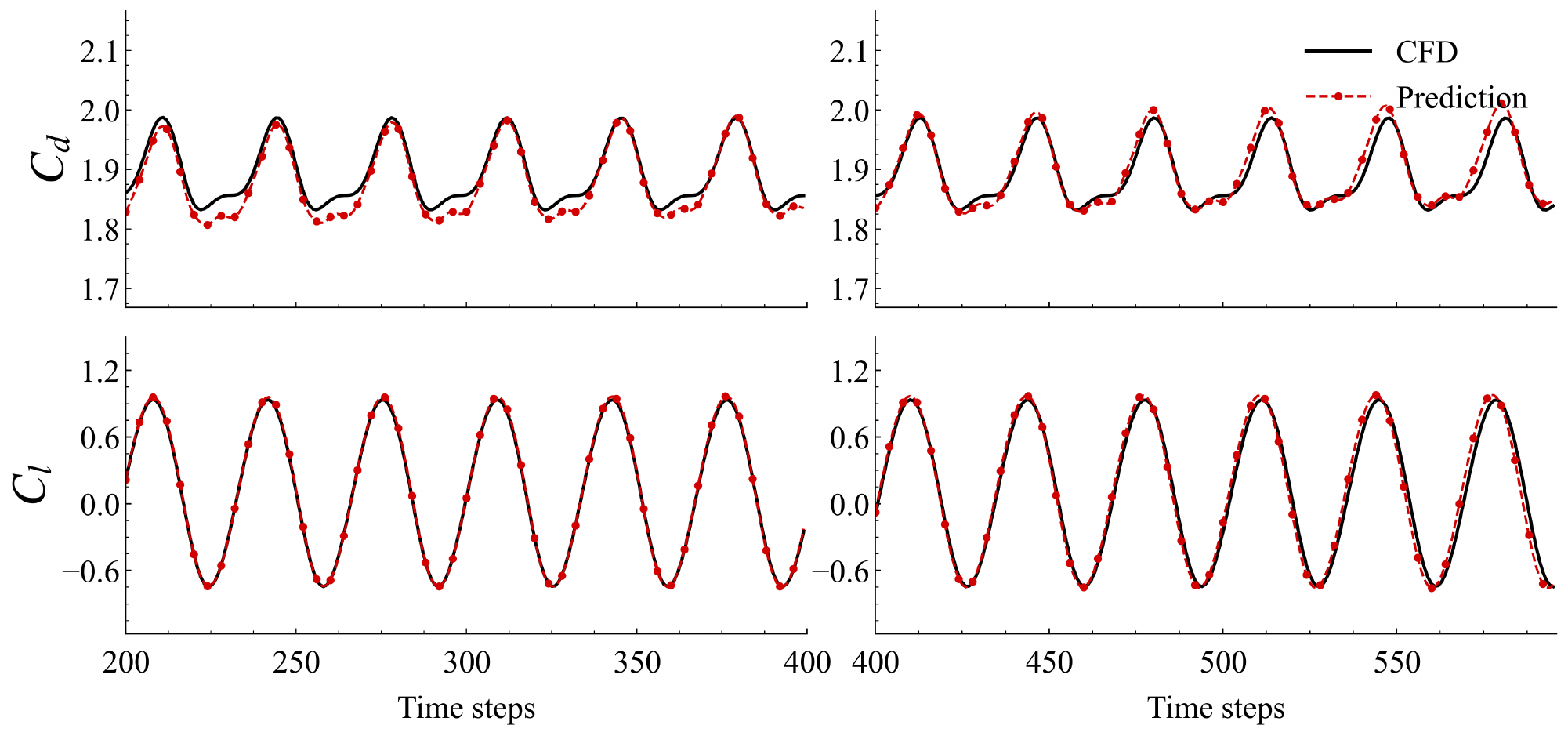}
    \end{minipage}
    \caption{Cylinder, Re=320 cylinder, Variation of $C_l$ and $C_d$ Curves at cylinder surface within the Time Steps <200, 600>}
    \label{unsteady cylinder 0 CL CD pressure distrubutaion}
\end{figure*}

\begin{figure*}
    \includegraphics[width=1\textwidth]{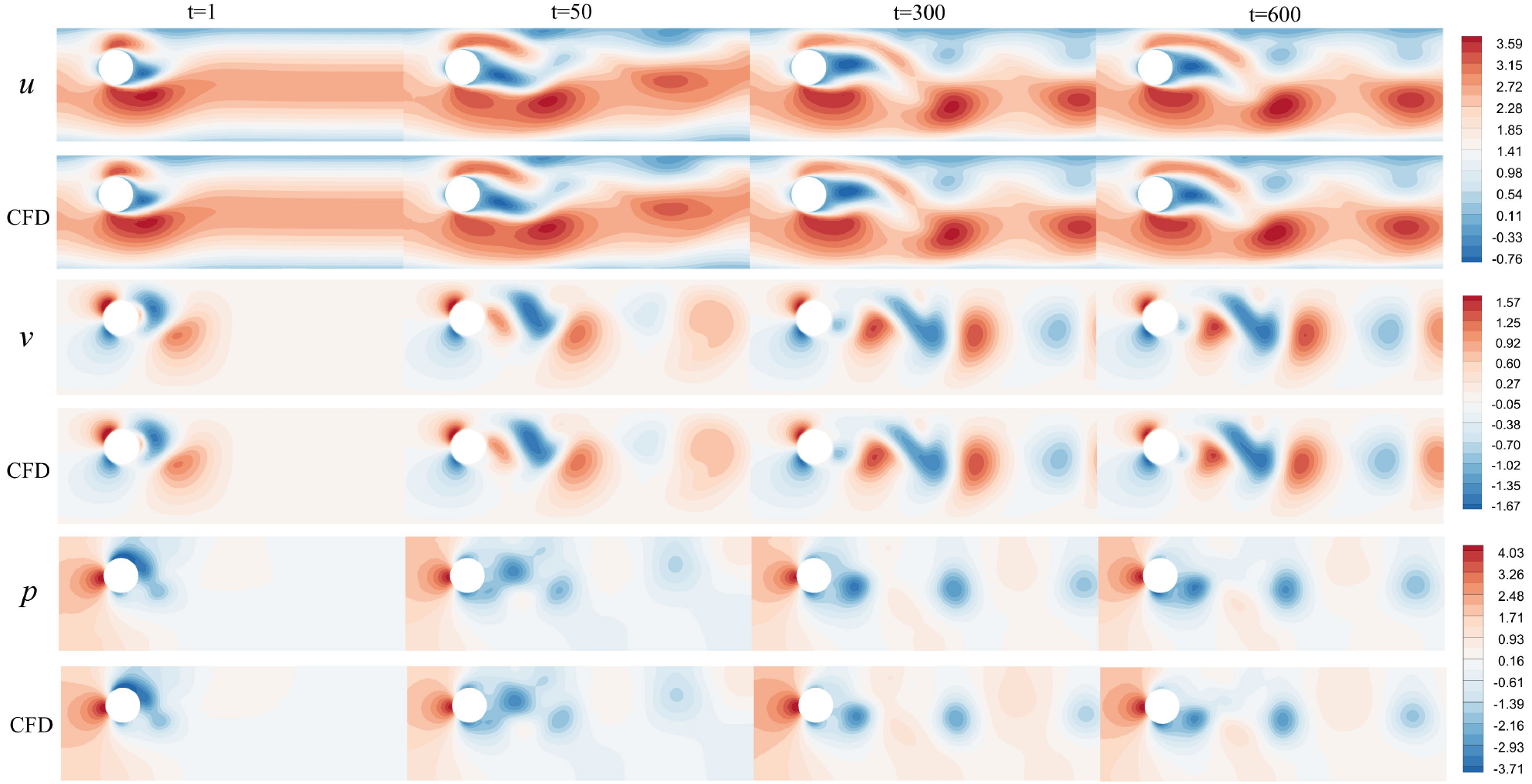}
    \caption{Cylinder, Re=320, Flow field predicted by FVGN}
    \label{unsteady cylinder re=320}
\end{figure*}

\begin{figure*}
\begin{minipage}{\linewidth}
    \includegraphics[width=1\textwidth]{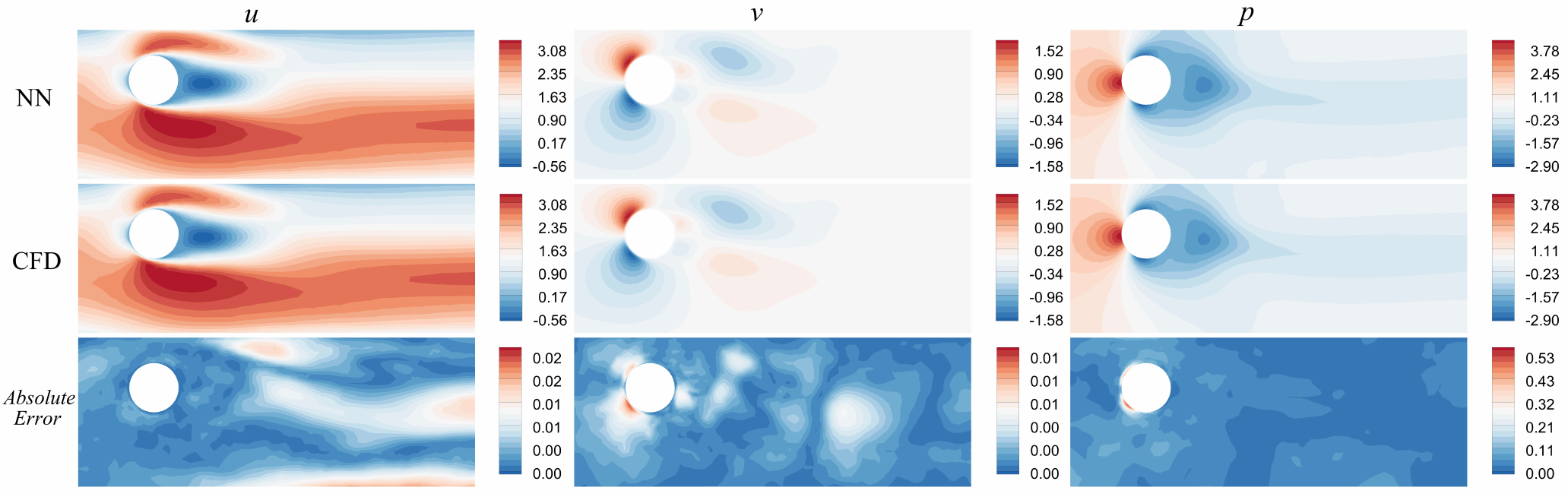}
    \end{minipage}
    \caption{Cylinder, Re=320, Comparison between FVGN and Truth in time-avg field, we use solution time step range of <200,600> to calculate time-averaged field, And Error was also calculated in $Absolute \ Error=\left | \overline{\hat{\mathbf{u}}}-\overline{\mathbf{u}^{GT}}\right | $}
    \label{unsteady cylinder avg re=450}
\end{figure*}

\subsection{Simulations of unsteady flow}\label{Simulations of unsteady flow}
In this section, we show three test results of training FVGN and Meshgraphnets with the HYBRIDFLOW dataset. FVGN on this dataset has a large improvement in both the accuracy of the final test prediction results and the training time consumption compared to MeshGraphNets. Although this dataset contains a wider Relnoyds number, more divergent obstacle geometry, and denser mesh compared to the CYLINDERFLOW dataset. Most importantly, all three test cases in the latter section were tested using an FVGN model state that was trained only on the first 400 time-steps data and a MGN model state that was trained on the full 600 time-steps data in the HYBRIDFLOW training set.

\subsubsection{Two-dimensional Cylinder Flow}\label{Two-dimensional cylinder flow}
The following test case with a Reynolds number (Re=320) and cylinder position combination is not present in the training set. In fact, 80\% of the training set samples in HYBRIDFLOW dataset are cylinder cases, so we mainly focused on testing the model's performance under conditions similar to the training set in this test case. We provided not only the Relative Mean Square Error (RMSE) for both types of models but also compared the predictive accuracy of FVGN and MGN in terms of lift and drag coefficients. The lift coefficient $C_l$ and drag coefficient $C_d$ can be derived from the following formulas: $C_l = L / (0.5 \times \rho \times V^{2} \times A)$, $C_d = D / (0.5 \times \rho \times V^{2} \times A)$, where the vertical lift $L$ and horizontal drag $D$ can be obtained by integrating the pressure and skin friction on the cylinder surface, and the parameter $A$ represents the diameter of the cylinder, serving as a reference length.
\begin{table}[H]
    \centering
    \renewcommand{\arraystretch}{1}
    \caption{Result comparison in all rollout time steps of unsteady cylinder flow, The values of $C_l$ and $C_d$ can be derived from the following formulas $C_l = L / (0.5 \times \rho \times V^{2} \times A)$, $C_d = D / (0.5 \times \rho \times V^{2} \times A)$}
    \label{tab: result comparison cylinder flow}
    \begin{ruledtabular}
        \scalebox{1}{
        \begin{tabular}{ccccc|cc}
        \multirow{2}{*}{Model} &
          \multicolumn{2}{c}{\begin{tabular}[c]{@{}c@{}}RMSE $\times10^{-3}$\\ \textless{}0,400\textgreater{}\end{tabular}} &
          \multicolumn{2}{c|}{\begin{tabular}[c]{@{}c@{}}RMSE $\times10^{-3}$\\ \textless{}0,600\textgreater{}\end{tabular}} &
          \multirow{2}{*}{$C_d$} &
          \multirow{2}{*}{$C_l$} \\
                                                                 & UV             & P               & UV             & P               &       &        \\ \hline
        FVGN                                                     & \textbf{0.134} & \textbf{11.825} & \textbf{0.166} & 17.064          & 1.883 & 0.122  \\
        \begin{tabular}[c]{@{}c@{}}MGN\\ (directed)\end{tabular} & 10.667         & 71.673          & 26.796         & 143.103         & 2.095 & -1.761 \\
        MGN                                                      & 0.763          & 12.233          & 0.919          & \textbf{12.497} & 1.892 & 0.115  \\ \hline
        CFD                                                      & -              & -               & -              & -               & 1.893 & 0.114 
        \end{tabular}
    }
    \end{ruledtabular}
    \renewcommand{\arraystretch}{1}
\end{table}

We not only compared the transient contour diagrams at various moments (Fig.\ref{unsteady cylinder re=320}) but also calculated the time-averaged field of the last 400 time steps (Fig.\ref{unsteady cylinder avg re=450}). From Tab.\ref{tab: result comparison cylinder flow}, it can be seen that within the first [200,400] timesteps, FVGN has a lower RMSE for both the velocity field (UV) and pressure field predictions, demonstrating its accuracy in capturing complex unsteady separated flows around the cylinder. Notably, when extending the rollout timestep length to 600, MeshGraphNets exhibited the highest precision in pressure prediction, reaching $12.497 \times 10^{-3}$. Moreover, its error did not significantly increase compared to the first 400 steps, making MeshGraphNets' predictions of lift and drag coefficients almost identical to CFD results. And, as seen from Fig.\ref{unsteady_cylinder_Z-Vorticity_re=320}, FVGN more accurately captured the phase of vortex shedding under unsteady flow, whereas the phase predicted by MGN(directed) did not align with CFD results, explaining the large relative mean square error for MeshGraphNets(directed) show in Tab.\ref{tab: result comparison cylinder flow}. This error includes not only the amplitude error of the predicted physical field but also the phase error of vortex shedding. In Fig.\ref{unsteady_cylinder_pressure_distribution}, we can see that the accuracy of FVGN's pressure distribution prediction at the 400th moment is lower than that at the 600th moment. Additionally, from the drag curve changes in Fig.\ref{unsteady cylinder 0 CL CD pressure distrubutaion}, it appears that the model becomes more accurate as the prediction timestep increases. However, as seen from the lift curve, starting from the 500th timestep, the frequency of the model's prediction results changes, leading to the illusion of increasing accuracy. This phenomenon demonstrates that in simulating unsteady flow with deep learning models, one should not rely solely on RMSE and contour plots for performance comparison. It's challenging to determine whether the model's errors are due to amplitude predictions or predictions of shedding frequency or phase based on the physical field differences at specific timesteps. And, the lift and drag curves provide a clearer judgment of the aspects in which the deep learning model is erring.

\begin{figure*}
\begin{minipage}{\linewidth}
    \includegraphics[width=0.9\textwidth]{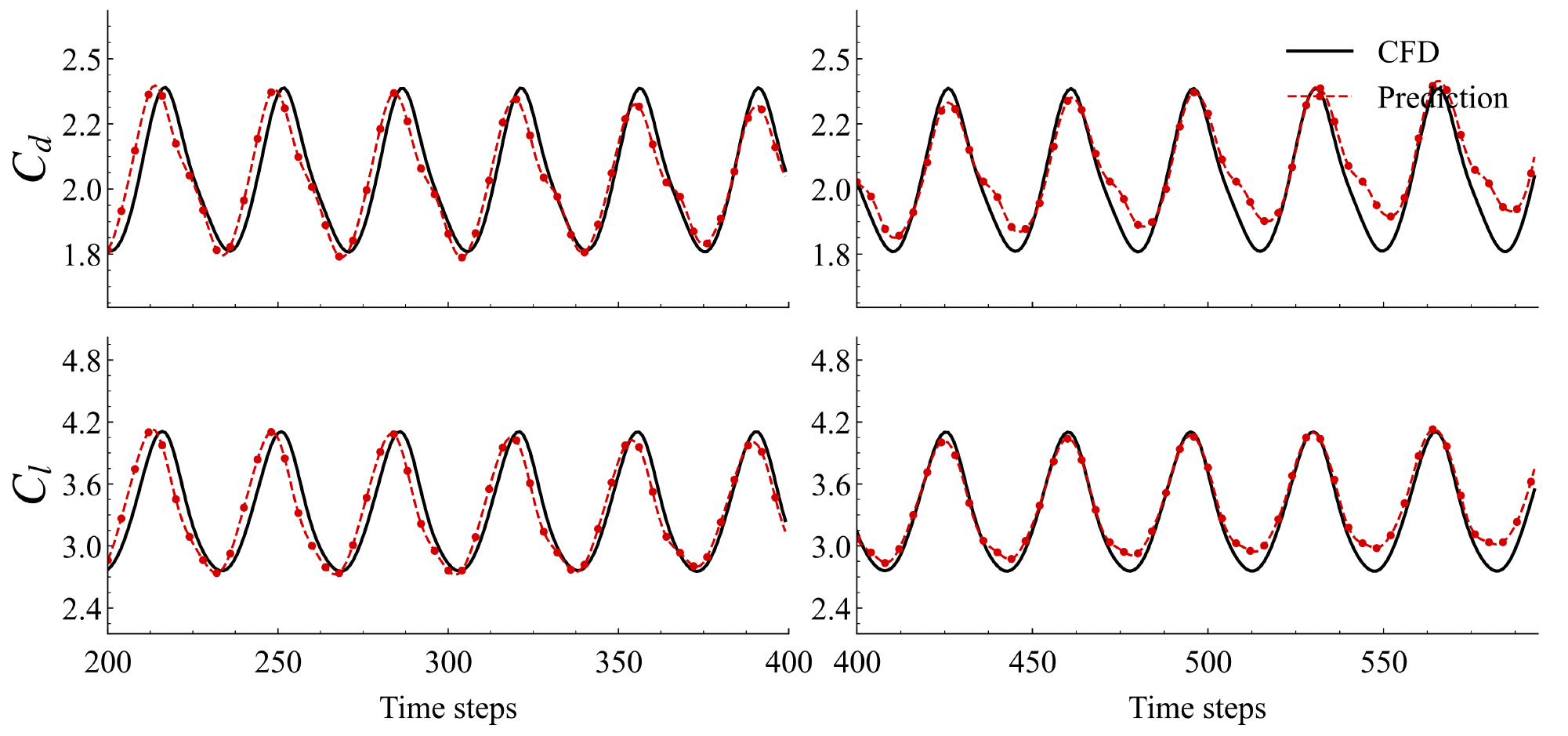}
    \end{minipage}
    \caption{NACA0012 airfoil, Re=900, Variation of $C_l$ and $C_d$ Curves at airfoil surface within the Time Steps <200, 600>}
    \label{unsteady_airfoil_cl_cd}
\end{figure*}

\begin{figure*}
\begin{minipage}{\linewidth}
    \includegraphics[width=0.87\textwidth]{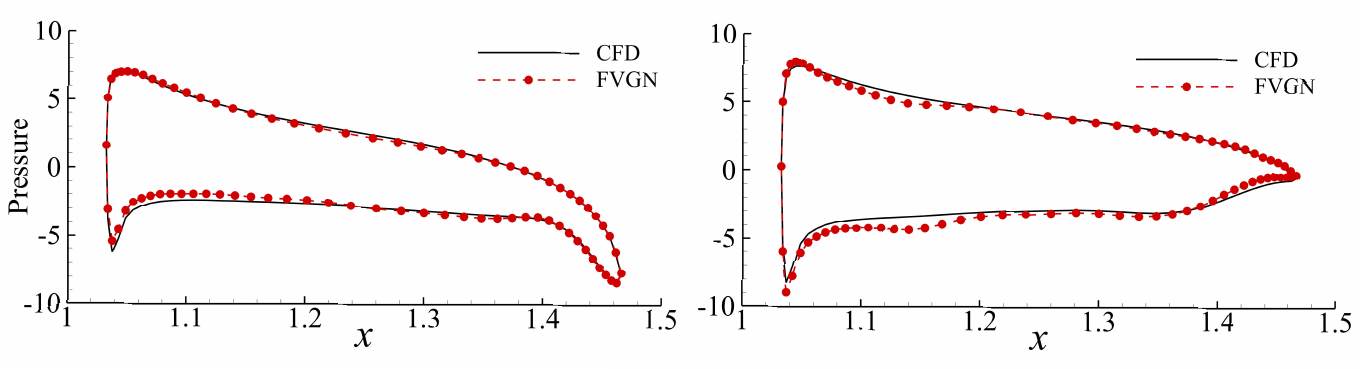}
    \end{minipage}
    \caption{NACA0012 airfoil, Re=900, Left: Pressure distribution of airfoil surface at t=400; Right: Pressure distribution of cylinder surface at t=600}
    \label{unsteady_airfoil_pressure_distrubutaion}
\end{figure*}

\subsubsection{NACA0012 Airfoil}\label{NACA0012 Airfoil}
In this section, we demonstrate a test case using the NACA0012 airfoil in the HYBRIDFLOW dataset. The combination of Reynolds number(Re=900) and angle of attack parameters that constitute this test case has also never appeared in the training set. Therefore, we want to verify whether FVGN can also exhibit good performance under the condition that only a minority of airfoil cases (accounting for only 10\% of the total number of training cases) appear in the training set. The inflow conditions are briefly described as follows: the Reynolds number is 900, and the angle of attack is $30^\circ$. We chose this angle of attack mainly to evaluate the performance of FVGN in facing more complex unsteady flows, which poses very high demands on the model's generalization ability. At the same time, FVGN also only used the data from the first 400 steps in the training set for training, while both MeshGraphNets and MeshGraphNets (directed) were trained using the complete 600 steps. This makes the case also a test of FVGN's generalization ability in the temporal dimension, all of which pose extremely high demands on FVGN.

\begin{table}[H]
    \centering
    \renewcommand{\arraystretch}{1}
    \caption{Result comparison in all rollout time steps of unsteady airfoil flow, The values of $C_l$ and $C_d$ can be derived from the following formulas $C_l = L / (0.5 \times \rho \times V^{2} \times A)$, $C_d = D / (0.5 \times \rho \times V^{2} \times A)$}
    \label{tab: result comparison airfoil flow}
    \begin{ruledtabular}
    \scalebox{1}{
    \begin{tabular}{ccccc|cc}
    \multirow{2}{*}{Model} &
      \multicolumn{2}{c}{\begin{tabular}[c]{@{}c@{}}RMSE $\times10^{-3}$\\ \textless{}0,400\textgreater{}\end{tabular}} &
      \multicolumn{2}{c|}{\begin{tabular}[c]{@{}c@{}}RMSE $\times10^{-3}$\\ \textless{}0,600\textgreater{}\end{tabular}} &
      \multirow{2}{*}{$C_d$} &
      \multirow{2}{*}{$C_l$} \\
                                                             & UV             & P               & UV             & P               &       &       \\ \hline
    FVGN                                                     & \textbf{5.934} & \textbf{18.453} & \textbf{9.444} & \textbf{24.339} & 2.045 & 3.345 \\
    \begin{tabular}[c]{@{}c@{}}MGN\\ (directed)\end{tabular} & 277.880        & 207.715         & 330.535        & 366.032         & 0.647 & 1.091 \\
    MGN                                                      & 107.717        & 172.383         & 125.983        & 197.493         & 2.261 & 3.702 \\ \hline
    CFD                                                      & -              & -               & -              & -               & 2.042 & 3.352
    \end{tabular}
    }
    \end{ruledtabular}
    \renewcommand{\arraystretch}{1}
\end{table}

\begin{figure*}
\begin{minipage}{\linewidth}
    \includegraphics[width=0.9\textwidth]{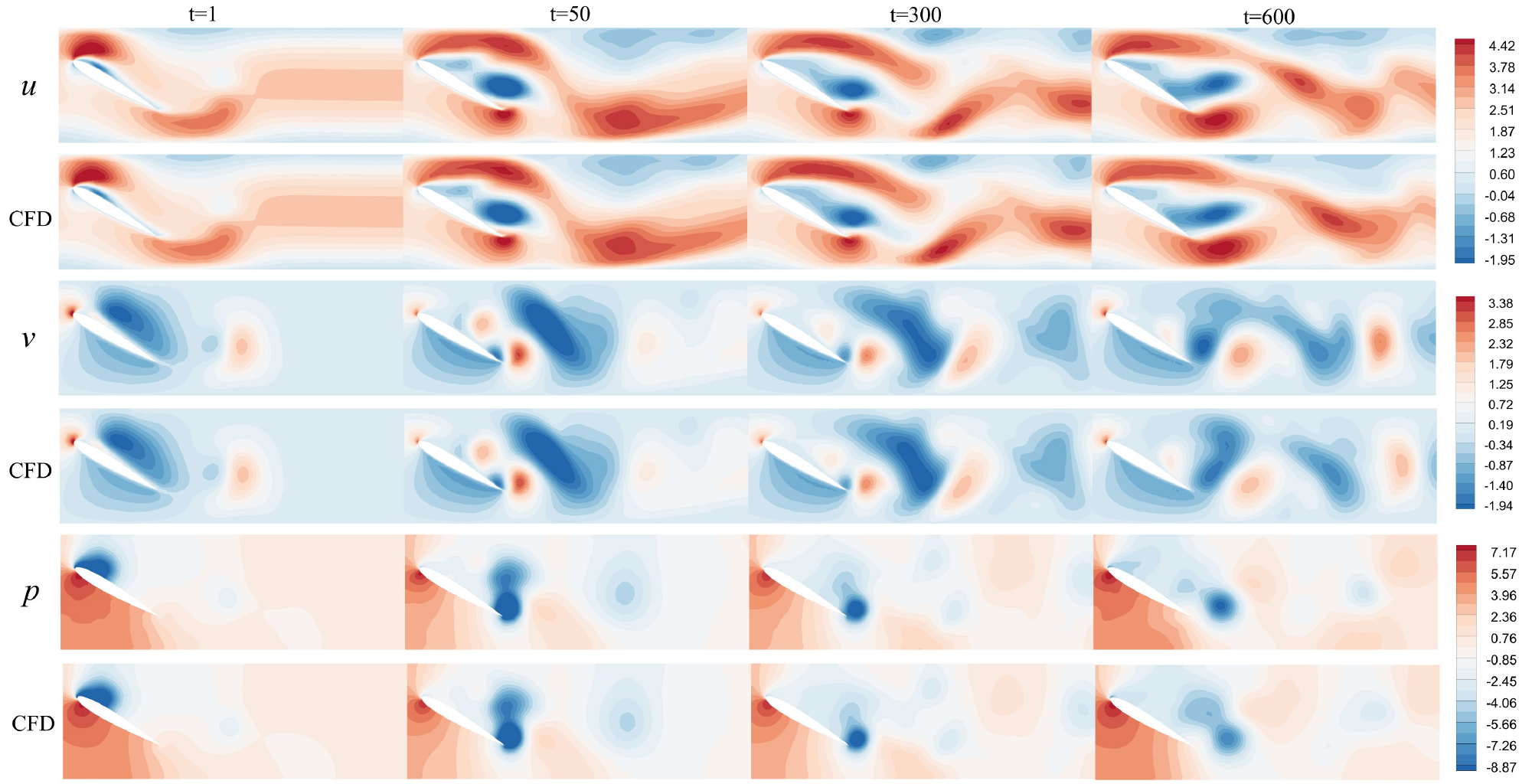}
    \end{minipage}
    \caption{NACA0012 airfoil, Re=900, FVGN's flow field prediction of velocity and pressure in unseen Reynolds number(Re=900) and attack angle combination situation}
    \label{airfoil-Re=900-timeseries}
\end{figure*}

From the velocity field contour diagram in the x direction (Fig.\ref{airfoil-Re=900-timeseries}), it can be seen that at the last moment, there is a relatively obvious difference between the velocity field in the x direction predicted by FVGN and the ground truth. This difference is concentrated in the vortex-shedding area. At the same time, It can be seen from Tab.\ref{tab: result comparison airfoil flow} that the performance of FVGN significantly outperforms MeshGraphNets, effectively validating that FVGN has higher prediction accuracy than MeshGraphNets at nearly every step, ensuring that its prediction accuracy is significantly higher than that of MeshGraphNets at the 600th time step. Both its directed and undirected forms highlight the superiority of FVGN. Particularly noteworthy is the comparison of lift and drag coefficients in Tab.\ref{tab: result comparison airfoil flow}, where FVGN shows a closer approximation to CFD results. The lift-drag curve in Fig.\ref{unsteady_airfoil_cl_cd} also indicates that FVGN can exhibit good performance near the boundary layer. This not only proves the model's accuracy in simulating the unsteady separated flow of the airfoil wake but also showcases its outstanding performance in predicting the flow characteristics near the boundary layer and the overall aerodynamic characteristics of the airfoil, which is a very crucial aspect in unsteady flow analysis.

\begin{figure*}
\begin{minipage}{\linewidth}
    \includegraphics[width=0.9\textwidth]{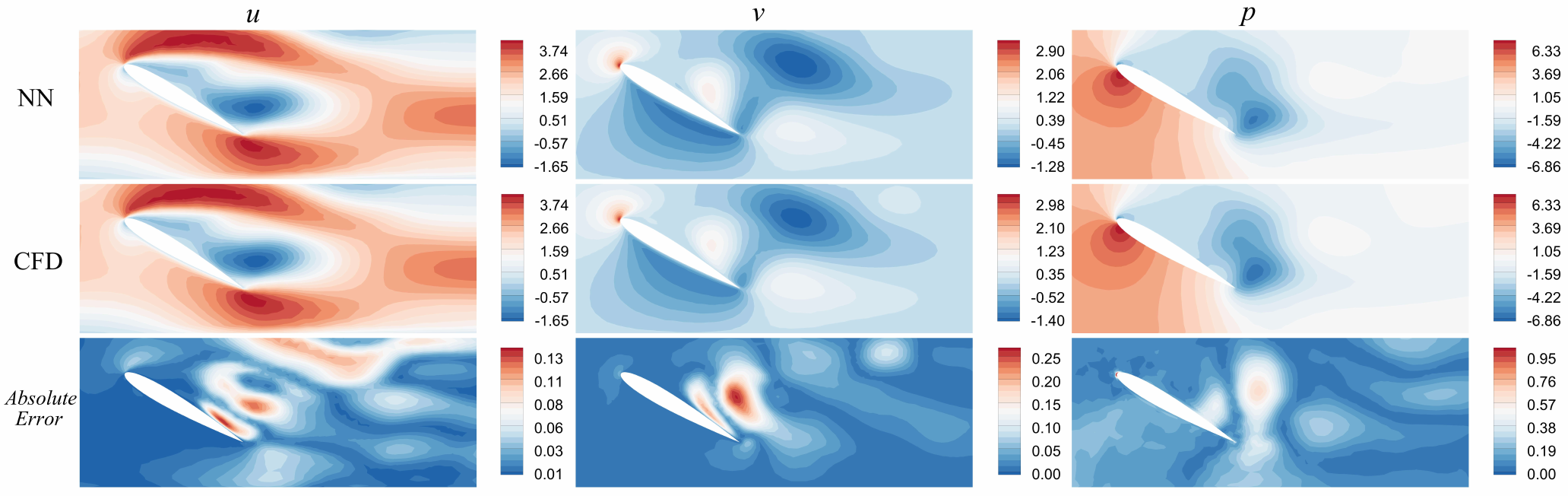}
    \end{minipage}
    \caption{NACA0012 airfoil, Re=900, time-avg field of velocity and pressure prediction in unseen Reynolds number and attack angle combination situation, and we use solution time step range of <200,600> to calculate time-averaged field, And Error was also calculated in $Absolute \ Error =\left | \overline{\hat{\mathbf{u}}}-\overline{\mathbf{u}^{GT}}\right | $}
    \label{airfoil unsteady re=890_time_avg_field}
\end{figure*}

However, we can still observe from Fig.\ref{unsteady_airfoil_cl_cd} that FVNG exhibits a situation similar to the previous cylinder flow case; After the 400th time step, the model produces a larger error in the prediction of lift and drag coefficients. Although we can see from Fig.\ref{unsteady_airfoil_pressure_distrubutaion} that the pressure distribution at the 600th time step is still relatively accurate, the model actually produces amplitude errors at the troughs of the lift-drag curve, while maintaining relatively high accuracy at other times. We interpret this phenomenon as follows: the model produces errors in the physical field at every time step, and these errors tend to concentrate in the wake region (as shown in Fig.\ref{airfoil unsteady re=890_time_avg_field}), so during each message-passing process, the errors generated in the wake region affect the prediction accuracy near the airfoil. Looking at the final time-averaged field results, the lift and drag predictions provided by FVGN in Tab.\ref{tab: result comparison airfoil flow} are still very close to CFD, meaning that the model proposed in this study can still provide accurate predictions for the actual force situation on the wings.

\begin{figure*}
\begin{minipage}{\linewidth}
    \includegraphics[width=0.9\textwidth]{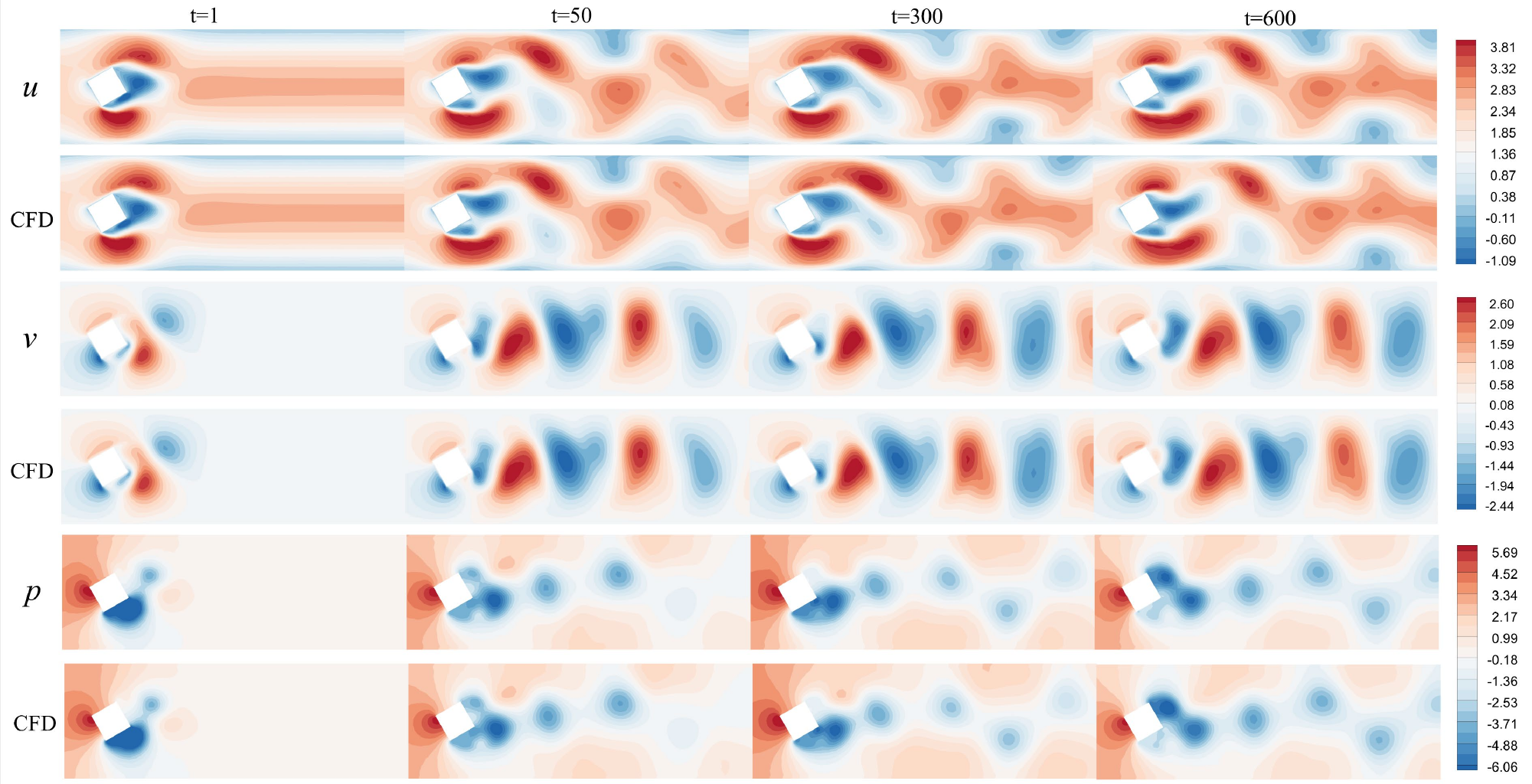}
    \end{minipage}
    \caption{Rectangular Column, Re=370, FVGN's flow field prediction of velocity and pressure in unseen obstacle position, Reynolds number and rotation angle combination situation, And Error was also calculated in $Absolute \ Error=\left | \overline{\hat{\mathbf{u}}}-\overline{\mathbf{u}^{GT}}\right | $}
    \label{square unsteady re=370_field}
\end{figure*}

\begin{figure*}
\begin{minipage}{\linewidth}
    \includegraphics[width=0.9\textwidth]{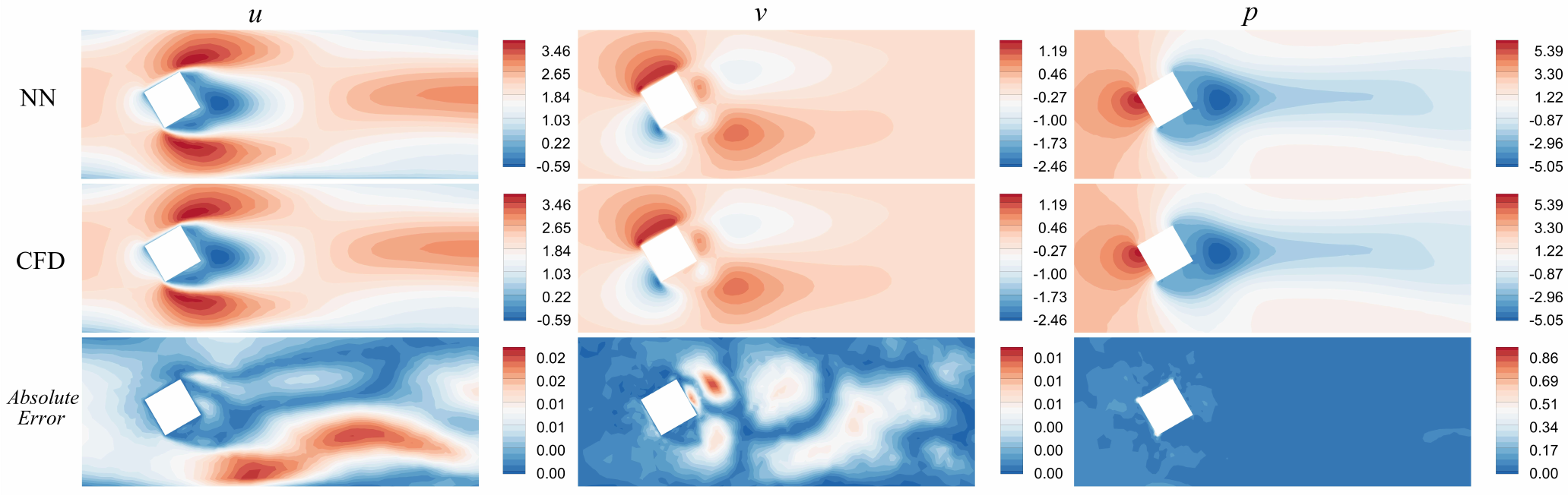}
    \end{minipage}
    \caption{Rectangular Column, Re=370, time-avg field of velocity and pressure prediction in unseen obstacle position, Reynolds number, and rotation angle combination. We use the solution time step range of <200,600> to calculate the time-averaged field, And Error was also calculated in $Absolute \ Error= \left | \overline{\hat{\mathbf{u}}}-\overline{\mathbf{u}^{GT}}\right | $}
    \label{square unsteady re=370_time_avg}
\end{figure*}

\begin{figure*}
\begin{minipage}{\linewidth}
    \includegraphics[width=0.9\textwidth]{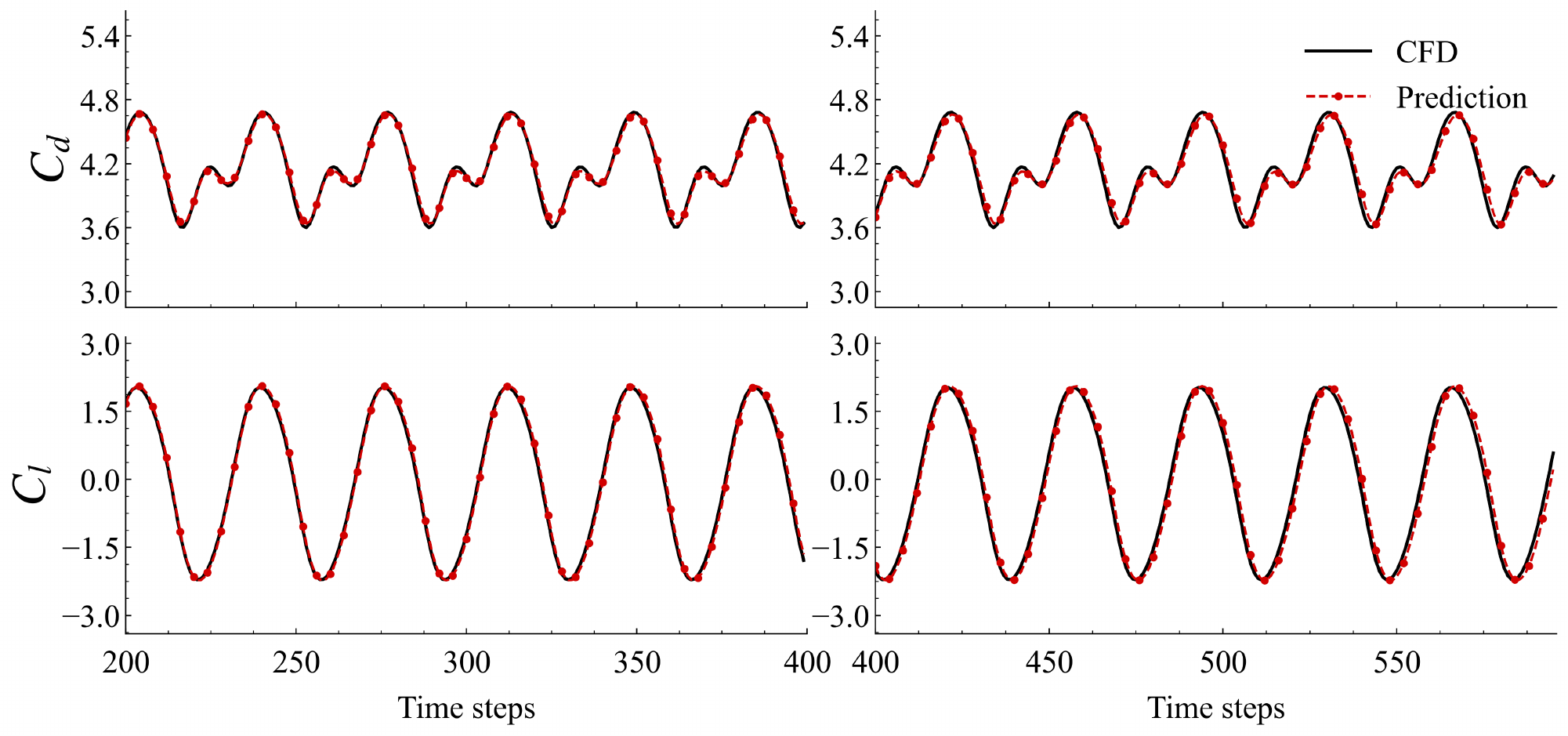}
    \end{minipage}
    \caption{Rectangular Column, Re=370, Variation of $C_l$ and $C_d$ Curves at airfoil surface within the Time Steps <200, 600>}
    \label{unsteady_square_cl_cd}
\end{figure*}

\begin{figure*}
\begin{minipage}{\linewidth}
    \includegraphics[width=0.9\textwidth]{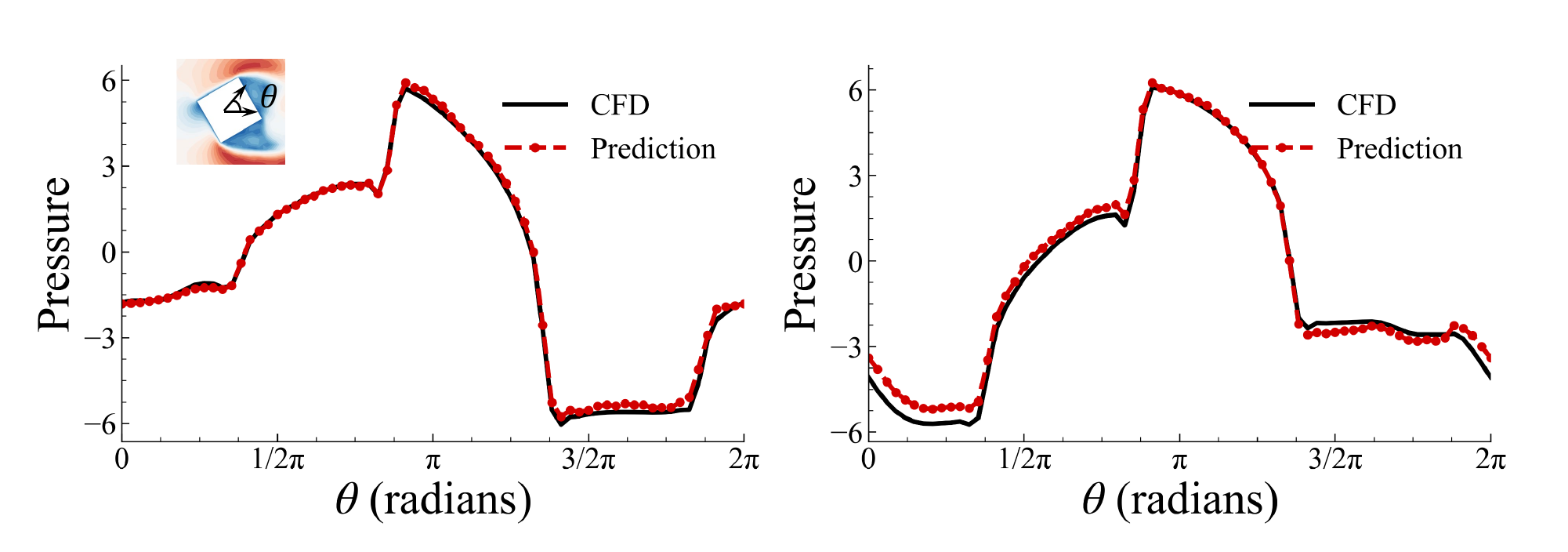}
    \end{minipage}
    \caption{Rectangular Column, Re=370, left: Pressure distribution of objective surface at t=400; right: Pressure distribution of objective surface at t=600}
    \label{unsteady_square_pressure_distribution}
\end{figure*}

\subsubsection{Rectangular Column}\label{Rectangle Column}

In the rectangular column test case within the HYBRIDFLOW dataset, the performance of FVGN was closely compared with MeshGraphNets in scenarios involving obstacles with sharp-angled boundaries. In the training set of this dataset, the examples of rectangular columns still only account for 10\% of the total number of training examples. The test case has a Reynolds number of 370 and a rotation angle of 25 degrees, and this combination of conditions also did not appear in the training set. FVGN only used the first 400 steps of data from the training set, while MeshGraphNets and MeshGraphNets(directed) still used the complete 600 time step training data. The above test conditions provided both models with a challenging environment.

\begin{table}[H]
    \centering
    \renewcommand{\arraystretch}{1}
    \caption{Result comparison in all rollout time steps of rectangular column flow, The values of $C_l$ and $C_d$ can be derived from the following formulas $C_l = L / (0.5 \times \rho \times V^{2} \times A)$, $C_d = D / (0.5 \times \rho \times V^{2} \times A)$}
    \label{tab: result comparison square column flow}
    \begin{ruledtabular}
    \begin{tabular}{ccccc|cc}
    \multirow{2}{*}{Model} &
      \multicolumn{2}{c}{\begin{tabular}[c]{@{}c@{}}RMSE $\times10^{-3}$\\ \textless{}0,400\textgreater{}\end{tabular}} &
      \multicolumn{2}{c|}{\begin{tabular}[c]{@{}c@{}}RMSE $\times10^{-3}$\\ \textless{}0,600\textgreater{}\end{tabular}} &
      \multirow{2}{*}{$C_d$} &
      \multirow{2}{*}{$C_l$} \\
                                                             & UV             & P              & UV             & P              &       &        \\ \hline
    FVGN                                                     & \textbf{0.160} & \textbf{3.472} & \textbf{0.242} & \textbf{3.488} & 4.153 & -0.066 \\
    \begin{tabular}[c]{@{}c@{}}MGN\\ (directed)\end{tabular} & 66.219         & 488.196        & 157.698        & 663.812        & 1.236 & 0.214  \\
    MGN                                                      & 2.688          & 7.169          & 4.445          & 8.549          & 4.237 & -0.098 \\ \hline
    CFD                                                      & -              & -              & -              & -              & 4.153 & -0.068
    \end{tabular}
    \renewcommand{\arraystretch}{1}
    \end{ruledtabular}
\end{table}

From the comparison of the contour diagrams at instantaneous moments (Fig.\ref{square unsteady re=370_field})
, it can be seen that FVGN has captured the vortex shedding phenomenon in the flow field well, while as shown in Tab.\ref{tab: result comparison square column flow}, it can be seen that FVGN outperforms MeshGraphNets in accurately predicting both the velocity field and the pressure field. Moreover, under the aforementioned training conditions, MeshGraphNets(directed) demonstrated considerable inaccuracy in predicting the lift and drag coefficients $C_l, \ C_d$. This observation once again indicates that reducing the size of the model's hidden space not only impacts the prediction accuracy for cases that predominantly occupy the training set (as mentioned in Sec.\ref{Two-dimensional cylinder flow}) but also significantly reduces the model's generalizability. Additionally, both FVGN and MeshGraphNets have higher accuracy in predicting aerodynamic parameters. As can be seen from Fig.\ref{unsteady_square_cl_cd}, as the number of predicted time steps increases, the predictions from FVGN gradually phase shift from the CFD results. This can also be seen from Fig.\ref{unsteady_square_pressure_distribution}, where FVGN's prediction of the surface pressure distribution at the last moment has some errors. And this trend is also consistent with the previous two sections, and we believe the reason is the same as with the NACA0012 case. As seen from Fig.\ref{square unsteady re=370_time_avg}, the main concentration of velocity field errors is in the wake area. The errors in the wake inevitably spread throughout the field due to the message-passing mechanism of the graph neural network. Overall, both FVGN and MeshGraphNets have provided relatively accurate results for aerodynamic parameters, but considering the final training time cost, the engineering application potential of FVGN is still greater than that of MeshGraphNets.

\subsection{Geometric Generalisation Capability}\label{Generalisation capability}

Thanks to the combinatorial generalization capabilities of Graph Networks (GN), networks based on GN usually have a certain degree of generalization ability. The previous section mainly demonstrated the performance of the model in test cases where combinations of Reynolds number-object position or Reynolds number-angle of attack not encountered in the training set were used. This section introduces the generalization test results on unseen elliptical shapes and airfoil types, using models trained on the HYBRIDFLOW dataset while keeping all hyperparameters consistent with those mentioned earlier. 
We tested a NACA2412 wing and an ellipse with a major-to-minor axis ratio of 1.33. The domain aspect ratios of these three test cases are consistent with all cases in the training set (i.e., 3.6:0.6), and their average cell count is maintained within the range of $<7,000-8,000>$. The airfoil test case was conducted at an angle of attack of 30 degrees and a Reynolds number of 800, while the ellipse was tested at a rotation angle of 30 degrees and a Reynolds number of 320.
\begin{table}[H]
\centering
\caption{Result comparison between FVGN and MGN. For the calculation of the Strouhal number for individual cases, We utilized the lift curve to calculate the vortex shedding frequency.}
\label{tab: generalization_comp}
    \scalebox{0.95}{
    \begin{ruledtabular}
        \begin{tabular}{ccccc|ccc}
        Geo. &
          Model &
          \begin{tabular}[c]{@{}c@{}}Rollout\\ Time\\ Range\end{tabular} &
          \begin{tabular}[c]{@{}c@{}}UV \\ RMSE\end{tabular} &
          \begin{tabular}[c]{@{}c@{}}P\\ RMSE\end{tabular} &
          \begin{tabular}[c]{@{}c@{}}St\\ Number\end{tabular} &
          $C_d$ &
          $C_l$ \\ \hline
        \multirow{3}{*}{\begin{tabular}[c]{@{}c@{}}Ellipse\\ Re=300\end{tabular}} &
          FVGN &
          \textless{}0,600\textgreater{} &
          \textbf{0.001} &
          \textbf{0.047} &
          0.251 &
          2.936 &
          -0.427 \\
         &
          MGN &
          \textless{}0,600\textgreater{} &
          0.095 &
          0.184 &
          0.251 &
          3.250 &
          -0.486 \\ \cline{2-8} 
         &
          CFD &
          \textless{}0,600\textgreater{} &
          - &
          - &
          0.251 &
          2.766 &
          -0.484 \\ \hline
        \multirow{3}{*}{\begin{tabular}[c]{@{}c@{}}NACA2412\\ Re=800\\ aoa=30°\end{tabular}} &
          FVGN &
          \textless{}0,600\textgreater{} &
          \textbf{0.034} &
          \textbf{0.221} &
          0.591 &
          1.938 &
          3.537 \\
         &
          MGN &
          \textless{}0,600\textgreater{} &
          0.127 &
          1.165 &
          0.709 &
          1.931 &
          3.578 \\ \cline{2-8} 
         &
          CFD &
          \textless{}0,600\textgreater{} &
          - &
          - &
          0.591 &
          2.081 &
          3.702 
        \end{tabular}
        \end{ruledtabular}
    }
\end{table}

\begin{figure*}
\centering
\begin{minipage}{\linewidth}
    \centering
    \includegraphics[width=1\textwidth]{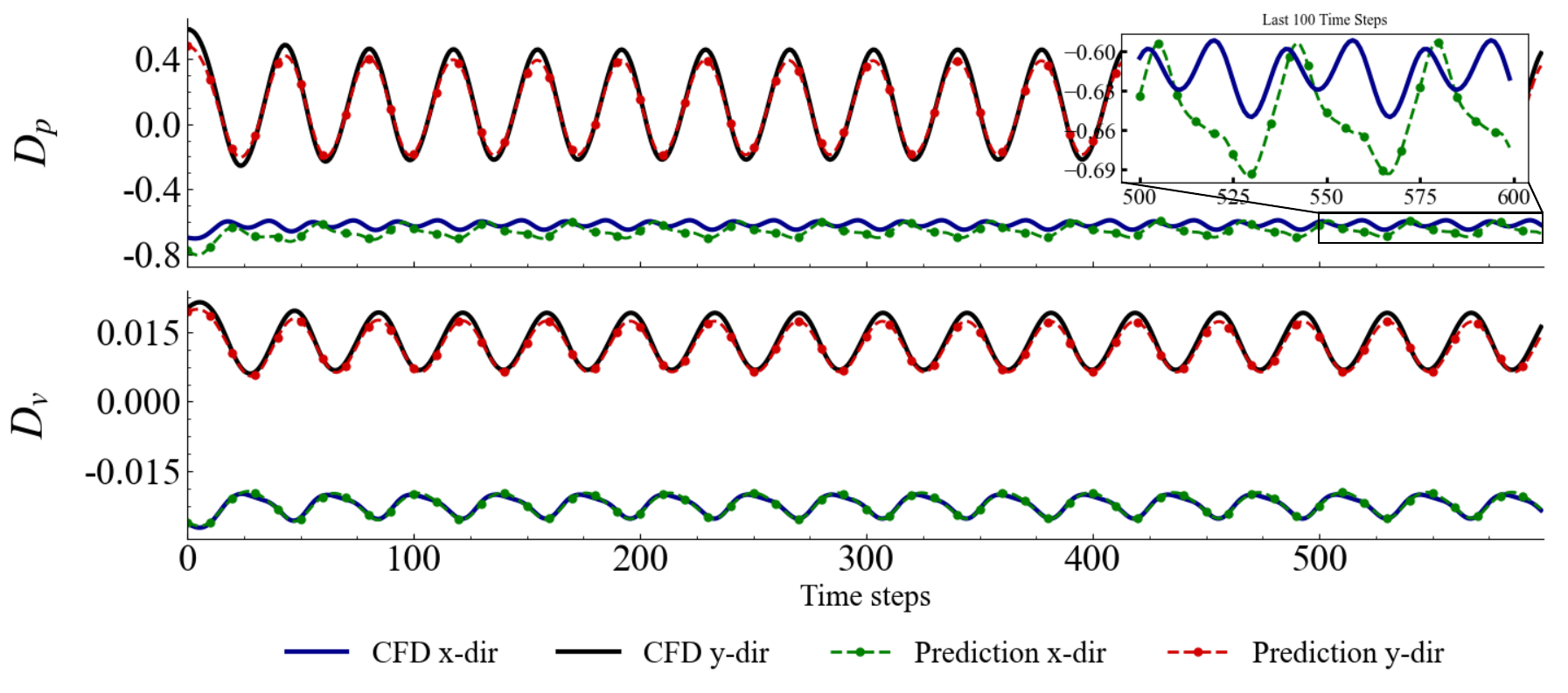}
    \end{minipage}
    \caption{Viscousity Force $D_v = \int_{\partial B } \tau_{w} \, dS
$ and Pressure Force $D_p = \int_{\partial B } p \mathbf{n} \, dS
$ Comparison of Ellipse test case}
    \label{generalization_force}
\end{figure*}

\noindent \textbf{Ellipse} \quad The Tab.\ref{tab: generalization_comp} shows the performance of FVGN and MeshGraphNets in these two test cases. We primarily conducted comparisons of the instantaneous pressure field, Strouhal number, and lift and drag coefficients. Among these, we used the lift coefficient to calculate the frequency of the Strouhal number. Additionally, for incompressible fluids, predicting the pressure field is the most challenging, and thus the accuracy of the pressure field predictions almost directly determines the ultimate performance of the model. This effectively verifies that our model, compared to directly data-driven models, possesses a superior generalization advantage in terms of geometric shapes. In fact, in the HYBRIDFLOW dataset, most cases are about flow around a cylinder, so theoretically, FVGN should have stronger generalization for elliptical shapes. Tab.\ref{tab: generalization_comp} also corroborates this, however, if we delve into the aerodynamic force analysis on the surface of the ellipse, we can see that FVGN fails to maintain the initial phase consistency in $D_p$ prediction with the solver's results in Fig.\ref{generalization_force}.  Espically, that there are significant errors in the prediction of the pressure force in the X direction, which directly resulted in a discrepancy of up to $6\%$ between the predicted drag coefficient and the CFD results.

\begin{figure*}
\centering
\begin{minipage}{\linewidth}
    \centering
    \includegraphics[width=0.9\textwidth]{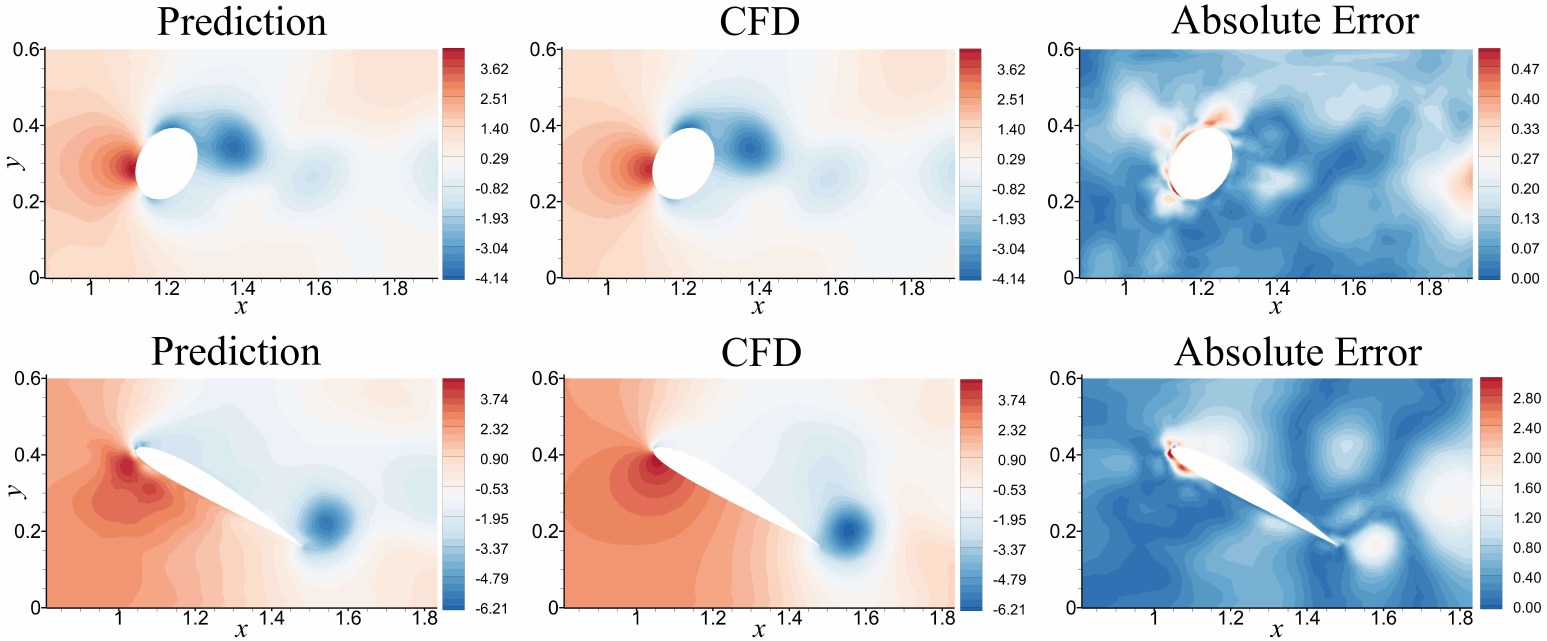}
    \end{minipage}
    \caption{Ellipse Re=300; NACA2412 aoa=$30^\circ$ Re=800; Pressure field Comparison of different obstacles with FVGN at rollout time step=400}
    \label{generalization_p_field}
    
\end{figure*}

\noindent \textbf{NACA2412} \quad From Fig.\ref{generalization_p_field}, it can be seen that the FVGN has produced significant errors in predicting the pressure field of the NACA2412 airfoil, especially in the nose section where FVGN incorrectly predicted the general structure of the pressure field. However, as shown in Tab.\ref{tab: generalization_comp}, FVGN has actually succeeded in capturing the vortex shedding frequency of the NACA2412 airfoil case, and the predictions of lift and drag coefficients on the wing surface are also relatively accurate.

From the test results of the above cases, it is evident that solely relying on RMSE or using contour plots to evaluate the performance of a model is inadequate. In our practical experience, models based on Graph Nerual Networks do not typically "diverge" when faced with unseen geometric shapes; instead, they provide seemingly reasonable results from the perspective of flow field contour plots, despite significant discrepancies with the actual flow. Therefore, enhancing the geometric shape generalization ability of FVGN will be a major focus in future work.


\section{Method Effectiveness Verification}\label{Method Effectiveness Verification}

\begin{table*}
\centering
\caption{Result comparison between FVGN and MeshGraphNets in all test rollout samples of CYLINDERFLOW dataset, For the calculation of the UV RMSE for individual cases, we utilized the following formula: $RMSE({UV}^{prediction},{UV}^{Truth})=\frac{\sum [(\hat{u}_{i}-u_{i})^{2}+(\hat{v}_{i}-v_{i})^{2}]}{{\sum{u}_{i}}^{2}+{{v}_{i}}^{2}}$. Meanwhile, for the computation of the P RMSE, we employed the subsequent formula: $RMSE({P}^{prediction},{P}^{Truth})=\frac{\sum (\hat{p}_{i}-p_{i})^{2}}{{\sum{p}_{i}}^{2}}$. We tested 100 cases, which encompassed a variety of Reynolds numbers and cylinder positions. The overall mean RMSE for these 100 cases was determined over 600 time steps, denoted as $RMSE({UV})=\frac {\sum RMSE({UV}^{prediction}_{i},{UV}^{Truth}_{i})}{Rollout \ Samples}$. The computational time consumption for model training was assessed on a single machine equipped with an 8-core CPU (workstation) and a single V100S GPU. FVGN(MA=1) and FVGN(ReLU) were both tested under the conditions of $\alpha=1, \beta=10$.}
\label{tab: cylinder flow result comparison}
\scalebox{0.98}{
\renewcommand{\arraystretch}{1.05} 
    \begin{ruledtabular}
    \begin{tabular}{ccccccccc}
    \begin{tabular}[c]{@{}c@{}}CYLINDERFLOW\\ \end{tabular} &
      \begin{tabular}[c]{@{}c@{}}Train\\ Time steps\\ Range\end{tabular} &
      \begin{tabular}[c]{@{}c@{}}Train\\ Epochs\end{tabular} &
      \begin{tabular}[c]{@{}c@{}}Rollout\\ Samples\end{tabular} &
      \begin{tabular}[c]{@{}c@{}}Rollout\\ Time steps\\ Range\end{tabular} &
      \begin{tabular}[c]{@{}c@{}}UV \\ RMSE\\ $\times10^{-3}$\end{tabular} &
      \begin{tabular}[c]{@{}c@{}}P\\ RMSE\\ $\times10^{-3}$\end{tabular} &
      \begin{tabular}[c]{@{}c@{}}Train\\ Time\\ Cost\\ (days)\end{tabular} &
      \begin{tabular}[c]{@{}c@{}}Model\\ Parameters\end{tabular} \\ \hline
    FVGN &
       \textbf{\textless{}0,300\textgreater{}}  &
      \textbf{23} &
      100 &
      \textless{}0,600\textgreater{} &
      \textbf{1.85} &
      17.46 &
      \textbf{3.2} &
      \textbf{2.2M} \\ \hline
    \begin{tabular}[c]{@{}c@{}}MGN\\ (directed)\end{tabular} &
      \textless{}0,600\textgreater{} &
      33 &
      100 &
      \textless{}0,600\textgreater{} &
      4.65 &
      28.05 &
      5.2 &
      2.4M \\
    MGN &
      \textless{}0,600\textgreater{} &
      33 &
      100 &
      \textless{}0,600\textgreater{} &
      2.01 &
      \textbf{15.83} &
      7.8 &
      2.4M 
    \end{tabular}
    \end{ruledtabular}
    \renewcommand{\arraystretch}{1.0} 
    }
\end{table*}

 We test the performance of FVGN and MeshGraphNets both on the CYLINDERFLOW dataset and HYBRIDFLOW dataset, and the following Tab.\ref{tab: cylinder flow result comparison} and Tab.\ref{tab: hybrid dataset result comparison} mainly presents a comparison of some key performance parameters. In the HYBRIDFLOW dataset comparison, it can be observed that as the average number of cells/nodes increases, both models experience an increase in training time and testing errors. However, the training time consumption of FVGN only increased by approximately 1 day, while that of MeshGraphNets increased by almost 4 days. This is mainly because MeshGraphNets encodes the features of bidirectional edges, whereas FVGN only encodes the features of unidirectional edges. With the increase in the number of cells, it is evident that MeshGraphNets will introduce greater computational cost. At the same time, from the perspective of test error accuracy comparison, the performance of both models on this dataset is inferior to that of the CYLINDERFLOW dataset. The main reason for this is that the HYBRIDFLOW dataset has a broader range of Reynolds number distributions. In the HYBRIDFLOW dataset, the characteristic scale of the cylinder case is taken as the diameter of the cylinder, the airfoil case is taken as the chord length of the airfoil, and the rectangular column case is taken as the diagonal length. Because the chord length of the airfoil is significantly larger than the diameter of the cylinder, the main error in this dataset is concentrated in the high Reynolds number airfoil cases. The aforementioned NACA0012 test case also proved that.

  
\begin{table*}
\centering
\caption{Result comparison in all test case rollouts of HYBRIDFLOW dataset, In this dataset, the FVGN model only used the first 400 time steps of the flow field from 1000 cases as training data. But test in 600 rollout steps.}
\label{tab: hybrid dataset result comparison}
\scalebox{0.98}{
\renewcommand{\arraystretch}{1.05} 
\begin{ruledtabular}
\begin{tabular}{ccccccccc}
HYBRIDFLOW &
  \begin{tabular}[c]{@{}c@{}}Train\\ Time Step\\ Range\end{tabular} &
  \begin{tabular}[c]{@{}c@{}}Train\\ Epochs\end{tabular} &
  \begin{tabular}[c]{@{}c@{}}Rollout\\ Samples\end{tabular} &
  \begin{tabular}[c]{@{}c@{}}Rollout\\ Time Step\\ Range\end{tabular} &
  \begin{tabular}[c]{@{}c@{}}UV\\ RMSE\\ $\times10^{-3}$\end{tabular} &
  \begin{tabular}[c]{@{}c@{}}P\\ RMSE\\ $\times10^{-3}$\end{tabular} &
  \begin{tabular}[c]{@{}c@{}}Train\\ Time\\ Cost\\ (days)\end{tabular} &
  \begin{tabular}[c]{@{}c@{}}Model\\ Parameters\end{tabular} \\ \hline
\begin{tabular}[c]{@{}c@{}}FVGN\\ ($\alpha=1$, $\beta=1$)\end{tabular} &
  \textless{}0,400\textgreater{} &
  23 &
  100 &
  \textless{}0,600\textgreater{} &
  31.74 &
  63.99 &
  4.8 &
  2.2M \\
\begin{tabular}[c]{@{}c@{}}FVGN\\ ($\alpha=1$, $\beta=10$)\end{tabular} &
  \textless{}0,400\textgreater{} &
  23 &
  100 &
  \textless{}0,600\textgreater{} &
  \textbf{13.27} &
  36.39 &
  4.8 &
  2.2M \\
\begin{tabular}[c]{@{}c@{}}FVGN\\ ($\alpha=0$, $\beta=10$)\end{tabular} &
  \textless{}0,400\textgreater{} &
  23 &
  100 &
  \textless{}0,600\textgreater{} &
  14.72 &
  \textbf{35.57} &
  4.7 &
  2.2M \\ \hline
\begin{tabular}[c]{@{}c@{}}FVGN\\ (MA=1)\end{tabular} &
  \textless{}0,400\textgreater{} &
  23 &
  100 &
  \textless{}0,600\textgreater{} &
  71.62 &
  132.40 &
  4.5 &
  2.2M \\
\begin{tabular}[c]{@{}c@{}}FVGN\\ (ReLU)\end{tabular} &
  \textless{}0,400\textgreater{} &
  23 &
  100 &
  \textless{}0,600\textgreater{} &
  29.26 &
  55.19 &
  4.6 &
  2.2M \\ \hline
\begin{tabular}[c]{@{}c@{}}MGN\\ (directed)\end{tabular} &
  \textless{}0,600\textgreater{} &
  33 &
  100 &
  \textless{}0,600\textgreater{} &
  86.71 &
  365.91 &
  7.5 &
  2.4M \\
\begin{tabular}[c]{@{}c@{}}MGN\\ (origin)\end{tabular} &
  \textless{}0,600\textgreater{} &
  33 &
  100 &
  \textless{}0,600\textgreater{} &
  57.54 &
  126.34 &
  10.8 &
  2.4M \\ 
\end{tabular}
    \end{ruledtabular}
    }
\renewcommand{\arraystretch}{1} 
\end{table*}

In this paper, we primarily propose two improved algorithms for unsteady flow field prediction on unstructured grids using Graph Networks (GNs). The first improvement involves introducing a physics-constrained loss to attempt to reduce the error at each rollout step, thereby mitigating the cumulative error phenomenon in the model over long-time rollouts. The second improvement is the transformation of cell-centered graph into vertex-centered graph when incorporating the finite volume method, allowing graph networks to perform message passing between the edge sets of these two mesh types. This enhances the efficiency of message passing and reduces the bidirectional edge features in MeshGraphNets to unidirectional edge features. Consequently, this section mainly demonstrates the performance improvements achieved by applying either one of the aforementioned methods individually or both methods simultaneously in comparison to the existing techniques.

\begin{figure*}
\centering
\begin{minipage}{\linewidth}
    \centering
    \includegraphics[width=1.0\textwidth]{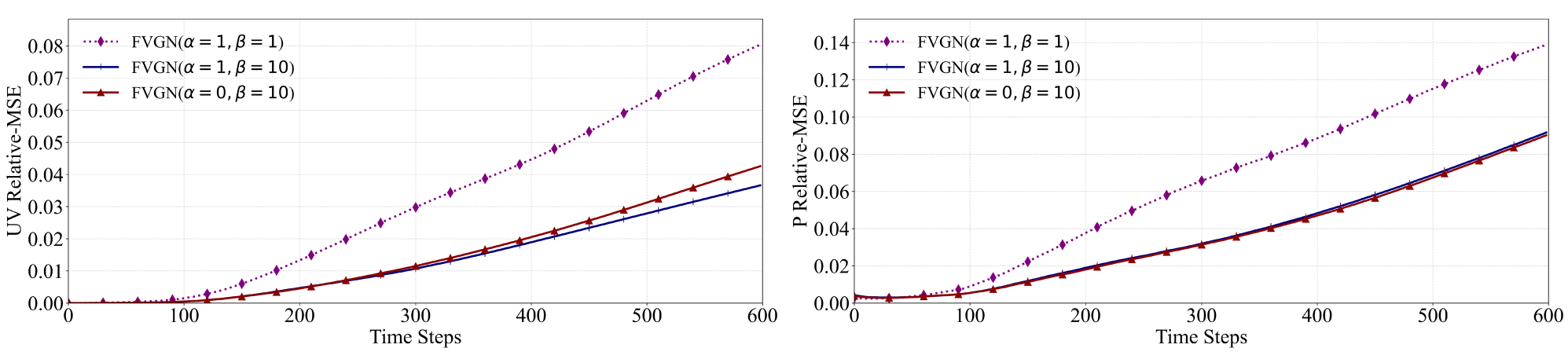}
    \end{minipage}
    \caption{Different hyperparameters configuration test results comparison on HYBRIDFLOW dataset over 600 rollout time-steps; left: velocity field comparison; right: pressure field comparison.}
    \label{Hybrid_UV_comp}
\end{figure*}

\noindent$\textbf{Physics Constraint}$ \quad It is necessary to introduce physics constraints based on the finite volume method in Meshgraphnets (directed), and the test results of 200 cases from two datasets (Tab.\ref{tab: cylinder flow result comparison} and Tab.\ref{tab: hybrid dataset result comparison}) indicate that this is the fundamental reason for the improvement in model prediction accuracy. At the same time, the physical constraints in FVGN can actually be divided into two categories: whether to introduce the constraints of the continuity equation. Fig.\ref{Hybrid_UV_comp} and Tab.\ref{tab: hybrid dataset result comparison} show the performance of FVGN (Without Cont. Eq) when the continuity equation constraint is removed (i.e., setting $\alpha$ to 0 in Eq.\eqref{total_loss_function}). It can be seen that without the constraints of the continuity equation, FVGN exhibits a more severe trend of error accumulation in predicting the velocity field, while it hardly affects the prediction of the pressure field. The reason for this phenomenon is understandable. We previously used solvers to calculate high-precision numerical solutions of the velocity field and set them as the constraints for the right-hand side of the momentum equation Eq.\eqref{eq physics Constrained Loss momtem}, which definitely satisfies the continuity equation. However, these numerical solutions are stored at grid points or cell centers, and we need to use geometric interpolation methods to interpolate them to the faces of the control volumes. This interpolation algorithm will inevitably introduce additional errors, thus reducing the accuracy of FVGN's prediction of Eq.\eqref{face_U} and consequently affecting the prediction accuracy of the convection term $A=u_fu_f$. Overall, considering the computational cost from Tab.\ref{tab: hybrid dataset result comparison}, introducing the continuity equation as an additional constraint does not add much computational overhead. Therefore, a complete FVGN should include both the constraints based on the momentum equation and the continuity equation.

Meanwhile, we also presented experiments with all hyperparameters set to 1, as shown in Fig.\ref{Hybrid_UV_comp} and Tab.\ref{tab: hybrid dataset result comparison}. It can be seen that both the velocity field and the pressure field exhibit a significant trend of error accumulation. The reason is that our model uses only the velocity field as the input for each round of prediction, and the low weight of the momentum equation leads to a large deviation in the prediction results of the compensation term $\mathbf{Q}$, ultimately resulting in low prediction accuracy of $\bigtriangleup \mathbf{u}$, causing the model's cumulative error to increase over time. In summary, from our practical results, the weight of the momentum equation must be higher than that of other supervision terms, and introducing the constraint of the continuity equation can improve the model's prediction accuracy without significantly increasing the computational cost.

\begin{figure*}
\centering
\begin{minipage}{\linewidth}
    \centering
    \includegraphics[width=1.0\textwidth]{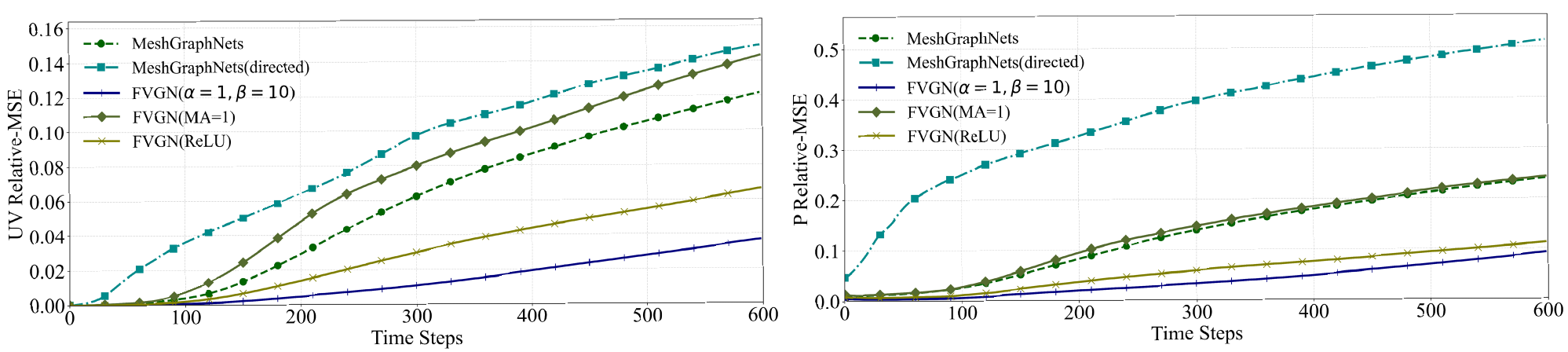}
    \end{minipage}
    \caption{Different model architecture test results comparison on HYBRIDFLOW dataset over 600 rollout time-steps; left: velocity field comparison; right: pressure field comparison}
    \label{Hybrid_P_comp}
\end{figure*}

\noindent$\textbf{Twice message aggregating mechanism}$ \quad In the entire forward process of FVGN, encoding only directed edge features significantly reduces the gradient computation cost of learnable parameters during the backward propagation process compared to the bidirectional edge encoding approach of MeshGraphNets. This is precisely the key to FVGN's improved training efficiency. 
However, simply replacing bidirectional edges with directed edges will inevitably affect the efficiency of message passing in the graph neural network's message-passing layer, leading to reduced final prediction accuracy. As shown in Fig.\ref{Message Passing Process Comparsion}, we first attempted to maintain the directed edge encoding approach while allowing the Cell Block of FVGN to perform message aggregation only once.
That is, in the Cell Block process shown in Fig.\ref{Message Passing Process Comparsion}, we directly aggregate all the edge features $e_{i,j}$ constituting a single cell to that cell, then concatenate the original features of that cell and enter into the Cell-MLP $\phi^{cp}$ for feature update. This operation is very similar to how the Node Block in MeshGraphnets aggregates shared edge features to the vertex and concatenates the original features at the vertex before entering the Node-MLP for feature updating (as shown in Fig.\ref{Message Passing Process Comparsion} d.). From both the error accumulation curve in Fig.\ref{Hybrid_P_comp} for FVGN(MA=1) and the performance results in Tab.\ref{tab: hybrid dataset result comparison} for FVGN(MA=1), it's evident that the performance is significantly inferior compared to FVGN using the twice message aggregating mechanism. At the same time, FVGN(MA=1) in Tab.\ref{tab: hybrid dataset result comparison} has only a slight improvement in training efficiency, which is negligible compared to the significant loss in accuracy. Therefore, it is necessary to introduce a twice-message aggregating mechanism in the message-passing layer.

\subsection{Computational Efficiency Comparision}\label{Computation efficency}
In this section, we mainly compare the efficiency differences between the message-passing methods in the forward process of FVGN and MeshGraphNets and mainly show the details of the twice-message aggregating process proposed in this research.

\begin{figure}[H]
\begin{minipage}{\linewidth}
    \centering
    \includegraphics[width=1\textwidth]{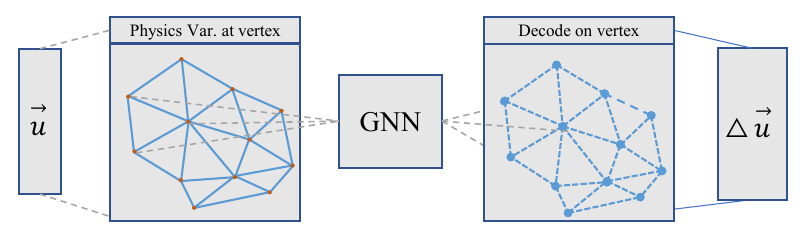}
    \caption{MeshGraphNets Forward Process}
    \label{meshgraphnet-forward-process}
\end{minipage}
\end{figure}

As seen in Fig.\ref{FVGN-foward-process}, FVGN incorporates an additional operation, "Integrate to center," compared to MeshGraphNets(Fig.\ref{meshgraphnet-forward-process}). We refer to this process as the Spatial Integration Layer (abbreviated as SIL). The introduction of the SIL layer significantly enhances the model's performance, and we conduct ablation experiments in Sec.\ref{tab: hybrid dataset result comparison} to verify its effectiveness. Furthermore, the inclusion of the SIL layer inevitably increases the model's time complexity; however, this increase is acceptable when implemented on a GPU. The time complexity of the SIL layer implemented on a CPU can be expressed as O(2N+3N), which approximates O(n). Nonetheless, on a GPU, this time complexity becomes O(1) during an ideal situation, as the SIL layer only involves multiplying each face's flux by the current surface vector and summing the results. The computation for each face in every control volume is executed simultaneously on a GPU. Therefore, the forward-time consumption of our model compared to MeshGraphNets experiences virtually increases linearly.

\begin{table}
\caption{The approximate ratio of Vertex to elements and faces in unstructured triangular meshes}
\label{tab: ratio of Vertex to elements and faces}
\begin{ruledtabular}
\begin{tabular}{ccc}
\multicolumn{3}{c}{Triangular Mesh} \\
Vertex    & Cell    & Face   \\ \hline
$N$       & $2N$    & $3N$ 
\end{tabular}
\end{ruledtabular}
\end{table}

\begin{table}
\centering
\caption{FVGN's forward time-consuming during inferencing, Inference timings of our model per step on a single GPU, for pure neural network inference and including graph recomputation (tfull). Our model has a significantly lower running cost compared to the ground truth simulation (tGT) but not much beyond MeshGraphNets. }
\label{Forward time consuming during inferencing}
\begin{ruledtabular}
\begin{tabular}{ccccccc}
\textbf{Model} &
  \textbf{Dataset} &
  \textbf{\begin{tabular}[c]{@{}c@{}}\#nodes/cells\\ (avg.)\end{tabular}} &
  \textbf{\#steps} &
  \begin{tabular}[c]{@{}c@{}}GPU\\ tfull\\ ms/step\end{tabular} &
  \begin{tabular}[c]{@{}c@{}}CPU \\ tfull \\ ms/step\end{tabular} &
  \begin{tabular}[c]{@{}c@{}}CPU\\ tGT\\ ms/step\end{tabular} \\ \hline
\multirow{2}{*}{MGN}  & \begin{tabular}[c]{@{}c@{}}CYLINDER\\ FLOW\end{tabular} & 1885 & 600 & 23 & 168 & 820  \\
                      & \begin{tabular}[c]{@{}c@{}}HYBRID\\ FLOW\end{tabular}   & 4000 & 600 & 30 & 224 & 1520 \\ \hline
\multirow{2}{*}{FVGN} & \begin{tabular}[c]{@{}c@{}}CYLINDER\\ FLOW\end{tabular} & 3500 & 600 & 57 & 354 & 820  \\
                      & \begin{tabular}[c]{@{}c@{}}HYBRID\\ FLOW\end{tabular}   & 7500 & 600 & 88 & 455 & 1520
\end{tabular}
\end{ruledtabular}
\end{table}

Tab.\ref{Forward time consuming during inferencing} specifically demonstrates the GPU time consumed by the FVGN during a single forward pass after the additional introduction of the spatial integration layer. As can be seen from the data in the table, the forward process of FVGN does not show a significant increase compared to MeshGraphNets. Moreover, all the GPU data presented here were obtained on a single V100S GPU. For MeshGraphNets' GPU/CPU time consumption on the CYLINDERFLOW dataset, we directly adopt the data from \cite{pfaff2020learning}. It is also worth noting that the input to MeshGraphNets is only a single mesh $G_{v}$, so the decoding target for MeshGraphNets is limited to vertices. In triangular mesh, the number of vertices is often less than the number of cells, with the ratio for the same mesh being approximately 1:2 and the ratio of vertices: edges is 1:3 (in Tab.\ref{tab: ratio of Vertex to elements and faces}). Therefore, it is reasonable for FVGN to consume more decoding time than MeshGraphNets. However, Tab.\ref{Forward time consuming during inferencing} also presents the time consumed when testing our model on a CPU. We are only demonstrating a trend in time consumption here, which may not accurately reflect the actual time consumed during real-world applications, as most numerical solvers are based on CPU computation. Finally, from the table, it can be observed that both our method and MeshGraphNets have a time consumption advantage of about five times compared to traditional numerical solvers in the CFD domain.

\section{Conclusion}\label{conclusion}
In this study, by integrating the concept of integration in the finite volume method, we propose a novel Spatial Integration Layer (S.I.L) based on graph neural networks to implement physical constraints. 
This significantly improves the predictive accuracy of graph neural networks in scenarios of unsteady flow prediction. 
At the same time, we utilized the extended stencil method in the finite volume method to construct a twice-message aggregation process in the message-passing layer of the graph neural network. 
This greatly enhances the efficiency of message passing in the case of unidirectional edges, thereby compensating for the performance loss due to unidirectional edges.
We can summarize all the improvements in this study as follows: the improvement in training efficiency is due to our transformation of the undirected graph with bidirectional edges into a directed graph with unidirectional edges, thereby reducing computational overhead during the training process. The increase in prediction accuracy is mainly due to the introduction of physical constraints based on the finite volume method and the introduction of the twice-message aggregation process. On this basis, due to the higher message passing efficiency, we only used the data from the first 50\% of time steps of the CYLINDERFLOW train-set and the first 66\% of time steps of the HYBRIDFLOW train-set, as well as a lower number of training rounds in the training process of FVGN, further significantly reducing the training overhead. Ultimately, these improvements have enabled the model proposed in this study to make significant progress in both accuracy and efficiency.

To verify the effectiveness of the aforementioned improvements, this study tested the performance of FVGN on two datasets, comprising 2000 cases with different Reynolds numbers and geometric shapes, and compared it with MeshGraphNets based on fully data-driven methods. The results show that when trained with a limited number of time-step data, FVGN outperforms MeshGraphNets in terms of training time overhead and final prediction accuracy. In the HYBRIDFLOW test results, FVGN demonstrated a 77\% improvement in accuracy in velocity field prediction compared to MeshGraphNets, and the training time overhead was reduced by up to 56\%. This study also selected three representative test cases from the HYBRIDFLOW test set for comparative analysis of surface pressure distribution and lift-drag coefficient curves. From these results, FVGN is superior to MeshGraphNets in the majority of cases.

Finally, this study also conducted tests on the generalizability of geometric shapes to verify that models incorporating physical inductive biases have better generalization capabilities compared to fully data-driven models. From the final results, FVGN was more accurate in capturing the shedding frequency compared to MeshGraphNets, but there were still significant errors in predicting the lift-drag coefficients. Especially in the elliptical cases, FVGN's large prediction error in the pressure field led to poor predictions of pressure differential drag in the x-direction, which directly affected the final drag coefficient, differing by nearly 6\% from the CFD results. However, this result is still significantly better than the 18\% drag coefficient prediction error of MeshGraphNets.

When we conducted the generality tests on the geometry, we controlled the mesh density and time step length of the test cases to be essentially consistent with those in the training set. This is because, based on our practice, if the number of grid cells used during testing deviates too much from the average mesh density of the training set or if different time steps are taken, then the test results tend to be poor. Therefore, this will also be a focus of our future work.

\blueText{From the results above, it is evident that although FVGN has made progress in terms of accuracy and efficiency compared to existing works, it still faces two major challenges in practical applications. Firstly, due to the reduction in the size of the latent space and under the training conditions used in this paper, when facing the more complex HYRBRIDFLOW dataset, if the velocity field at the initial moment of flow field initialization is used as the input at time t=0, the flow field predicted by FVGN at the critical moments of generating unsteady states may differ from the CFD results, thus leading to a phase difference. The presence of a phase difference makes the calculation of RMSE less ideal. Therefore, in constructing the HYRBRIDFLOW dataset, we actually used the flow fields at physical time steps <15,614> as training data, that is, we allowed FVGN to start predicting from the critical moment of generating unsteady phenomena, rather than training and predicting FVGN from the physical time step t=0. Although choosing a larger latent space size and extending the number of training steps might alleviate this issue, it would inevitably lead to greater time and spatial complexity, thereby deviating from the goal of this paper to "balance model performance with computational costs." Secondly, FVGN and most data-driven models share a core problem in the task of predicting unsteady flow fields: the acquisition of high-accuracy datasets. Especially, generating high-accuracy unsteady datasets under unstructured grid discretization is very time-consuming. This challenge not only affects the effectiveness of model training but may also limit its applicability in a wider range of scenarios. Therefore, in future research, we plan to attempt to introduce the complete discretization process of the finite volume method on unstructured grids into graph neural networks to achieve fully unsupervised training and prediction. Thus, achieving the goal of truly getting rid of the dependency on high-accuracy datasets while improving FVGN's ability to predict unsteady flow fields at any initial moment across a wider range of Reynolds numbers.}

\section*{Acknowledgments}
This research is funded by the National Key Project of China (Grant No. GJXM92579) and is also supported by the Sichuan Science and Technology Program (Project No. 2023YFG0158).

\section*{Supplementary Material}
The model configuration details are provided in this section. It outlines the setup of the model architecture, and a pseudo code is also presented. It details the training process, which involves accumulating samples, injecting noise into training data for robustness, normalizing node and edge features, and using an encoder-processor-decoder architecture.

Finally, it also mentions the use of a noise injection strategy for training, similar to that in Meshgraphnets, to enhance model robustness in long-term prediction scenarios. This approach is particularly noted for its application in the HYBRIDFLOW and CYLINDERFLOW datasets.

\section*{Data Availability Statement}

The data that support the ﬁndings of this study are available from the corresponding author upon reasonable request.

\section*{REFERENCES}
\bibliography{reference}

\begin{thebibliography}{39}%
\makeatletter
\providecommand \@ifxundefined [1]{%
 \@ifx{#1\undefined}
}%
\providecommand \@ifnum [1]{%
 \ifnum #1\expandafter \@firstoftwo
 \else \expandafter \@secondoftwo
 \fi
}%
\providecommand \@ifx [1]{%
 \ifx #1\expandafter \@firstoftwo
 \else \expandafter \@secondoftwo
 \fi
}%
\providecommand \natexlab [1]{#1}%
\providecommand \enquote  [1]{``#1''}%
\providecommand \bibnamefont  [1]{#1}%
\providecommand \bibfnamefont [1]{#1}%
\providecommand \citenamefont [1]{#1}%
\providecommand \href@noop [0]{\@secondoftwo}%
\providecommand \href [0]{\begingroup \@sanitize@url \@href}%
\providecommand \@href[1]{\@@startlink{#1}\@@href}%
\providecommand \@@href[1]{\endgroup#1\@@endlink}%
\providecommand \@sanitize@url [0]{\catcode `\\12\catcode `\$12\catcode `\&12\catcode `\#12\catcode `\^12\catcode `\_12\catcode `\%12\relax}%
\providecommand \@@startlink[1]{}%
\providecommand \@@endlink[0]{}%
\providecommand \url  [0]{\begingroup\@sanitize@url \@url }%
\providecommand \@url [1]{\endgroup\@href {#1}{\urlprefix }}%
\providecommand \urlprefix  [0]{URL }%
\providecommand \Eprint [0]{\href }%
\providecommand \doibase [0]{http://dx.doi.org/}%
\providecommand \selectlanguage [0]{\@gobble}%
\providecommand \bibinfo  [0]{\@secondoftwo}%
\providecommand \bibfield  [0]{\@secondoftwo}%
\providecommand \translation [1]{[#1]}%
\providecommand \BibitemOpen [0]{}%
\providecommand \bibitemStop [0]{}%
\providecommand \bibitemNoStop [0]{.\EOS\space}%
\providecommand \EOS [0]{\spacefactor3000\relax}%
\providecommand \BibitemShut  [1]{\csname bibitem#1\endcsname}%
\let\auto@bib@innerbib\@empty
\bibitem [{\citenamefont {Kochkov}\ \emph {et~al.}(2021)\citenamefont {Kochkov}, \citenamefont {Smith}, \citenamefont {Alieva}, \citenamefont {Wang}, \citenamefont {Brenner},\ and\ \citenamefont {Hoyer}}]{kochkov2021machine}%
  \BibitemOpen
  \bibfield  {author} {\bibinfo {author} {\bibfnamefont {D.}~\bibnamefont {Kochkov}}, \bibinfo {author} {\bibfnamefont {J.~A.}\ \bibnamefont {Smith}}, \bibinfo {author} {\bibfnamefont {A.}~\bibnamefont {Alieva}}, \bibinfo {author} {\bibfnamefont {Q.}~\bibnamefont {Wang}}, \bibinfo {author} {\bibfnamefont {M.~P.}\ \bibnamefont {Brenner}}, \ and\ \bibinfo {author} {\bibfnamefont {S.}~\bibnamefont {Hoyer}},\ }\bibfield  {title} {\enquote {\bibinfo {title} {Machine learning--accelerated computational fluid dynamics},}\ }\href@noop {} {\bibfield  {journal} {\bibinfo  {journal} {Proceedings of the National Academy of Sciences}\ }\textbf {\bibinfo {volume} {118}},\ \bibinfo {pages} {e2101784118} (\bibinfo {year} {2021})}\BibitemShut {NoStop}%
\bibitem [{\citenamefont {Raissi}, \citenamefont {Perdikaris},\ and\ \citenamefont {Karniadakis}(2019)}]{raissi2019physics}%
  \BibitemOpen
  \bibfield  {author} {\bibinfo {author} {\bibfnamefont {M.}~\bibnamefont {Raissi}}, \bibinfo {author} {\bibfnamefont {P.}~\bibnamefont {Perdikaris}}, \ and\ \bibinfo {author} {\bibfnamefont {G.~E.}\ \bibnamefont {Karniadakis}},\ }\bibfield  {title} {\enquote {\bibinfo {title} {Physics-informed neural networks: A deep learning framework for solving forward and inverse problems involving nonlinear partial differential equations},}\ }\href@noop {} {\bibfield  {journal} {\bibinfo  {journal} {Journal of Computational physics}\ }\textbf {\bibinfo {volume} {378}},\ \bibinfo {pages} {686--707} (\bibinfo {year} {2019})}\BibitemShut {NoStop}%
\bibitem [{\citenamefont {Rao}, \citenamefont {Sun},\ and\ \citenamefont {Liu}(2020)}]{rao_physics-informed_2020}%
  \BibitemOpen
  \bibfield  {author} {\bibinfo {author} {\bibfnamefont {C.}~\bibnamefont {Rao}}, \bibinfo {author} {\bibfnamefont {H.}~\bibnamefont {Sun}}, \ and\ \bibinfo {author} {\bibfnamefont {Y.}~\bibnamefont {Liu}},\ }\bibfield  {title} {\enquote {\bibinfo {title} {Physics-informed deep learning for incompressible laminar flows},}\ }\href@noop {} {\bibfield  {journal} {\bibinfo  {journal} {Theoretical and Applied Mechanics Letters}\ }\textbf {\bibinfo {volume} {10}},\ \bibinfo {pages} {207--212} (\bibinfo {year} {2020})}\BibitemShut {NoStop}%
\bibitem [{\citenamefont {Jin}\ \emph {et~al.}(2021)\citenamefont {Jin}, \citenamefont {Cai}, \citenamefont {Li},\ and\ \citenamefont {Karniadakis}}]{jin_nsfnets_2021}%
  \BibitemOpen
  \bibfield  {author} {\bibinfo {author} {\bibfnamefont {X.}~\bibnamefont {Jin}}, \bibinfo {author} {\bibfnamefont {S.}~\bibnamefont {Cai}}, \bibinfo {author} {\bibfnamefont {H.}~\bibnamefont {Li}}, \ and\ \bibinfo {author} {\bibfnamefont {G.~E.}\ \bibnamefont {Karniadakis}},\ }\bibfield  {title} {\enquote {\bibinfo {title} {{NSFnets} ({Navier}-{Stokes} flow nets): {Physics}-informed neural networks for the incompressible {Navier}-{Stokes} equations},}\ }\href {\doibase 10.1016/j.jcp.2020.109951} {\bibfield  {journal} {\bibinfo  {journal} {Journal of Computational Physics}\ }\textbf {\bibinfo {volume} {426}},\ \bibinfo {pages} {109951} (\bibinfo {year} {2021})}\BibitemShut {NoStop}%
\bibitem [{\citenamefont {Li}\ \emph {et~al.}(2021)\citenamefont {Li}, \citenamefont {Zheng}, \citenamefont {Kovachki}, \citenamefont {Jin}, \citenamefont {Chen}, \citenamefont {Liu}, \citenamefont {Azizzadenesheli},\ and\ \citenamefont {Anandkumar}}]{li2021physics}%
  \BibitemOpen
  \bibfield  {author} {\bibinfo {author} {\bibfnamefont {Z.}~\bibnamefont {Li}}, \bibinfo {author} {\bibfnamefont {H.}~\bibnamefont {Zheng}}, \bibinfo {author} {\bibfnamefont {N.}~\bibnamefont {Kovachki}}, \bibinfo {author} {\bibfnamefont {D.}~\bibnamefont {Jin}}, \bibinfo {author} {\bibfnamefont {H.}~\bibnamefont {Chen}}, \bibinfo {author} {\bibfnamefont {B.}~\bibnamefont {Liu}}, \bibinfo {author} {\bibfnamefont {K.}~\bibnamefont {Azizzadenesheli}}, \ and\ \bibinfo {author} {\bibfnamefont {A.}~\bibnamefont {Anandkumar}},\ }\bibfield  {title} {\enquote {\bibinfo {title} {Physics-informed neural operator for learning partial differential equations},}\ }\href@noop {} {\bibfield  {journal} {\bibinfo  {journal} {arXiv preprint arXiv:2111.03794}\ } (\bibinfo {year} {2021})}\BibitemShut {NoStop}%
\bibitem [{\citenamefont {Wandel}\ \emph {et~al.}(2022)\citenamefont {Wandel}, \citenamefont {Weinmann}, \citenamefont {Neidlin},\ and\ \citenamefont {Klein}}]{wandel_spline-pinn_2022}%
  \BibitemOpen
  \bibfield  {author} {\bibinfo {author} {\bibfnamefont {N.}~\bibnamefont {Wandel}}, \bibinfo {author} {\bibfnamefont {M.}~\bibnamefont {Weinmann}}, \bibinfo {author} {\bibfnamefont {M.}~\bibnamefont {Neidlin}}, \ and\ \bibinfo {author} {\bibfnamefont {R.}~\bibnamefont {Klein}},\ }\href {http://arxiv.org/abs/2109.07143} {\enquote {\bibinfo {title} {Spline-{PINN}: {Approaching} {PDEs} without {Data} using {Fast}, {Physics}-{Informed} {Hermite}-{Spline} {CNNs}},}\ } (\bibinfo {year} {2022}),\ \bibinfo {note} {arXiv:2109.07143 [physics]}\BibitemShut {NoStop}%
\bibitem [{\citenamefont {Arzani}, \citenamefont {Wang},\ and\ \citenamefont {D'Souza}(2021)}]{arzani2021uncovering}%
  \BibitemOpen
  \bibfield  {author} {\bibinfo {author} {\bibfnamefont {A.}~\bibnamefont {Arzani}}, \bibinfo {author} {\bibfnamefont {J.-X.}\ \bibnamefont {Wang}}, \ and\ \bibinfo {author} {\bibfnamefont {R.~M.}\ \bibnamefont {D'Souza}},\ }\bibfield  {title} {\enquote {\bibinfo {title} {Uncovering near-wall blood flow from sparse data with physics-informed neural networks},}\ }\href@noop {} {\bibfield  {journal} {\bibinfo  {journal} {Physics of Fluids}\ }\textbf {\bibinfo {volume} {33}} (\bibinfo {year} {2021})}\BibitemShut {NoStop}%
\bibitem [{\citenamefont {Baydin}\ \emph {et~al.}(2018)\citenamefont {Baydin}, \citenamefont {Pearlmutter}, \citenamefont {Radul},\ and\ \citenamefont {Siskind}}]{baydin2018automatic}%
  \BibitemOpen
  \bibfield  {author} {\bibinfo {author} {\bibfnamefont {A.~G.}\ \bibnamefont {Baydin}}, \bibinfo {author} {\bibfnamefont {B.~A.}\ \bibnamefont {Pearlmutter}}, \bibinfo {author} {\bibfnamefont {A.~A.}\ \bibnamefont {Radul}}, \ and\ \bibinfo {author} {\bibfnamefont {J.~M.}\ \bibnamefont {Siskind}},\ }\bibfield  {title} {\enquote {\bibinfo {title} {Automatic differentiation in machine learning: a survey},}\ }\href@noop {} {\bibfield  {journal} {\bibinfo  {journal} {Journal of Marchine Learning Research}\ }\textbf {\bibinfo {volume} {18}},\ \bibinfo {pages} {1--43} (\bibinfo {year} {2018})}\BibitemShut {NoStop}%
\bibitem [{\citenamefont {Pfaff}\ \emph {et~al.}(2020)\citenamefont {Pfaff}, \citenamefont {Fortunato}, \citenamefont {Sanchez-Gonzalez},\ and\ \citenamefont {Battaglia}}]{pfaff2020learning}%
  \BibitemOpen
  \bibfield  {author} {\bibinfo {author} {\bibfnamefont {T.}~\bibnamefont {Pfaff}}, \bibinfo {author} {\bibfnamefont {M.}~\bibnamefont {Fortunato}}, \bibinfo {author} {\bibfnamefont {A.}~\bibnamefont {Sanchez-Gonzalez}}, \ and\ \bibinfo {author} {\bibfnamefont {P.~W.}\ \bibnamefont {Battaglia}},\ }\bibfield  {title} {\enquote {\bibinfo {title} {Learning mesh-based simulation with graph networks},}\ }\href@noop {} {\bibfield  {journal} {\bibinfo  {journal} {arXiv preprint arXiv:2010.03409}\ } (\bibinfo {year} {2020})}\BibitemShut {NoStop}%
\bibitem [{\citenamefont {Ranade}, \citenamefont {Hill},\ and\ \citenamefont {Pathak}(2021)}]{ranade2021discretizationnet}%
  \BibitemOpen
  \bibfield  {author} {\bibinfo {author} {\bibfnamefont {R.}~\bibnamefont {Ranade}}, \bibinfo {author} {\bibfnamefont {C.}~\bibnamefont {Hill}}, \ and\ \bibinfo {author} {\bibfnamefont {J.}~\bibnamefont {Pathak}},\ }\bibfield  {title} {\enquote {\bibinfo {title} {Discretizationnet: A machine-learning based solver for navier--stokes equations using finite volume discretization},}\ }\href@noop {} {\bibfield  {journal} {\bibinfo  {journal} {Computer Methods in Applied Mechanics and Engineering}\ }\textbf {\bibinfo {volume} {378}},\ \bibinfo {pages} {113722} (\bibinfo {year} {2021})}\BibitemShut {NoStop}%
\bibitem [{\citenamefont {Wandel}, \citenamefont {Weinmann},\ and\ \citenamefont {Klein}(2020)}]{wandel2020learning}%
  \BibitemOpen
  \bibfield  {author} {\bibinfo {author} {\bibfnamefont {N.}~\bibnamefont {Wandel}}, \bibinfo {author} {\bibfnamefont {M.}~\bibnamefont {Weinmann}}, \ and\ \bibinfo {author} {\bibfnamefont {R.}~\bibnamefont {Klein}},\ }\bibfield  {title} {\enquote {\bibinfo {title} {Learning incompressible fluid dynamics from scratch--towards fast, differentiable fluid models that generalize},}\ }\href@noop {} {\bibfield  {journal} {\bibinfo  {journal} {arXiv preprint arXiv:2006.08762}\ } (\bibinfo {year} {2020})}\BibitemShut {NoStop}%
\bibitem [{\citenamefont {Wandel}, \citenamefont {Weinmann},\ and\ \citenamefont {Klein}(2021)}]{wandel_teaching_2021}%
  \BibitemOpen
  \bibfield  {author} {\bibinfo {author} {\bibfnamefont {N.}~\bibnamefont {Wandel}}, \bibinfo {author} {\bibfnamefont {M.}~\bibnamefont {Weinmann}}, \ and\ \bibinfo {author} {\bibfnamefont {R.}~\bibnamefont {Klein}},\ }\bibfield  {title} {\enquote {\bibinfo {title} {Teaching the {Incompressible} {Navier}-{Stokes} {Equations} to {Fast} {Neural} {Surrogate} {Models} in {3D}},}\ }\href {\doibase 10.1063/5.0047428} {\bibfield  {journal} {\bibinfo  {journal} {Physics of Fluids}\ }\textbf {\bibinfo {volume} {33}},\ \bibinfo {pages} {047117} (\bibinfo {year} {2021})},\ \bibinfo {note} {arXiv:2012.11893 [physics]}\BibitemShut {NoStop}%
\bibitem [{\citenamefont {Gao}, \citenamefont {Sun},\ and\ \citenamefont {Wang}(2021)}]{gao_phygeonet_2021}%
  \BibitemOpen
  \bibfield  {author} {\bibinfo {author} {\bibfnamefont {H.}~\bibnamefont {Gao}}, \bibinfo {author} {\bibfnamefont {L.}~\bibnamefont {Sun}}, \ and\ \bibinfo {author} {\bibfnamefont {J.-X.}\ \bibnamefont {Wang}},\ }\bibfield  {title} {\enquote {\bibinfo {title} {{PhyGeoNet}: {Physics}-informed geometry-adaptive convolutional neural networks for solving parameterized steady-state {PDEs} on irregular domain},}\ }\href {\doibase 10.1016/j.jcp.2020.110079} {\bibfield  {journal} {\bibinfo  {journal} {Journal of Computational Physics}\ }\textbf {\bibinfo {volume} {428}},\ \bibinfo {pages} {110079} (\bibinfo {year} {2021})}\BibitemShut {NoStop}%
\bibitem [{\citenamefont {Gao}, \citenamefont {Zahr},\ and\ \citenamefont {Wang}(2022)}]{gao_physics-informed_2022}%
  \BibitemOpen
  \bibfield  {author} {\bibinfo {author} {\bibfnamefont {H.}~\bibnamefont {Gao}}, \bibinfo {author} {\bibfnamefont {M.~J.}\ \bibnamefont {Zahr}}, \ and\ \bibinfo {author} {\bibfnamefont {J.-X.}\ \bibnamefont {Wang}},\ }\bibfield  {title} {\enquote {\bibinfo {title} {Physics-informed graph neural galerkin networks: A unified framework for solving pde-governed forward and inverse problems},}\ }\href@noop {} {\bibfield  {journal} {\bibinfo  {journal} {Computer Methods in Applied Mechanics and Engineering}\ }\textbf {\bibinfo {volume} {390}},\ \bibinfo {pages} {114502} (\bibinfo {year} {2022})}\BibitemShut {NoStop}%
\bibitem [{\citenamefont {Gao}, \citenamefont {Deo},\ and\ \citenamefont {Jaiman}(2022)}]{gao2022finite}%
  \BibitemOpen
  \bibfield  {author} {\bibinfo {author} {\bibfnamefont {R.}~\bibnamefont {Gao}}, \bibinfo {author} {\bibfnamefont {I.~K.}\ \bibnamefont {Deo}}, \ and\ \bibinfo {author} {\bibfnamefont {R.~K.}\ \bibnamefont {Jaiman}},\ }\bibfield  {title} {\enquote {\bibinfo {title} {A finite element-inspired hypergraph neural network: Application to fluid dynamics simulations},}\ }\href@noop {} {\bibfield  {journal} {\bibinfo  {journal} {Available at SSRN 4462715}\ } (\bibinfo {year} {2022})}\BibitemShut {NoStop}%
\bibitem [{\citenamefont {Moukalled}, \citenamefont {Mangani},\ and\ \citenamefont {Darwish}(2016)}]{moukalled2016finite}%
  \BibitemOpen
  \bibfield  {author} {\bibinfo {author} {\bibfnamefont {F.}~\bibnamefont {Moukalled}}, \bibinfo {author} {\bibfnamefont {L.}~\bibnamefont {Mangani}}, \ and\ \bibinfo {author} {\bibfnamefont {M.}~\bibnamefont {Darwish}},\ }\bibfield  {title} {\enquote {\bibinfo {title} {The finite volume method},}\ }in\ \href@noop {} {\emph {\bibinfo {booktitle} {The finite volume method in computational fluid dynamics}}}\ (\bibinfo  {publisher} {Springer},\ \bibinfo {year} {2016})\ pp.\ \bibinfo {pages} {103--135}\BibitemShut {NoStop}%
\bibitem [{\citenamefont {Praditia}\ \emph {et~al.}(2021)\citenamefont {Praditia}, \citenamefont {Karlbauer}, \citenamefont {Otte}, \citenamefont {Oladyshkin}, \citenamefont {Butz},\ and\ \citenamefont {Nowak}}]{praditia_finite_2021}%
  \BibitemOpen
  \bibfield  {author} {\bibinfo {author} {\bibfnamefont {T.}~\bibnamefont {Praditia}}, \bibinfo {author} {\bibfnamefont {M.}~\bibnamefont {Karlbauer}}, \bibinfo {author} {\bibfnamefont {S.}~\bibnamefont {Otte}}, \bibinfo {author} {\bibfnamefont {S.}~\bibnamefont {Oladyshkin}}, \bibinfo {author} {\bibfnamefont {M.~V.}\ \bibnamefont {Butz}}, \ and\ \bibinfo {author} {\bibfnamefont {W.}~\bibnamefont {Nowak}},\ }\bibfield  {title} {\enquote {\bibinfo {title} {Finite volume neural network: Modeling subsurface contaminant transport},}\ }\href@noop {} {\bibfield  {journal} {\bibinfo  {journal} {arXiv preprint arXiv:2104.06010}\ } (\bibinfo {year} {2021})}\BibitemShut {NoStop}%
\bibitem [{\citenamefont {Karlbauer}\ \emph {et~al.}(2022)\citenamefont {Karlbauer}, \citenamefont {Praditia}, \citenamefont {Otte}, \citenamefont {Oladyshkin}, \citenamefont {Nowak},\ and\ \citenamefont {Butz}}]{karlbauer_composing_2022}%
  \BibitemOpen
  \bibfield  {author} {\bibinfo {author} {\bibfnamefont {M.}~\bibnamefont {Karlbauer}}, \bibinfo {author} {\bibfnamefont {T.}~\bibnamefont {Praditia}}, \bibinfo {author} {\bibfnamefont {S.}~\bibnamefont {Otte}}, \bibinfo {author} {\bibfnamefont {S.}~\bibnamefont {Oladyshkin}}, \bibinfo {author} {\bibfnamefont {W.}~\bibnamefont {Nowak}}, \ and\ \bibinfo {author} {\bibfnamefont {M.~V.}\ \bibnamefont {Butz}},\ }\href {http://arxiv.org/abs/2111.11798} {\enquote {\bibinfo {title} {Composing {Partial} {Differential} {Equations} with {Physics}-{Aware} {Neural} {Networks}},}\ } (\bibinfo {year} {2022}),\ \bibinfo {note} {arXiv:2111.11798 [cs]}\BibitemShut {NoStop}%
\bibitem [{\citenamefont {Chen}\ and\ \citenamefont {Thuerey}(2023)}]{chen_towards_2021}%
  \BibitemOpen
  \bibfield  {author} {\bibinfo {author} {\bibfnamefont {L.-W.}\ \bibnamefont {Chen}}\ and\ \bibinfo {author} {\bibfnamefont {N.}~\bibnamefont {Thuerey}},\ }\bibfield  {title} {\enquote {\bibinfo {title} {Towards high-accuracy deep learning inference of compressible flows over aerofoils},}\ }\href@noop {} {\bibfield  {journal} {\bibinfo  {journal} {Computers \& Fluids}\ }\textbf {\bibinfo {volume} {250}},\ \bibinfo {pages} {105707} (\bibinfo {year} {2023})}\BibitemShut {NoStop}%
\bibitem [{\citenamefont {Brahmachary}\ and\ \citenamefont {Thuerey}(2023)}]{brahmachary2023unsteady}%
  \BibitemOpen
  \bibfield  {author} {\bibinfo {author} {\bibfnamefont {S.}~\bibnamefont {Brahmachary}}\ and\ \bibinfo {author} {\bibfnamefont {N.}~\bibnamefont {Thuerey}},\ }\bibfield  {title} {\enquote {\bibinfo {title} {Unsteady cylinder wakes from arbitrary bodies with differentiable physics-assisted neural network},}\ }\href@noop {} {\bibfield  {journal} {\bibinfo  {journal} {arXiv preprint arXiv:2308.04296}\ } (\bibinfo {year} {2023})}\BibitemShut {NoStop}%
\bibitem [{\citenamefont {Gilmer}\ \emph {et~al.}(2017)\citenamefont {Gilmer}, \citenamefont {Schoenholz}, \citenamefont {Riley}, \citenamefont {Vinyals},\ and\ \citenamefont {Dahl}}]{gilmer_neural_nodate}%
  \BibitemOpen
  \bibfield  {author} {\bibinfo {author} {\bibfnamefont {J.}~\bibnamefont {Gilmer}}, \bibinfo {author} {\bibfnamefont {S.~S.}\ \bibnamefont {Schoenholz}}, \bibinfo {author} {\bibfnamefont {P.~F.}\ \bibnamefont {Riley}}, \bibinfo {author} {\bibfnamefont {O.}~\bibnamefont {Vinyals}}, \ and\ \bibinfo {author} {\bibfnamefont {G.~E.}\ \bibnamefont {Dahl}},\ }\bibfield  {title} {\enquote {\bibinfo {title} {Neural message passing for quantum chemistry},}\ }in\ \href@noop {} {\emph {\bibinfo {booktitle} {International conference on machine learning}}}\ (\bibinfo {organization} {PMLR},\ \bibinfo {year} {2017})\ pp.\ \bibinfo {pages} {1263--1272}\BibitemShut {NoStop}%
\bibitem [{\citenamefont {Battaglia}\ \emph {et~al.}(2018)\citenamefont {Battaglia}, \citenamefont {Hamrick}, \citenamefont {Bapst}, \citenamefont {Sanchez-Gonzalez}, \citenamefont {Zambaldi}, \citenamefont {Malinowski}, \citenamefont {Tacchetti}, \citenamefont {Raposo}, \citenamefont {Santoro}, \citenamefont {Faulkner} \emph {et~al.}}]{battaglia2018relational}%
  \BibitemOpen
  \bibfield  {author} {\bibinfo {author} {\bibfnamefont {P.~W.}\ \bibnamefont {Battaglia}}, \bibinfo {author} {\bibfnamefont {J.~B.}\ \bibnamefont {Hamrick}}, \bibinfo {author} {\bibfnamefont {V.}~\bibnamefont {Bapst}}, \bibinfo {author} {\bibfnamefont {A.}~\bibnamefont {Sanchez-Gonzalez}}, \bibinfo {author} {\bibfnamefont {V.}~\bibnamefont {Zambaldi}}, \bibinfo {author} {\bibfnamefont {M.}~\bibnamefont {Malinowski}}, \bibinfo {author} {\bibfnamefont {A.}~\bibnamefont {Tacchetti}}, \bibinfo {author} {\bibfnamefont {D.}~\bibnamefont {Raposo}}, \bibinfo {author} {\bibfnamefont {A.}~\bibnamefont {Santoro}}, \bibinfo {author} {\bibfnamefont {R.}~\bibnamefont {Faulkner}},  \emph {et~al.},\ }\bibfield  {title} {\enquote {\bibinfo {title} {Relational inductive biases, deep learning, and graph networks},}\ }\href@noop {} {\bibfield  {journal} {\bibinfo  {journal} {arXiv preprint arXiv:1806.01261}\ } (\bibinfo {year} {2018})}\BibitemShut {NoStop}%
\bibitem [{\citenamefont {Zhou}\ \emph {et~al.}(2020)\citenamefont {Zhou}, \citenamefont {Cui}, \citenamefont {Hu}, \citenamefont {Zhang}, \citenamefont {Yang}, \citenamefont {Liu}, \citenamefont {Wang}, \citenamefont {Li},\ and\ \citenamefont {Sun}}]{zhou_graph_2020}%
  \BibitemOpen
  \bibfield  {author} {\bibinfo {author} {\bibfnamefont {J.}~\bibnamefont {Zhou}}, \bibinfo {author} {\bibfnamefont {G.}~\bibnamefont {Cui}}, \bibinfo {author} {\bibfnamefont {S.}~\bibnamefont {Hu}}, \bibinfo {author} {\bibfnamefont {Z.}~\bibnamefont {Zhang}}, \bibinfo {author} {\bibfnamefont {C.}~\bibnamefont {Yang}}, \bibinfo {author} {\bibfnamefont {Z.}~\bibnamefont {Liu}}, \bibinfo {author} {\bibfnamefont {L.}~\bibnamefont {Wang}}, \bibinfo {author} {\bibfnamefont {C.}~\bibnamefont {Li}}, \ and\ \bibinfo {author} {\bibfnamefont {M.}~\bibnamefont {Sun}},\ }\bibfield  {title} {\enquote {\bibinfo {title} {Graph neural networks: {A} review of methods and applications},}\ }\href {\doibase 10.1016/j.aiopen.2021.01.001} {\bibfield  {journal} {\bibinfo  {journal} {AI Open}\ }\textbf {\bibinfo {volume} {1}},\ \bibinfo {pages} {57--81} (\bibinfo {year} {2020})}\BibitemShut {NoStop}%
\bibitem [{\citenamefont {Horie}\ and\ \citenamefont {Mitsume}(2023)}]{horie_physics-embedded_2023}%
  \BibitemOpen
  \bibfield  {author} {\bibinfo {author} {\bibfnamefont {M.}~\bibnamefont {Horie}}\ and\ \bibinfo {author} {\bibfnamefont {N.}~\bibnamefont {Mitsume}},\ }\href {http://arxiv.org/abs/2205.11912} {\enquote {\bibinfo {title} {Physics-{Embedded} {Neural} {Networks}: {Graph} {Neural} {PDE} {Solvers} with {Mixed} {Boundary} {Conditions}},}\ } (\bibinfo {year} {2023}),\ \bibinfo {note} {arXiv:2205.11912 [cs]}\BibitemShut {NoStop}%
\bibitem [{\citenamefont {Li}\ and\ \citenamefont {Farimani}(2022)}]{li_graph_2022}%
  \BibitemOpen
  \bibfield  {author} {\bibinfo {author} {\bibfnamefont {Z.}~\bibnamefont {Li}}\ and\ \bibinfo {author} {\bibfnamefont {A.~B.}\ \bibnamefont {Farimani}},\ }\bibfield  {title} {\enquote {\bibinfo {title} {Graph neural network-accelerated {Lagrangian} fluid simulation},}\ }\href {\doibase https://doi.org/10.1016/j.cag.2022.02.004} {\bibfield  {journal} {\bibinfo  {journal} {Computers \& Graphics}\ }\textbf {\bibinfo {volume} {103}},\ \bibinfo {pages} {201--211} (\bibinfo {year} {2022})}\BibitemShut {NoStop}%
\bibitem [{\citenamefont {Peng}\ \emph {et~al.}(2023)\citenamefont {Peng}, \citenamefont {Hua}, \citenamefont {Li}, \citenamefont {Chen}, \citenamefont {Wu},\ and\ \citenamefont {Aubry}}]{peng2023physics}%
  \BibitemOpen
  \bibfield  {author} {\bibinfo {author} {\bibfnamefont {J.-Z.}\ \bibnamefont {Peng}}, \bibinfo {author} {\bibfnamefont {Y.}~\bibnamefont {Hua}}, \bibinfo {author} {\bibfnamefont {Y.-B.}\ \bibnamefont {Li}}, \bibinfo {author} {\bibfnamefont {Z.-H.}\ \bibnamefont {Chen}}, \bibinfo {author} {\bibfnamefont {W.-T.}\ \bibnamefont {Wu}}, \ and\ \bibinfo {author} {\bibfnamefont {N.}~\bibnamefont {Aubry}},\ }\bibfield  {title} {\enquote {\bibinfo {title} {Physics-informed graph convolutional neural network for modeling fluid flow and heat convection},}\ }\href@noop {} {\bibfield  {journal} {\bibinfo  {journal} {Physics of Fluids}\ }\textbf {\bibinfo {volume} {35}} (\bibinfo {year} {2023})}\BibitemShut {NoStop}%
\bibitem [{\citenamefont {Sanchez-Gonzalez}\ \emph {et~al.}(2018)\citenamefont {Sanchez-Gonzalez}, \citenamefont {Heess}, \citenamefont {Springenberg}, \citenamefont {Merel}, \citenamefont {Riedmiller}, \citenamefont {Hadsell},\ and\ \citenamefont {Battaglia}}]{sanchez2018graph}%
  \BibitemOpen
  \bibfield  {author} {\bibinfo {author} {\bibfnamefont {A.}~\bibnamefont {Sanchez-Gonzalez}}, \bibinfo {author} {\bibfnamefont {N.}~\bibnamefont {Heess}}, \bibinfo {author} {\bibfnamefont {J.~T.}\ \bibnamefont {Springenberg}}, \bibinfo {author} {\bibfnamefont {J.}~\bibnamefont {Merel}}, \bibinfo {author} {\bibfnamefont {M.}~\bibnamefont {Riedmiller}}, \bibinfo {author} {\bibfnamefont {R.}~\bibnamefont {Hadsell}}, \ and\ \bibinfo {author} {\bibfnamefont {P.}~\bibnamefont {Battaglia}},\ }\bibfield  {title} {\enquote {\bibinfo {title} {Graph networks as learnable physics engines for inference and control},}\ }in\ \href@noop {} {\emph {\bibinfo {booktitle} {International Conference on Machine Learning}}}\ (\bibinfo {organization} {PMLR},\ \bibinfo {year} {2018})\ pp.\ \bibinfo {pages} {4470--4479}\BibitemShut {NoStop}%
\bibitem [{\citenamefont {Sanchez-Gonzalez}\ \emph {et~al.}(2020)\citenamefont {Sanchez-Gonzalez}, \citenamefont {Godwin}, \citenamefont {Pfaff}, \citenamefont {Ying}, \citenamefont {Leskovec},\ and\ \citenamefont {Battaglia}}]{sanchez2020learning}%
  \BibitemOpen
  \bibfield  {author} {\bibinfo {author} {\bibfnamefont {A.}~\bibnamefont {Sanchez-Gonzalez}}, \bibinfo {author} {\bibfnamefont {J.}~\bibnamefont {Godwin}}, \bibinfo {author} {\bibfnamefont {T.}~\bibnamefont {Pfaff}}, \bibinfo {author} {\bibfnamefont {R.}~\bibnamefont {Ying}}, \bibinfo {author} {\bibfnamefont {J.}~\bibnamefont {Leskovec}}, \ and\ \bibinfo {author} {\bibfnamefont {P.}~\bibnamefont {Battaglia}},\ }\bibfield  {title} {\enquote {\bibinfo {title} {Learning to simulate complex physics with graph networks},}\ }in\ \href@noop {} {\emph {\bibinfo {booktitle} {International Conference on Machine Learning}}}\ (\bibinfo {organization} {PMLR},\ \bibinfo {year} {2020})\ pp.\ \bibinfo {pages} {8459--8468}\BibitemShut {NoStop}%
\bibitem [{\citenamefont {Seo*}, \citenamefont {Meng*},\ and\ \citenamefont {Liu}(2020)}]{seo2019physics}%
  \BibitemOpen
  \bibfield  {author} {\bibinfo {author} {\bibfnamefont {S.}~\bibnamefont {Seo*}}, \bibinfo {author} {\bibfnamefont {C.}~\bibnamefont {Meng*}}, \ and\ \bibinfo {author} {\bibfnamefont {Y.}~\bibnamefont {Liu}},\ }\bibfield  {title} {\enquote {\bibinfo {title} {Physics-aware difference graph networks for sparsely-observed dynamics},}\ }in\ \href {https://openreview.net/forum?id=r1gelyrtwH} {\emph {\bibinfo {booktitle} {International Conference on Learning Representations}}}\ (\bibinfo {year} {2020})\BibitemShut {NoStop}%
\bibitem [{\citenamefont {Han}\ \emph {et~al.}(2022)\citenamefont {Han}, \citenamefont {Gao}, \citenamefont {Pffaf}, \citenamefont {Wang},\ and\ \citenamefont {Liu}}]{han2022predicting}%
  \BibitemOpen
  \bibfield  {author} {\bibinfo {author} {\bibfnamefont {X.}~\bibnamefont {Han}}, \bibinfo {author} {\bibfnamefont {H.}~\bibnamefont {Gao}}, \bibinfo {author} {\bibfnamefont {T.}~\bibnamefont {Pffaf}}, \bibinfo {author} {\bibfnamefont {J.-X.}\ \bibnamefont {Wang}}, \ and\ \bibinfo {author} {\bibfnamefont {L.-P.}\ \bibnamefont {Liu}},\ }\bibfield  {title} {\enquote {\bibinfo {title} {Predicting physics in mesh-reduced space with temporal attention},}\ }\href@noop {} {\bibfield  {journal} {\bibinfo  {journal} {arXiv preprint arXiv:2201.09113}\ } (\bibinfo {year} {2022})}\BibitemShut {NoStop}%
\bibitem [{\citenamefont {Vaswani}\ \emph {et~al.}(2017)\citenamefont {Vaswani}, \citenamefont {Shazeer}, \citenamefont {Parmar}, \citenamefont {Uszkoreit}, \citenamefont {Jones}, \citenamefont {Gomez}, \citenamefont {Kaiser},\ and\ \citenamefont {Polosukhin}}]{vaswani_attention_nodate}%
  \BibitemOpen
  \bibfield  {author} {\bibinfo {author} {\bibfnamefont {A.}~\bibnamefont {Vaswani}}, \bibinfo {author} {\bibfnamefont {N.}~\bibnamefont {Shazeer}}, \bibinfo {author} {\bibfnamefont {N.}~\bibnamefont {Parmar}}, \bibinfo {author} {\bibfnamefont {J.}~\bibnamefont {Uszkoreit}}, \bibinfo {author} {\bibfnamefont {L.}~\bibnamefont {Jones}}, \bibinfo {author} {\bibfnamefont {A.~N.}\ \bibnamefont {Gomez}}, \bibinfo {author} {\bibfnamefont {{\L}.}~\bibnamefont {Kaiser}}, \ and\ \bibinfo {author} {\bibfnamefont {I.}~\bibnamefont {Polosukhin}},\ }\bibfield  {title} {\enquote {\bibinfo {title} {Attention is all you need},}\ }\href@noop {} {\bibfield  {journal} {\bibinfo  {journal} {Advances in neural information processing systems}\ }\textbf {\bibinfo {volume} {30}} (\bibinfo {year} {2017})}\BibitemShut {NoStop}%
\bibitem [{\citenamefont {Chen}, \citenamefont {Hachem},\ and\ \citenamefont {Viquerat}(2021)}]{chen2021graph}%
  \BibitemOpen
  \bibfield  {author} {\bibinfo {author} {\bibfnamefont {J.}~\bibnamefont {Chen}}, \bibinfo {author} {\bibfnamefont {E.}~\bibnamefont {Hachem}}, \ and\ \bibinfo {author} {\bibfnamefont {J.}~\bibnamefont {Viquerat}},\ }\bibfield  {title} {\enquote {\bibinfo {title} {Graph neural networks for laminar flow prediction around random two-dimensional shapes},}\ }\href@noop {} {\bibfield  {journal} {\bibinfo  {journal} {Physics of Fluids}\ }\textbf {\bibinfo {volume} {33}} (\bibinfo {year} {2021})}\BibitemShut {NoStop}%
\bibitem [{\citenamefont {He}, \citenamefont {Wang},\ and\ \citenamefont {Li}(2022)}]{he2022flow}%
  \BibitemOpen
  \bibfield  {author} {\bibinfo {author} {\bibfnamefont {X.}~\bibnamefont {He}}, \bibinfo {author} {\bibfnamefont {Y.}~\bibnamefont {Wang}}, \ and\ \bibinfo {author} {\bibfnamefont {J.}~\bibnamefont {Li}},\ }\bibfield  {title} {\enquote {\bibinfo {title} {Flow completion network: Inferring the fluid dynamics from incomplete flow information using graph neural networks},}\ }\href@noop {} {\bibfield  {journal} {\bibinfo  {journal} {Physics of Fluids}\ }\textbf {\bibinfo {volume} {34}} (\bibinfo {year} {2022})}\BibitemShut {NoStop}%
\bibitem [{\citenamefont {Wu}, \citenamefont {Ma},\ and\ \citenamefont {Zhou}(2015)}]{wu_vortical_2015}%
  \BibitemOpen
  \bibfield  {author} {\bibinfo {author} {\bibfnamefont {J.-Z.}\ \bibnamefont {Wu}}, \bibinfo {author} {\bibfnamefont {H.-Y.}\ \bibnamefont {Ma}}, \ and\ \bibinfo {author} {\bibfnamefont {M.-D.}\ \bibnamefont {Zhou}},\ }\href {\doibase 10.1007/978-3-662-47061-9} {\emph {\bibinfo {title} {Vortical {Flows}}}}\ (\bibinfo  {publisher} {Springer Berlin Heidelberg},\ \bibinfo {address} {Berlin, Heidelberg},\ \bibinfo {year} {2015})\BibitemShut {NoStop}%
\bibitem [{\citenamefont {Brandstetter}, \citenamefont {Worrall},\ and\ \citenamefont {Welling}(2022)}]{brandstetter2022message}%
  \BibitemOpen
  \bibfield  {author} {\bibinfo {author} {\bibfnamefont {J.}~\bibnamefont {Brandstetter}}, \bibinfo {author} {\bibfnamefont {D.}~\bibnamefont {Worrall}}, \ and\ \bibinfo {author} {\bibfnamefont {M.}~\bibnamefont {Welling}},\ }\bibfield  {title} {\enquote {\bibinfo {title} {Message passing neural pde solvers},}\ }\href@noop {} {\bibfield  {journal} {\bibinfo  {journal} {arXiv preprint arXiv:2202.03376}\ } (\bibinfo {year} {2022})}\BibitemShut {NoStop}%
\bibitem [{\citenamefont {Elfwing}, \citenamefont {Uchibe},\ and\ \citenamefont {Doya}(2018)}]{elfwing2018sigmoid}%
  \BibitemOpen
  \bibfield  {author} {\bibinfo {author} {\bibfnamefont {S.}~\bibnamefont {Elfwing}}, \bibinfo {author} {\bibfnamefont {E.}~\bibnamefont {Uchibe}}, \ and\ \bibinfo {author} {\bibfnamefont {K.}~\bibnamefont {Doya}},\ }\bibfield  {title} {\enquote {\bibinfo {title} {Sigmoid-weighted linear units for neural network function approximation in reinforcement learning},}\ }\href@noop {} {\bibfield  {journal} {\bibinfo  {journal} {Neural networks}\ }\textbf {\bibinfo {volume} {107}},\ \bibinfo {pages} {3--11} (\bibinfo {year} {2018})}\BibitemShut {NoStop}%
\bibitem [{\citenamefont {Chen}\ \emph {et~al.}(2020)\citenamefont {Chen}, \citenamefont {Lin}, \citenamefont {Li}, \citenamefont {Li}, \citenamefont {Zhou},\ and\ \citenamefont {Sun}}]{chen_measuring_2020}%
  \BibitemOpen
  \bibfield  {author} {\bibinfo {author} {\bibfnamefont {D.}~\bibnamefont {Chen}}, \bibinfo {author} {\bibfnamefont {Y.}~\bibnamefont {Lin}}, \bibinfo {author} {\bibfnamefont {W.}~\bibnamefont {Li}}, \bibinfo {author} {\bibfnamefont {P.}~\bibnamefont {Li}}, \bibinfo {author} {\bibfnamefont {J.}~\bibnamefont {Zhou}}, \ and\ \bibinfo {author} {\bibfnamefont {X.}~\bibnamefont {Sun}},\ }\bibfield  {title} {\enquote {\bibinfo {title} {Measuring and {Relieving} the {Over}-{Smoothing} {Problem} for {Graph} {Neural} {Networks} from the {Topological} {View}},}\ }\href {\doibase 10.1609/aaai.v34i04.5747} {\bibfield  {journal} {\bibinfo  {journal} {Proceedings of the AAAI Conference on Artificial Intelligence}\ }\textbf {\bibinfo {volume} {34}},\ \bibinfo {pages} {3438--3445} (\bibinfo {year} {2020})}\BibitemShut {NoStop}%
\bibitem [{\citenamefont {Rubanova}\ \emph {et~al.}(2021)\citenamefont {Rubanova}, \citenamefont {Sanchez-Gonzalez}, \citenamefont {Pfaff},\ and\ \citenamefont {Battaglia}}]{rubanova2021constraint}%
  \BibitemOpen
  \bibfield  {author} {\bibinfo {author} {\bibfnamefont {Y.}~\bibnamefont {Rubanova}}, \bibinfo {author} {\bibfnamefont {A.}~\bibnamefont {Sanchez-Gonzalez}}, \bibinfo {author} {\bibfnamefont {T.}~\bibnamefont {Pfaff}}, \ and\ \bibinfo {author} {\bibfnamefont {P.}~\bibnamefont {Battaglia}},\ }\bibfield  {title} {\enquote {\bibinfo {title} {Constraint-based graph network simulator},}\ }\href@noop {} {\bibfield  {journal} {\bibinfo  {journal} {arXiv preprint arXiv:2112.09161}\ } (\bibinfo {year} {2021})}\BibitemShut {NoStop}%
\bibitem [{\citenamefont {Thuerey}\ \emph {et~al.}(2020)\citenamefont {Thuerey}, \citenamefont {Weissenow}, \citenamefont {Prantl},\ and\ \citenamefont {Hu}}]{thuerey_deep_2020}%
  \BibitemOpen
  \bibfield  {author} {\bibinfo {author} {\bibfnamefont {N.}~\bibnamefont {Thuerey}}, \bibinfo {author} {\bibfnamefont {K.}~\bibnamefont {Weissenow}}, \bibinfo {author} {\bibfnamefont {L.}~\bibnamefont {Prantl}}, \ and\ \bibinfo {author} {\bibfnamefont {X.}~\bibnamefont {Hu}},\ }\bibfield  {title} {\enquote {\bibinfo {title} {Deep {Learning} {Methods} for {Reynolds}-{Averaged} {Navier}-{Stokes} {Simulations} of {Airfoil} {Flows}},}\ }\href {\doibase 10.2514/1.j058291} {\bibfield  {journal} {\bibinfo  {journal} {AIAA Journal}\ }\textbf {\bibinfo {volume} {58}},\ \bibinfo {pages} {25--36} (\bibinfo {year} {2020})},\ \bibinfo {note} {arXiv:1810.08217 [physics, stat]}\BibitemShut {NoStop}%
\end{thebibliography}%

\clearpage

\section*{Supplementary Material}
\subsection{ADDITIONAL MODEL DETAILS}

\subsubsection{Model Architecture Setup}

We maintained the same number of message passing layers (15) and hidden size (128) as used in Meshgraphnets. For FVGN, the encoder $\phi^{c}, \phi^{e}$, processor $\phi^{cp}, \phi^{ep}$, and decoder $\phi^{d}$ MLPs all use the SiLU activation function (MeshGraphNets use ReLU activation function, the effectiveness of this change can also be observed in Fig.\ref{Hybrid_UV_comp} and Fig.\ref{Hybrid_P_comp}.), with two hidden layers each. Except for the output size of $\phi^{d}$, which must match the predictions of $\mathbf{u}, p$, the remaining MLPs maintain a hidden space size of 128.

\begin{algorithm*}
\caption{Training Process}
\SetKwInOut{Input}{input}\SetKwInOut{Output}{output}
\KwData{A set $\mathfrak{\Huge  G}$ contains $\left \{ \left [ G_{v}^{k}(V,E_{v}),G_{c}^{k}(C,E_{c}) \right ],..., \left [G_{v}^{t}(V,E_{v}),G_{c}^{t}(C,E_{c}) \right ] \mid 0<k<t \right \}$, each graph pair belongs to a certain time step $k$ of a trajtory}
\Input{Randomly draw a batch of graph pair $\left \{\left [ G_{v}^{k}(V,E_{v}),G_{c}^{k}(C,E_{c}) \right ]...\right \}$ from $\mathfrak{\Huge  G }$, $G_{v}^{k}(V,E_{v}) $ represents vertex-centered graph, and $G_{c}^{k}(C,E_{c}) $ represents cell-centered graph, $\mathbf{c}_i \in C$ represents cell attributes, $\mathbf{e}_{c_{ij}} \in E_{c}$ represents neighboring cell indices,  feed them into an encoder-processor-decoder architecture $FVGN(\mathcal{G} ,\theta)$ }
\Output{$FVGN(\mathcal{G} ,\theta)$ with learned parameter $\theta$}
\While{Not Convergence of parameters $\theta$}{
    accumulate the amount $n$ of samples at current batch\;
     $\mathbf{v}\in \mathbb{R} ^{V\times c} $, $\overline{\mathbf{u}}^{GT}_{t}$ $ \leftarrow $ Inject normal noise which has specified mean value and variance to cell attributes $\mathbf{u}^{GT}_{t}$, and put it into node features $\mathbf{c}\in \mathbb{R} ^{C}$ in $G_{c}^{k}(C, E_{c})$\;
    \If{$n < a$ specified number of accumulation times }{
    accumulate the variance $\hat{x}$ and the mean value $\overline{x}$ at each channel dimension of node features \; and edge features, using $ G_{c}$ from current batch}

    Normalize $\mathbf{c}$ and $\mathbf{e_c}$ using $\hat{x}$ and $\overline{x}$\;
    $\mathcal{G}(\mathbf{c'},\mathbf{e'}_c)$ $\leftarrow $ Encoder($G_{c}^{k}(C,E_{c})$)\;
    
    \For{m < messpassing layers}{
        \For{$i \in \left \{1...E_{v}^{n} \right \}$}{
            \textbf{let} $E^{'} = \left\{(\mathbf{e}^{'}_{c},\mathbf{c}^{\prime}_{i},\mathbf{c}^{\prime}_{j})\right\}$\;
            $\overline{\mathbf{v}^{'}}\leftarrow \rho^{e\rightarrow{v}}(E^{'})$ \Comment{\begin{tabular}[t]{@{}l@{}}Aggregate edge attributes per node in $G_{v}^{k}$\end{tabular}}\\
                \For{$j \in \left \{ triangles 1...triangles C \right \}$}{
                    \textbf{let} $C^{'}_{j} = \left\{(v^{1}_{j},v^{2}_{j},v^{3}_{j})\right\}$\;
                    $\mathbf{c}^{'}\leftarrow \rho^{v\rightarrow{c}}(C_j^{'})$  \Comment{\begin{tabular}[t]{@{}l@{}}Aggregate node attributes per cell in $G_{v}^{k}$\end{tabular}}\\
                    $\overline{\mathbf{c}}^{'}_{i} \leftarrow \phi^{cp}(\mathbf{c}^{'})$ \Comment{\begin{tabular}[t]{@{}l@{}}Compute updated cell attributes in mesh $G_{v}^{k}$\end{tabular}}\\
                }
            }
         \For{$e \in E_{c} $}{
            $\overline{\mathbf{e}_{c}}{\prime} \leftarrow \phi^{ep}(\mathbf{e}^{\prime}_{c},\overline{\mathbf{c}}_{i}^{\prime},\overline{\mathbf{c}}^{\prime}_{j})$ \Comment{\begin{tabular}[t]{@{}l@{}}Compute updated edge attributes in $G_{c}^{k}$\end{tabular}}\\
            }
        }
        
    $\left \{ \mathbf{u}_{f},p_f,q_x,q_y \right \}_{e}$ $\leftarrow$ Decoder($G_{c}^{k}(C,E_{c})$)\;
    
    Continuity equation $loss_{cont}$ at each cell center $\leftarrow$ Integrate $ \mathbf{u}_f $ to cell center per cell using $G_{c}^{k}(C,E_{c})$ \;
    
    $\bigtriangleup \mathbf{u} $ at each cell center $\leftarrow$ Integrate $\left \{ \mathbf{u}_f,p_f,q_x,q_y \right \}_{e}$ to cell center per cell using $G_{c}^{k}(C,E_{c})$\;
    
    Calculate $loss_{cont}$ and $MSE( \left\{ \bigtriangleup \mathbf{u},\mathbf{u}^{GT}_{t+dt}-\overline{\mathbf{u}}^{GT}_{t}\right\} )$ \;
}
\end{algorithm*}

Additionally, except for $\phi^{d}$, all MLP outputs are normalized through LayerNorm. All input cell features $c$ and edge features $e_{i,j}$ for FVGN are normalized using the method proposed in MeshGraphNets. During training, both MeshGraphNets and FVGN use a one-step training mode and optimize parameters with the Adam optimizer. It is important to note that the MeshGraphNets model was trained on a single v100s GPU using the Adam optimizer for 10 million training steps, with an exponential learning rate decay from $10^{-4}$ to $10^{-6}$ at the 5M-th training step. All these hyperparameters related to the MeshGraphnets model are consistent with those provided by the original authors, and MeshGraphnets (directed) also uses the same settings.

\subsubsection{TRAINING PROCESS \& TRAINING NOISE}\label{TRAINING PROCESS TRAINING NOISE}

We used the same training noise injection strategy as \cite{sanchez2020learning} and MeshGraphNets to enhance the model's robustness in long-term prediction scenarios. Since the variance of the injected noise is closely related to the range of $u, v, p$ values in the dataset, in this study, the same method for finding noise variance as in \cite{sanchez2020learning} and MeshGraphNets was adopted when training with the second dataset HYBRIDFLOW. The final result was a noise variance of 2e-2, consistent with that used in CYLINDERFLOW dataset.

\end{document}